\documentclass[10pt, twocolumn]{article}
\usepackage[utf8]{inputenc}
\usepackage{amsmath}
\usepackage{multirow}
\usepackage{multicol}
\usepackage{mathtools}
\usepackage{amssymb}
\usepackage{booktabs}
\usepackage{bbm}
\usepackage{xurl}%
\usepackage{float}
\usepackage{tabularx}
\usepackage[style=nature]{biblatex}
\usepackage[font=footnotesize]{caption,subcaption} 
\usepackage{graphicx}
\usepackage[affil-it]{authblk}
\usepackage{array} %
\usepackage{pdflscape}
\usepackage{verbatim}
\usepackage{placeins} %
\usepackage[dvipsnames,table,xcdraw,svgnames]{xcolor}
\usepackage{pdfpages}
\usepackage{lineno} %
\usepackage[percent]{overpic} %
\usepackage{chngcntr} %
\usepackage{fullpage}
\newcolumntype{C}[1]{>{\centering\let\newline\\\arraybackslash\hspace{0pt}}m{#1}}

\renewcommand*{\thefootnote}{\fnsymbol{footnote}}
\usepackage{pdfpages}

\title{The unequal effects of the health-economy tradeoff 

during the COVID\nobreakdash-19 pandemic$^{\star}$}

\author[1,2]{Marco Pangallo$^{\star,}$}
\author[3]{Alberto Aleta}
\author[4]{R. Maria del Rio Chanona}
\author[4]{Anton Pichler}
\author[5]{David Mart\'{i}n-Corral}
\author[6,7]{Matteo Chinazzi}
\author[8]{Fran\c{c}ois Lafond}
\author[9]{Marco Ajelli}
\author[5,10]{Esteban Moro}
\author[2,3,4]{Yamir Moreno}
\author[6]{Alessandro Vespignani}
\author[8,11]{J. Doyne Farmer}

\affil[1]{Institute of Economics and Department EMbeDS, 
Sant'Anna School of Advanced Studies, Pisa, Italy}
\affil[2]{CENTAI Institute, Turin 10138, Italy}
\affil[3]{Institute for Biocomputation and Physics of Complex Systems (BIFI) and Department of Theoretical Physics, University of Zaragoza, Zaragoza 50018, Spain}
\affil[4]{Complexity Science Hub, Vienna, Austria}
\affil[5]{Department of Mathematics and GISC, Universidad Carlos III de Madrid, Leganes, Spain}
\affil[6]{MOBS Lab, Northeastern University, Boston (MA), US}
\affil[7]{The Roux Institute, Northeastern University, Portland (ME), US}
\affil[8]{Institute for New Economic Thinking and Mathematical Institute, Oxford, UK}
\affil[9]{Laboratory for Computational Epidemiology and Public Health, Department of Epidemiology and Biostatistics, Indiana University School of Public Health, Bloomington, IN, USA}
\affil[10]{Connection Science, Institute for Data Science and Society, MIT, Cambridge (MA), US}
\affil[11]{Santa Fe Institute, Santa Fe (NM), US}

\date{\today}

\defbibfilter{appendixOnlyFilter}{
  segment=1 %
  and not segment=0 %
}

\begin{document}

\twocolumn[
  \begin{@twocolumnfalse}
    \maketitle
    \begin{abstract}
      The potential tradeoff between health outcomes and economic impact has been a major challenge in the policy making process during the COVID\nobreakdash-19 pandemic. Epidemic-economic models designed to address this issue are either too aggregate to consider heterogeneous outcomes across socio-economic groups, or, when sufficiently fine-grained, not well grounded by empirical data. To fill this gap, we introduce a data-driven, granular, agent-based model that simulates epidemic and economic outcomes across industries, occupations, and income levels with geographic realism. The key mechanism coupling the epidemic and economic modules is the reduction in consumption demand due to fear of infection.  We calibrate the model to the first wave of COVID\nobreakdash-19 in the New York metropolitan area, showing that it reproduces key epidemic and economic statistics, and then examine counterfactual scenarios. We find that: (a) both high fear of infection and strict restrictions similarly harm the economy but reduce infections; (b) low-income workers bear the brunt of both the economic and epidemic harm; (c) closing non-customer-facing industries such as manufacturing and construction only marginally reduces the death toll while considerably increasing unemployment; and (d) delaying the start of protective measures does little to help the economy and worsens epidemic outcomes in all scenarios. We anticipate that our model will help designing effective and equitable non-pharmaceutical interventions that minimize disruptions in the face of a novel pandemic.
    \end{abstract}
    \vspace{1em}
  \end{@twocolumnfalse}
]

\footnotetext[1]{Author contributions: MP proposed the project; all authors designed research; MP and AA performed research, with contributions from MdRC, AP and DMC; MP and AA wrote the first draft. All authors contributed to the results interpretation, wrote and edited the paper. MP is the corresponding author.}

\renewcommand*{\thefootnote}{\arabic{footnote}}

Since the outset of the COVID\nobreakdash-19 pandemic, governments worldwide have successfully slowed down the transmission of SARS-CoV-2 by enacting non-pharmaceutical interventions (NPIs) \cite{flaxman2020estimating}. These interventions include the shutdown, to different degrees and extents, of non-essential customer-facing economic activities, e.g. entertainment and restaurants, and the mandating of work-from-home policies. Such protective measures have heterogeneous outcomes across socio-economic groups (``distributional effects''): workers who work in non-essential industries or can work from home become less likely to get exposed to the pathogen while essential, in-person workers remain at a higher risk of exposure. At the same time, NPIs have different distributional economic effects depending on the industry and occupation of the workers. For example, low-income workers are more likely to work in customer-facing industries and perform in-person occupations, leading to high risk of unemployment when these industries are closed \cite{del2020supply, chetty2020economic}.

Together with NPIs, the COVID\nobreakdash-19 pandemic induced very strong behavioral change, as individuals voluntarily reduced their contacts and their consumption of customer-facing services out of fear they would get infected. How effective exactly this mechanism is compared to NPIs remains debated \cite{goolsbee2021fear,farboodi2021internal,krueger2022macroeconomic}, and whether behavior change has distributional consequences like NPIs do, is an open problem.

Addressing the effectiveness of NPIs over behavior change, both at the aggregate and distributional level, requires building theoretical, mechanistic models that jointly simulate epidemic and economic dynamics at a fine-grained level. Most of the models that have been put forward \cite{ash2022disease,eichenbaum2021macroeconomics,kaplan2020great,alvarez2021simple,baqaee2020policies}, however, are quite aggregate on the epidemic and/or economic dimension, and so are not fully equipped to study distributional outcomes. There are a few Agent-Based Models (ABMs) that simulate epidemic spreading and economic decisions at the level of individual, heterogeneous agents \cite{delli2020abc, basurto2020economic,mellacher2020covid}, but these models are mostly meant to qualitatively evaluate different policies, with a basic parameter calibration that matches only a few aggregate moments of the data.

In this paper, we introduce an ABM that simulates epidemic and economic outcomes of a large synthetic population in a metropolitan area. The socio-economic characteristics of the agents and their consumption and contact patterns are initialized from detailed census, survey and mobility data, while the structure of the economy is initialized from input-output tables and national and regional accounts. Our joint epidemic-economic ABM merges and largely extends our former epidemic \cite{aleta2020modelling,aleta2022quantifying} and economic \cite{del2020supply,pichler2021and} COVID\nobreakdash-19 models.

\begin{figure}[!h]
    \centering
    \includegraphics[width = 0.48\textwidth]{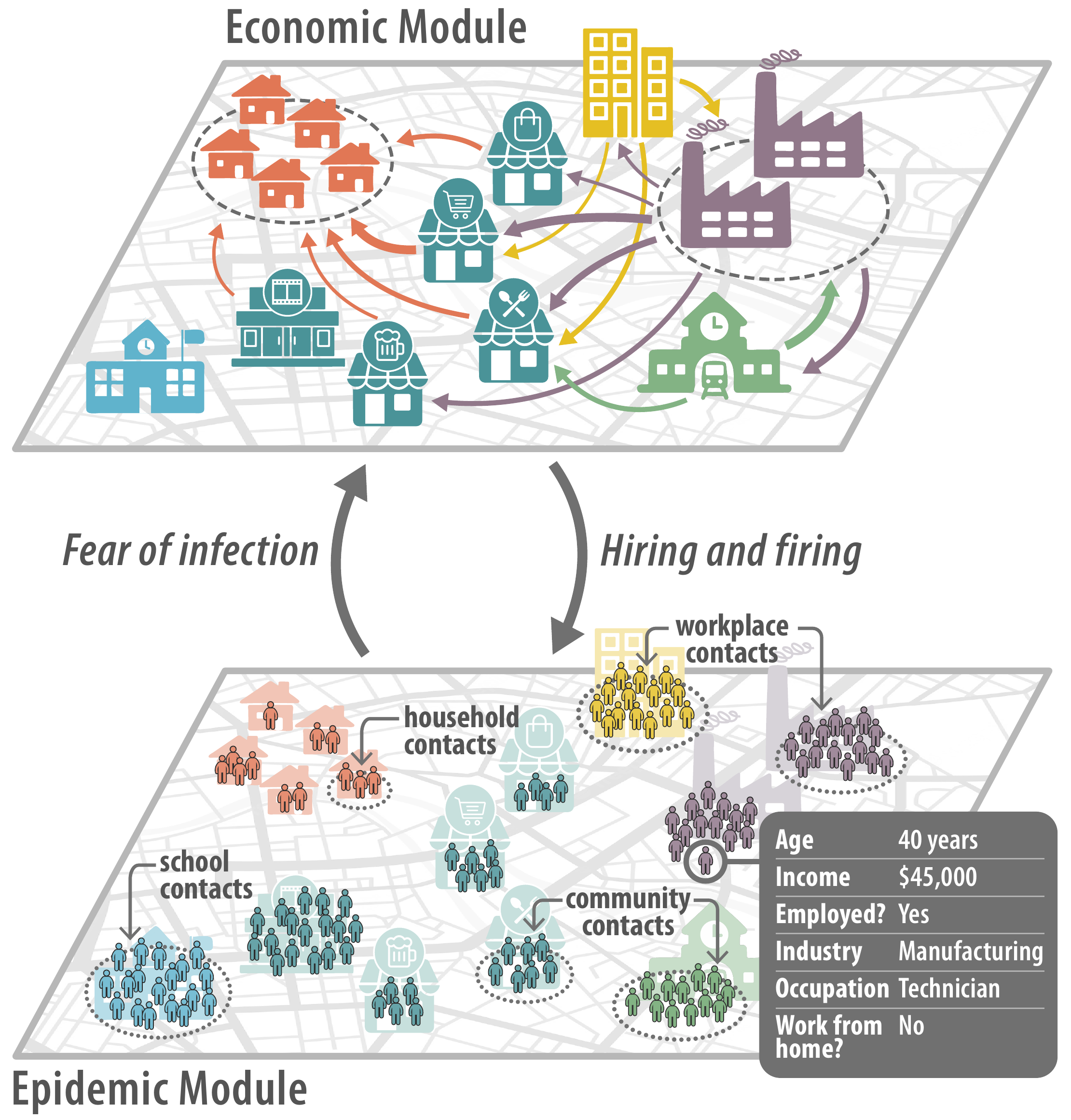}
    \caption{\textbf{Schematic representation of the joint epidemic-economic model.} In the economic module, goods and services flow between industries and to final consumers (input-output network). In the epidemic module, individual agents are exposed to the pathogen in the workplace, in community/consumption venues, in schools and households (contact network). Individuals are highly heterogeneous across the several socio-economic characteristics shown in the box. The economic and epidemic modules are tightly coupled: The epidemic module affects the economic module through reductions in consumption due to fear of infection. Conversely, the economic model affects the epidemic model due to changes in workplace and community contacts triggered by hiring and firing of workers from different industries.  }
        \label{fig:schematic_model}
\end{figure}

\paragraph{The epidemic-economic model.} We build a data-driven, granular, ABM of the New York-Newark-Jersey City, NY-NJ-PA Metro Area. The main agents of the model are the 416,442 individuals of a synthetic population that is representative of the real population across multiple socio-economic characteristics, including household composition, age, income, occupation and possibility to work from home (see Figure \ref{fig:schematic_model} for a schematic representation, and Materials and Methods and Supplementary Information for a detailed description of the model). 

The epidemic module of the ABM is built on the contact network that connects synthetic individuals. This network has multiple layers, where each layer captures interactions occurring (i) in the household; (ii) in school; (iii) in the workplace; (iv) in the community (during on-site consumption, such as in shops, restaurants or movie theaters). The epidemic propagates on these contact networks. We initialize contacts between individuals in the workplace and community layer by using a longitudinal database of detailed, privacy-enhanced mobility data from anonymized users who have opted-in to provide access to their (mobile phone-based) location data, through a GDPR-compliant framework provided by Cuebiq. At a daily level, we observe workplace visitation patterns among our panel of users. We are also able to estimate the probability of co-location of users in the same community places, as obtained from a dataset provided by Foursquare. On top of the contact networks and synthetic population, we run a stochastic, discrete-time transmission model in which individuals transition from one epidemic state to the other according to the distributions of key time-to-event intervals (e.g., incubation period, generation time) as per available data on SARS-CoV-2 transmission.

From an economic point of view, individuals play a role as workers and consumers. They work in one of multiple industries, producing goods and services that are either sold to other industries as intermediate products or sold to final consumers as consumption products. The economic module is particularly detailed on employment and consumption. Hiring and firing depend on industries' workforce needs, closures of economic activities and the possibility for workers to work from home. Consumption is heterogeneous for agents with different age and income, and changes depending on the state of the pandemic as households reduce the consumption demand for customer-facing industries due to fear of infection (customer-facing industries are: entertainment, accommodation-food, other services, retail, transportation, health, education; see Supplementary Section \ref{sec:apx_def_customer_contact}). The model also considers the input-output network of intermediates that industries use to produce final goods and services \cite{miller2009input}, thus considering the propagation of COVID\nobreakdash-19 shocks to the entire economy.

\begin{figure*}[!h]
    \centering
\includegraphics[width = 0.9\textwidth]{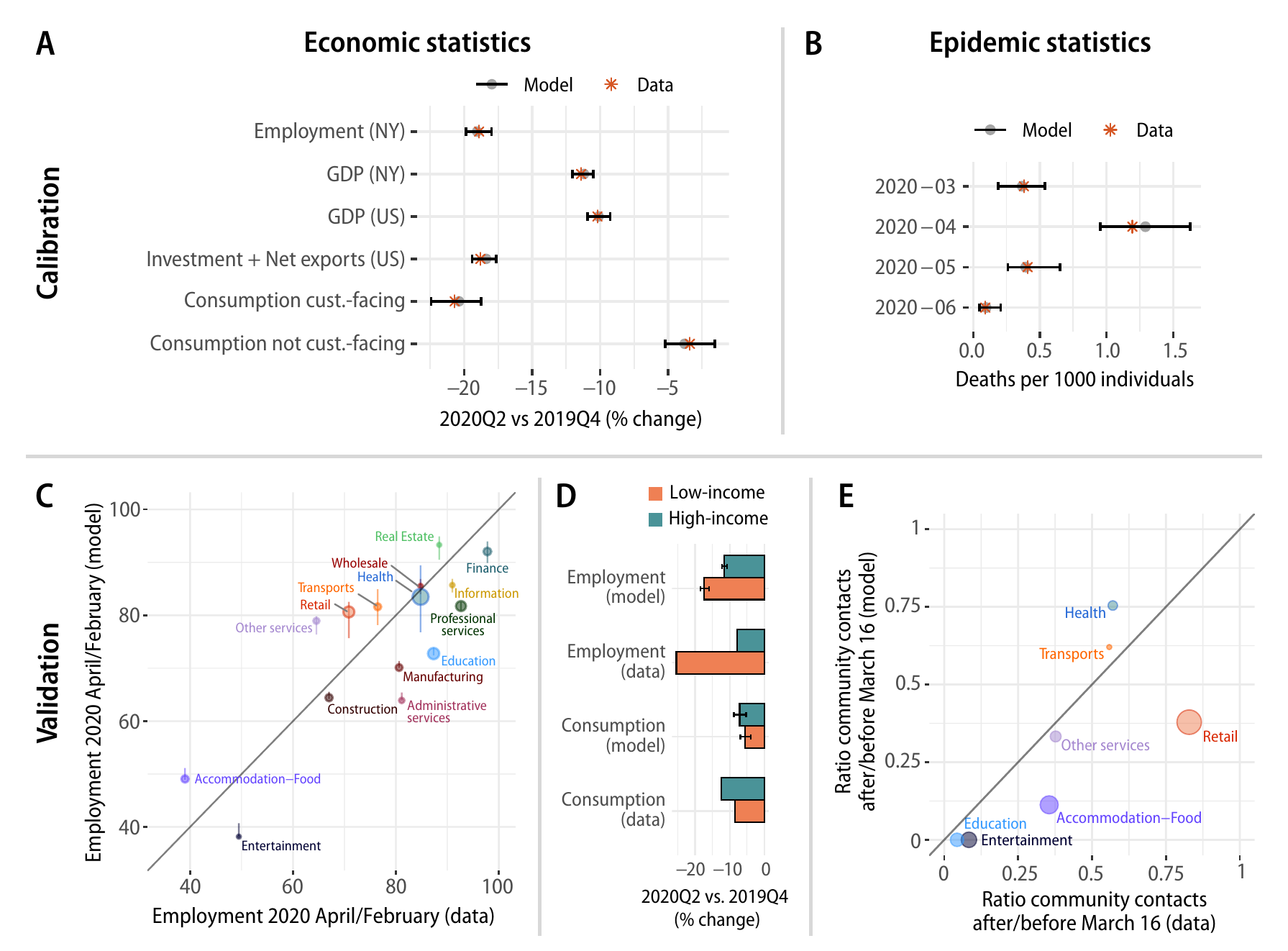}
   \caption{\textbf{The first wave of COVID\nobreakdash-19 in New York: empirical scenario.} The top panels show statistics that were directly targeted in the calibration, while the bottom panels show validation results for statistics that were not directly targeted. A: Percentage change from October-December 2019 (2019Q4) to April-May 2020 (2020Q2) across six official economic statistics, in the model and in the data (Supplementary Section \ref{apx:calibration}). Here, and throughout the paper, error bars indicate the 2.5-97.5 percentiles across simulation runs that differ by stochastic factors, see Materials and Methods. B: Comparison between model and data for the number of COVID\nobreakdash-19 deaths, which is the key epidemic statistic that we targeted. C: Employment in April 2020 as a percentage of employment in February 2020, across the main 2-digit NAICS industries, in the model and in the data. Circle size is proportional to employment in February 2020. D: Employment and consumption, in the model and in the data \cite{chetty2020economic}, among low-income and high-income households (low-income: $<\$27,000$; high-income: $>\$60,000$; these bands are chosen for comparison to real data, see Supplementary Section \ref{apx:validation_data}).  E: Ratio between community contacts with infectious individuals (Supplementary Section \ref{apx:empirical_epidemic_results}) before and after the imposition of protective measures, in the model and in the data, for the seven customer-facing industries (Supplementary Section \ref{sec:apx_def_customer_contact}). Circle size is proportional to the share of pre-pandemic contacts.}
        \label{fig:validation}
\end{figure*}

\paragraph{Modeling the first wave of COVID\nobreakdash-19 in New York.} We calibrate the key parameters of the model, including the strength of behavior change, which we name \textit{fear of infection}, so that the model fits some key epidemic and economic statistics of the first wave in the NY metro area (Supplementary Section \ref{apx:calibration}). We also calibrate the epidemiological parameters to the ancestral SARS-CoV-2 lineages (Supplementary Table \ref{table:parameters}). We start our simulations on February 12, 2020; impose a number of protective measures on March 16; relax them on May 15; end our simulations on June 30. As protective measures, we close schools, mandate work from home, and shut down all non-essential economic activities, such as entertainment and most of the accommodation-food industry, but also large parts of manufacturing and construction. We use the official NY regulations to estimate the degree to which a given industry is essential (Supplementary Section \ref{apx:supplyshocks}) and assume that workers who can work from home are not directly affected by these closures \cite{del2020supply}. We name this set of assumptions \textit{the empirical scenario}.

\paragraph{Economic validation.} Our model accurately matches the six official economic statistics we calibrated it on (Figure \ref{fig:validation}A). It correctly reproduces the fact that employment declined more strongly than Gross Domestic Product (GDP) (this is because industries most affected by shutdown orders produce less output per worker). It also correctly reproduces the fact that consumption of goods and services produced by customer-facing industries declined more strongly than consumption of goods and services produced by industries that are not customer-facing, ranging from manufacturing products to utilities and financial services (Supplementary Figure \ref{fig:supp_empirical_consumption_by_industry}). Our model also predicts some empirical properties on which it has not been calibrated (Figure \ref{fig:validation}C, D). First, thanks to our estimate of pandemic shocks \cite{del2020supply} and shock propagation model \cite{pichler2021and}, we are able to predict industry-specific changes in employment induced by the pandemic (Figure \ref{fig:validation}C), with a Pearson correlation coefficient of 0.82 (p-value:  $2 \cdot 10^{-4}$) between model and data. Second, thanks to our granular and data-driven characterization of employment and consumption patterns (Supplementary Figures \ref{fig:plots_joint_occ_ind} and \ref{fig:marginals_age_income}), we reproduce a key fact: low-income individuals were more likely to become unemployed but reduced consumption less than high-income individuals \cite{chetty2020economic,cox2020initial} (Figure \ref{fig:validation}D). This happens because low-income individuals are more likely to work in the industries and occupations most affected by closures (Supplementary Figure \ref{fig:supp_empirical_employment_by_occupation}), but they spend a larger share of their income on essential goods and services such as housing and utilities (we do not consider here the effect of the stimulus program, which would further increase the spending of low-income individuals). 

\paragraph{Epidemic validation.} On the epidemic side, our model correctly matches the death count data on which it has been calibrated, correctly replicating the spike in the number of reported deaths in April 2020 and the strong reduction in June (Figure \ref{fig:validation}D, Supplementary Figure \ref{fig:supp_empirical_epidemic_validation_time_series}). It also correctly predicts the changes in contact patterns that occurred after protective measures were implemented, although it was not calibrated on these data (Figure \ref{fig:validation}E). Both in the model and in the data, community contacts substantially reduced (Pearson: 0.75, p-value: 0.05), more in mostly non-essential industries such as entertainment and restaurants than in mostly essential industries such as retail and health. We also accurately predict the reduction in workplace contacts across industries (Supplementary Figure \ref{fig:alt_contacts_plot_workplace}; Pearson: 0.88, p-value: $5 \cdot 10^{-6}$), the temporal profile of reduction in contacts (Supplementary Figure \ref{fig:supp_empirical_contacts_validation}) and the increase in prevalence over time (Supplementary Figure \ref{fig:supp_empirical_epidemic_validation_time_series}). Finally, the model makes a number of predictions about how many infections happen across each layer and industry over time, as well as which occupation, income and age groups are most affected (Supplementary Figures \ref{fig:supp_empirical_epidemic} to \ref{fig:supp_empirical_infections_by_age_by_origin}). While we are not able to find data to quantitatively evaluate these predictions, our literature review provides some support to these findings (Supplementary Section \ref{apx:empirical_epidemic_evidence}).

\begin{figure*}[!h]
    \centering
\includegraphics[width = 1\textwidth]{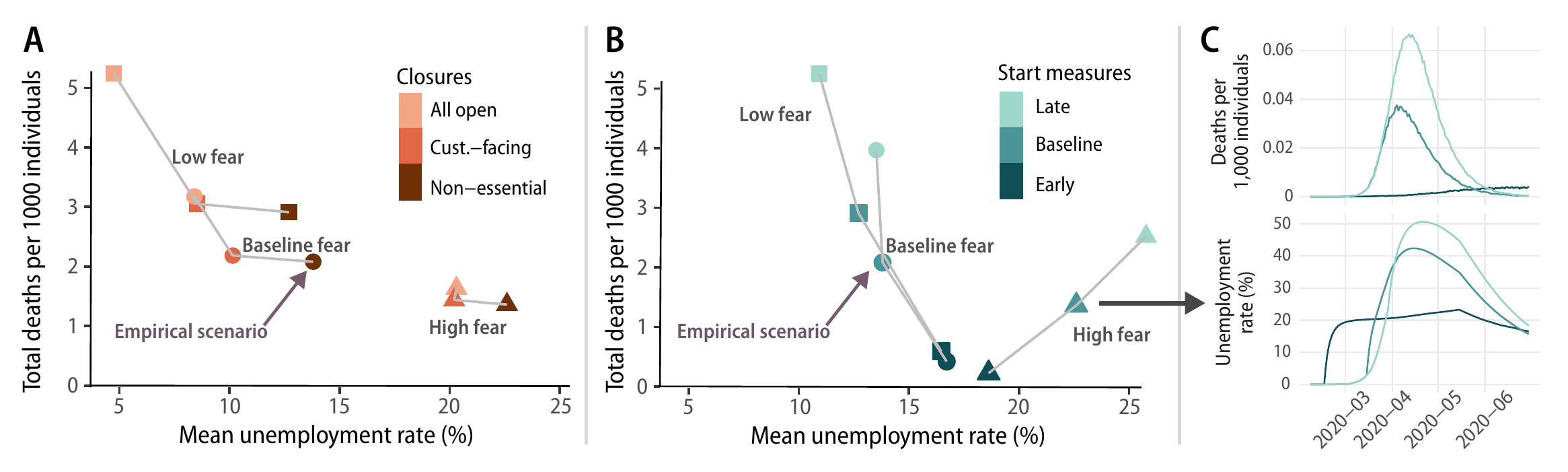}
    \caption{\textbf{Aggregate results on counterfactuals.} Deaths and unemployment across scenarios. For each scenario, we show the aggregate unemployment rate and the cumulative number of deaths, as averaged throughout the simulation period and the simulation runs (Supplementary Figure \ref{fig:supp_counterfactuals_aggregate_distributions} shows the variability across simulation runs and discusses its interpretation). The empirical scenario is highlighted to serve as a benchmark. Scenarios are distinguished by the strength of behavior change, as exemplified by the fear of infection parameter (square: low; circle: baseline; triangle: high). Panel A further distinguishes scenarios by the specific closure of economic activities (all non-essential industries, as occurred empirically; only customer-facing industries; and no closures), keeping the start of protective measures fixed at the baseline, empirically observed date. Panel B distinguishes the cases the start of protective measures (baseline: 2020-03-16, as empirically; early: 2020-02-17; late: 2020-03-30), keeping closures fixed at all non-essential industries. For the specific combination of high fear of infection and three different starts of protective measures, panel C shows time series of unemployment and deaths corresponding to the three scenarios.
    }
        \label{fig:main_counterfactuals_aggregate}
\end{figure*}

\paragraph{Counterfactual  scenarios.} We vary three factors that characterize the behavioral and policy response to the first wave of the COVID\nobreakdash-19 pandemic. Throughout we use the term ``baseline'' to refer to the estimated parameter values that match the empirical data.  First, we vary the strength of behavior change, which is represented by the fear of infection parameter. Our baseline calibration gives us a distribution for this parameter (Supplementary Figure \ref{fig:posterior_pars}) that implies that consumption demand of customer-facing industries was reduced by 14\% exclusively due to fear of infection at the epidemic peak. We cannot give a causal interpretation to this calibrated value, because it conflates the effect of NPIs with the effect of behavior change --- if there were no NPIs, consumption demand would have needed to decline much more for the model to explain observed consumption changes. To address this issue, we examine two counterfactuals, in which we set the fear of infection parameter distribution to 0.1 (``low'') and 10 (``high'') times the baseline. These values imply that, at the epidemic peak, consumption demand was reduced respectively by 1\% and 77\% exclusively due to fear of infection, representing reasonable lower and upper bounds. 

We then vary two factors related to policy. First, we explore the effect of different closures of economic activities. In addition to the closure scenario that we used in our baseline calibration, in which all non-essential industries were closed, we consider two milder closure scenarios: (i) only non-essential customer-facing industries are closed, and (ii) there are no closures, i.e. all economic activities are open. Second, we examine what would have happened if protective measures started four weeks earlier (2020-02-17) or two weeks later (2020-03-30). (In the Supplementary Information we also consider more counterfactuals that include partial closure of customer-facing industries, as well as not imposing work from home and not closing schools, see Supplementary Figures \ref{fig:supp_counterfactuals_aggregate_all_closures} and \ref{fig:supp_counterfactuals_aggregate_all_schools_wfh}.)

\begin{figure*}[!h]
    \centering
\includegraphics[width = 1\textwidth]{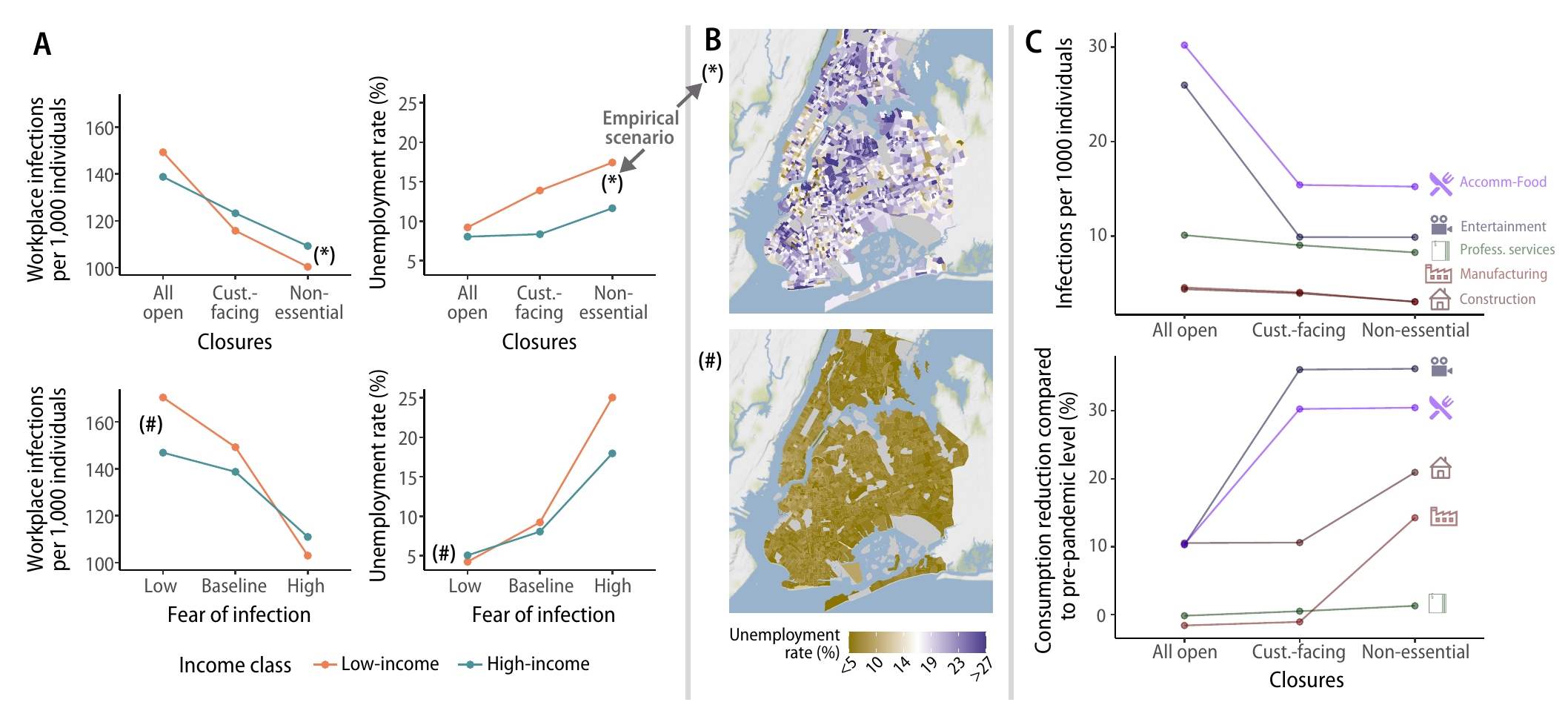}
    \caption{\textbf{Results on counterfactuals disaggregated by income, geography and industry.} A: Workplace infections and unemployment across income classes. In the top panels, we vary the level of closures, keeping fear of infection and the start of protective measures to their baseline values (so the case with all non-essential activities closed is the empirical scenario). In the bottom panels, we vary fear of infection keeping all economic activities open and starting protective measures on the baseline date.  B: maps of unemployment across census tracts in New York City, corresponding to two scenarios in panel A, including the empirical scenario (asterisk) and the counterfactual with no closures and low fear (hash). C: Infections and reduction in consumption across five selected industries and three levels of closures, for baseline fear and start of protective measures. }
        \label{fig:main_counterfactuals_disaggregate}
\end{figure*}

\paragraph{Behavior change vs. NPIs.} Aggregate economic and epidemic results are shown in Figure \ref{fig:main_counterfactuals_aggregate}, while results disaggregated by income, geography and industry are shown in Figure \ref{fig:main_counterfactuals_disaggregate} (see also Supplementary Figures \ref{fig:supp_counterfactuals_disaggregate_employment_consumption_by_income} to \ref{fig:supp_counterfactuals_disaggregate_infections_by_occupation}). Figure \ref{fig:main_counterfactuals_aggregate}A conveys our first key result: stricter closure of economic activities leads to more unemployment and fewer COVID\nobreakdash-19 deaths, and the same effect takes place with higher fear of infection. To see this, start from the combination with baseline fear of infection and opening of all economic activities, corresponding to the light-colored circle. Holding fear of infection constant at the baseline value, closure of all non-essential economic activities (as in the baseline scenario) leads to 64\% higher unemployment rate, but to a 35\% lower number of deaths. At the same time, keeping the level of closures to the empirical baseline but assuming high fear of infection (dark triangle) leads to 40\% higher unemployment and 50\% lower deaths compared the empirical scenario. The situation across other scenarios is similar. (Variability of total deaths and mean unemployment across simulation runs can be substantial, but the relative effects of different policies are robust; see Supplementary Figure \ref{fig:supp_counterfactuals_aggregate_distributions}.)

At the distributional level, both higher fear of infection and stricter closures lead to saving lives at the expense of jobs, for both low and high income workers  (Figure \ref{fig:main_counterfactuals_disaggregate}A). However, for low income workers, higher fear of infection or stricter closures have a more dramatic effect, leading to more lives saved and more jobs lost, compared to high-income workers, for whom the effects of these changes are milder. As we will show later, outside the household setting, most infections occur in customer-facing industries, where most low-income workers are concentrated. Thus, mandated closure or spontaneous avoidance of these industries leads to both more unemployment and fewer workplace infections among low-income workers.

 This also leads to dramatic spatial disparities. Focusing on unemployment, Figure \ref{fig:main_counterfactuals_disaggregate}B shows two maps of unemployment in Manhattan in the empirical scenario (asterisk) and in a counterfactual with low fear and no closures (hash). We see that in the counterfactual the unemployment rate is very evenly spatially distributed, while in the empirical scenario low-income areas such as the Queens and the Bronx have a high unemployment rate of more than 20\%,  compared to high-income areas such as Manhattan, with unemployment rates around 15\%.

Overall, our work addresses the controversial debate around the effectiveness of behavior change vs. NPIs in saving both lives and the economy. Our results indicate a qualitative similarity between strong behavior change and strict closure of economic activities: spontaneously avoiding consumption of services provided by customer-facing industries, like mandated closure of these industries, increases unemployment but saves lives, and this is especially true among low-income individuals. Therefore, media campaigns aimed at increasing fear of infection are likely to have qualitatively similar effects to explicit restrictions of economic activities.

\paragraph{Industry-specific closures.} Another question that our model can address is the effectiveness of closing all non-essential economic activities (including large parts of manufacturing and construction) vs. just closing customer-facing industries. Our key finding is that closing all non-essential activities leads to marginally fewer deaths than closing only customer-facing industries, while greatly increasing unemployment. Indeed, a counterfactual with the empirical baseline fear of infection but closure of only customer-facing industries has a 4\% higher death rate but a much lower 36\% unemployment rate. To explain these results, consider Figure \ref{fig:main_counterfactuals_disaggregate}C. Among the five industries shown, most infections are concentrated in the customer-facing ``Entertainment'' and ``Accommodation-Food''. When these industries are closed, infections decrease substantially at the cost of a strong decrease in consumption (and employment). Conversely, closing ``Manufacturing'' greatly reduces consumption of manufacturing goods, but since the share of infections in manufacturing is small, closing manufacturing only marginally decreases infections. The result is similar for ``Construction''. ``Professional services'', in contrast, are not much affected by different closures because they are either essential or can be largely performed from home. 

From a methodological point of view, we could obtain industry-specific results because we linked each consumption venue as recorded in the mobility data to an economic activity (Supplementary Section \ref{apx:subPOI}), quantifying contacts across industries. This way, our granular, data-driven model leads to insights that would have been missed in more aggregate, qualitative models.

\paragraph{Timing of interventions.} The final question that we consider is the effectiveness of starting protective measures earlier (four weeks before) or later (two weeks later) than in the empirical scenario. As we show in Figure \ref{fig:main_counterfactuals_aggregate}B, starting protective measures later than the empirical baseline leaves unemployment only about 2\% lower while greatly increasing the number of deaths by 50\%. Surprisingly, with high fear of infection, starting protective measures late leads to both 46\% more deaths \textit{and} 12\% more unemployment than the baseline. The mechanism for these results is suggested in Figure \ref{fig:main_counterfactuals_aggregate}C, where we show time series across the three counterfactuals with high fear of infection. We see that an early start of protective measures prevents an epidemic wave, leading to no further increase in unemployment due to fear of infection. Conversely, with a baseline or late start, strong behavior change leads to reduced consumption, and this in turn leads industries to fire their employees, increasing unemployment.

\paragraph{Discussion}
Mitigating the health outcomes of the COVID\nobreakdash-19 pandemic entailed immense societal and economic costs, spurring heated debates. According to some, restrictions and protective measures were essential to curb the epidemic and there was no tradeoff between economy and health, because there could be no economic recovery without reducing virus circulation to very low levels. According to others, behavioral responses could be more effective than imposed protective measures to control epidemic dynamics; letting individuals spontaneously reduce their risk of exposure when the epidemic situation worsened would have lead to optimal epidemic and economic outcomes. 

 Our results suggest an equivalence between behavioral response and closure of economic activities: both strong behavioral response and strict closures increase unemployment and reduce infections. This effect is more pronounced among low-income workers than among high-income workers.  Our results also show that in most scenarios there is some tradeoff between health and the economy, but this tradeoff depends strongly on which economic activities are closed. Closing industries that are not customer-facing, such as manufacturing and construction, leads to a substantial increase in unemployment while  marginally reducing deaths. Finally, when fear of infection is high, starting protective measures late leads to \textit{both} more deaths and more unemployment, and in the other cases starting late increases deaths a lot while marginally reducing unemployment.

A huge body of scientific work has been studying the effectiveness and equitability of pandemic control policies since the outset of the COVID\nobreakdash-19 pandemic. Our paper studies both distributional epidemic and economic outcomes while being firmly grounded in real-world data. This makes it possible to credibly address questions that other models could not consider. For instance, the result showing that high fear of infection affects low-income workers the most is possible through our data-driven mapping between the incomes, occupations and industries of workers. 

Our results have the usual limitations pertaining to modeling studies. In this paper we exclusively focus on the first wave of COVID\nobreakdash-19 in one specific metropolitan area. It could also be important to consider other  aspects of the COVID\nobreakdash-19 pandemic that became relevant after the first wave of infections such as masks, test, trace and quarantining, variants, vaccination, and waning of immunity. However, we expect our key results to hold, and we view our model as mostly applicable to the short-term management of emerging/re-emerging infectious diseases. Another important limitation is that the matching between synthetic individuals and mobility traces is probabilistic, as we do not have socio-economic information about specific Cuebiq users. Nonetheless, our privacy-preserving matching algorithm based on census tracts is likely to be accurate given the strong socio-economic disparities in different parts of the New York metro area. From the epidemiological standpoint, we assume the same per-contact risk of infection in different occupational settings. If empirical epidemiological data were collected about the contribution of these settings to SARS-CoV-2 transmission, our estimates could be further refined. Moreover, we did not consider differential risk of severe disease and death for individuals with different socio-economic status. The inclusion of this factor into the model could further exacerbate the highlighted heterogeneity in the health and economic impact of the pandemic and adopted policies on different segments of the population. From the economic standpoint, we consider industries located over the entire metro area, rather than heterogeneous firms at specific geographical locations. Although this is a limitation of our analysis, we believe that this is the right level of aggregation for the questions considered here, but we acknowledge that a more detailed representation of the production sector may be needed to address questions such as the effectiveness of spatially-targeted lockdowns. Finally, the infection transmission and economic models are combined through a ``fear'' mechanism that was modeled in a simple manner (i.e., as a function of the number of reported deaths on the previous day). For example, individuals may not retrieve information on a daily basis and media may amplify some information (e.g., a spike in the number of deaths) at certain times, thus altering the perception of the population \cite{poletti2011effect}; different segments of the population may have a different risk perception \cite{hodbod2021covid}. This highlights the importance of conducting future studies to better characterize the relation between risk perception and human behavior during epidemic outbreaks. 

Building on the model introduced in this paper could enrich both the epidemics literature and economic impact studies. From an epidemiological point of view, we introduce industries, occupations and possibility to work from home in a fine-grained transmission model, showing the usefulness of adding an economic dimension to models of epidemic spreading \cite{tizzoni2022addressing}. Studies on the economic impact of disasters often just focus on industries \cite{hallegatte2008adaptive,guan2020global}, or consider an aggregate representative household \cite{pichler2021and}. Having a synthetic population that is a highly-detailed representation of the real population, as we do in this paper, would make it possible to address a new set of distributional questions, e.g. the distributional effects of hurricanes.  

From a policy perspective, the results presented in this paper highlight the importance of targeted policies. Closure of customer-facing industries is highly effective at reducing epidemic spreading--especially when enacted early. With such a policy, income-support schemes could target specific occupational categories, such as food preparation and serving or personal care and services, rather than workers in general, such as those engaged in construction, maintenance, production, extraction and repair occupations.  Enhancing surveillance and contact tracing activities in industries where these low-income workers operate could also be particularly beneficial from both the health and economic perspectives. Crucially, these policies are not only needed when customer-facing industries are closed by the government, but also when spontaneous behavior change reduces consumption of goods and services produced by these industries. Our results could be instrumental to design policies aimed at reducing the health and economic impact of pandemics and epidemics as well as reducing inequalities by protecting low-income segments of the population.

\paragraph{Acknowledgements.}
MP, MdRC, and AP acknowledge funding from the James S. Mc Donnell Foundation Postdoctoral Fellowship Award. FL and JDF acknowledge funding from Baillie Gifford and the Institute for New Economic Thinking at the Oxford Martin School. AA acknowledges support through the grant RYC2021‐033226‐I funded by MCIN/AEI/10.13039/501100011033 and the European Union ``NextGenerationEU/PRTR''. Y.M was partially supported by the Government of Aragon, Spain and ``ERDF A way of making Europe'' through grant E36‐20R (FENOL), by Ministerio de Ciencia e Innovación, Agencia Espa\~nola de Investigación (MCIN/AEI/10.13039/501100011033) Grant No. PID2020‐115800GB‐I00,and by Soremartec S.A. and Soremartec Italia, Ferrero Group.  MA, MC and AV acknowledge the support of the from the HHS/CDC 6U01IP001137, HHS/CDC 5U01IP0001137. We acknowledge the use of the computational resources of COSNET Lab at Institute BIFI, funded by Banco Santander through grant Santander‐UZ 2020/0274 and by the Government of Aragon (FONDO–COVID19-UZ-164255).

\AtNextBibliography{\scriptsize}
\printbibliography[segment=0]
\newrefsegment

\clearpage

\begin{figure*}[!h]
    \centering
\includegraphics[width = 1\textwidth]{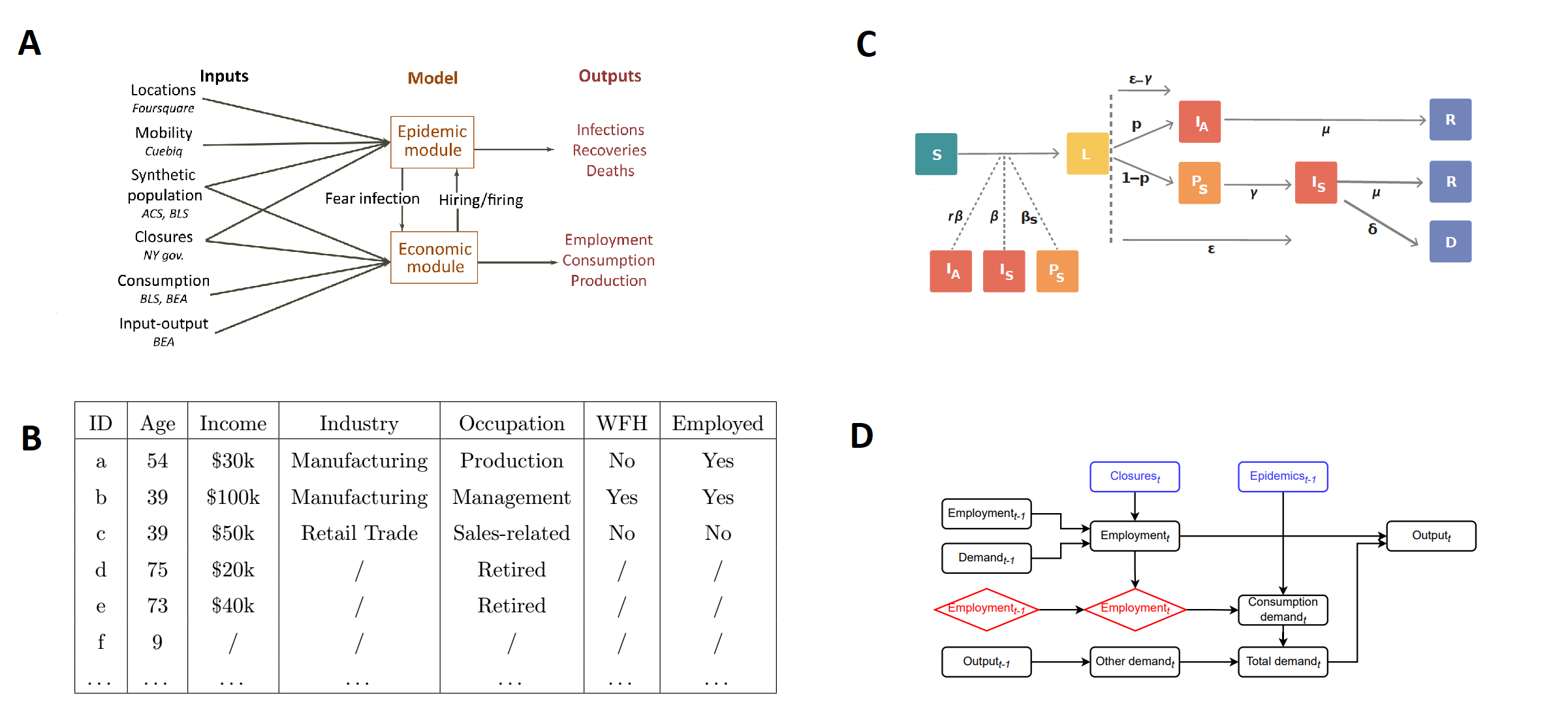}
    \caption{\textbf{Schematic of the methodology.} A:  Inputs and outputs of the epidemic and economic modules. For each group of input data, we list the data source. ACS=American Community Survey; BLS=Bureau of Labor Statistics; BEA=Bureau of Economic Analysis. The feedback from the epidemic to the economic module is through fear of infection, while a reverse feedback occurs through hiring and firing. B: Socio-economic characteristics of six example synthetic individuals, including age, income, industry, occupation, possibility to Work From Home (WFH), employment status. C: Schematic of the epidemic module. The different compartments and transition times from one compartment to another are shown. See the text for the meaning of the symbols. D: Causal diagram of the economic module. Blue rectangles show processes exogenous to the economic module; black rectangles represent industry-level endogenous variables; red diamonds are individual-level endogenous variables (in this case, the employment status of each individual). A link between variables indicates that the linking variable influences the linked variable. We only show the key variables. On the left we show lagged values of these variables at time $t-1$, while on the right we show the relations between these variables at time $t$.  }
        \label{fig:figureMaM}
\end{figure*}

\section*{Materials and methods}
\label{apx:overview}

{\footnotesize

This section provides an overview of the economic and epidemic modules, how they are coupled, which data we use to initialize them, and what outputs they produce (Figure \ref{fig:figureMaM}A). A longer description that gives all details and justifies all the assumptions can be found in the Supplementary Information.

\paragraph{Geography.} The epidemic-economic model focuses on the New York–Newark–Jersey City, NY–NJ–PA Metropolitan Statistical Area (FIPS code C3562), which we will often abbreviate as NY MSA. The NY MSA includes the highly urbanized area of New York City (i.e., the five boroughs of Manhattan, Bronx, Queens, Brooklyn and Staten Island), but also some rural and industrial areas. The total population, as of 2019, is 19,216,182 individuals, making up about 6\% of the U.S. population. Total GDP, as of 2019, is around \$1.5 trillion, a bit more than 7\% of U.S. GDP. A map of the NY MSA is shown in Supplementary Figure \ref{fig:tracts_by_income}. The epidemic module exclusively considers the NY MSA. This is because, while case importation from other parts of the U.S. and from abroad is an important factor in the early phase of a pandemic, it has little impact once the local incidence grows \cite{Russell2021Jan}. Our model is initialized in a period when local incidence was already important (see Section \ref{apx:epicalibration}), and so it is a valid approximation to only consider the NY MSA. The economic module models in detail the NY MSA area and also features a simplified model of the Rest of the U.S.: The NY MSA economy is deeply integrated with that of the US, and this integration must be taken into account throughout the pandemic to properly estimate economic impacts.

\paragraph{Agents.} The main agents of the epidemic-economic ABM are the 416,442 individuals of a synthetic population that is representative of the NY MSA (Supplementary Section \ref{apx:summary_synthpop}). The population size is determined by the availability of mobility data (see below). The agents are heterogeneous by several socio-economic characteristics (Figure \ref{fig:figureMaM}B), including age, income, employment status, occupation, possibility to work from home, and the census tract where they live. Individuals are grouped into 153,547 households, whose composition is consistent with census microdata. We derive socio-economic characteristics of synthetic individuals from tables provided by the American Community Survey (ACS) and the Bureau of Labor Statistics (BLS), trying to get as many joint distributions of variables as we can. For instance, to study distributional outcomes, it is important that synthetic individuals have the correct joint distribution of income, occupation and industry; this is important to replicate the empirical fact that managers working in the finance industry earn high income, while food preparation workers in the restaurants industry earn low income. To achieve this objective, we combine US-level BLS tables reporting incomes by industry-occupation pair with spatially detailed ACS tables giving incomes down to the census tract level. More details on the synthetic population building algorithm and validation tests can be found in Supplementary Section \ref{apx:synthpop}.

In the economic module we also treat industries as agents, considering a single representative firm per industry. We use the 2-digit NAICS level of aggregation, giving 20 different industries. Industries are mainly dependent on one another through the input-output network of consumption of intermediate goods. Since no official data for the NY MSA exist, we downloaded national data from the Bureau of Economic Analysis (BEA) and then used a regionalization method known as Flegg Location Quotient \cite{flegg2000regional} to obtain an input-output table that distinguishes between the NY MSA and the Rest of the US. The main idea behind this method is that a region that is more specialized in some good or service (such as information or finance in NY) can just rely on itself to source that good or service, but if it is less specialized (as for manufacturing in NY) it is more likely to import that good or service from the other(s) region. More details on our reconstruction of the input-output table can be found in Supplementary Section \ref{apx:io}.

\paragraph{Epidemic module.} The epidemic module is a standard epidemiological model that runs on top of a contact network extracted from GPS location data. Except for the integration with the economic module, the basic model is the same as the one described in \cite{aleta2022quantifying}. In this model, individuals interact through a contact network composed of four layers: (i) the community layer captures occasional interactions between individuals, for instance occurring in consumption venues; (ii) the workplace layer captures interactions between workers; (iii) the household layer captures interactions between household members; (iv) the school layer captures interactions between children attending the same school. To initialize contacts in the community and workplace layers to the pre-pandemic situation, we use privacy-preserving location intelligence data provided by Cuebiq, merging information about visits to Points of Interest (POI) with a large database by Foursquare that characterizes POIs. We devised a privacy-preserving algorithm to match Cuebiq users to synthetic individuals, mainly based on the census tract where they live (Supplementary Section \ref{apx:matching}). Our approach to reconstructing contacts is probabilistic: because we cannot observe co-location of individuals reliably in the data, we use mobility data to estimate the probability that any pair of individuals is in contact in a given day and in a certain venue (Supplementary Section \ref{apx:epi_model}). The contact networks are initialized using data on pre-pandemic mobility, and modified over time due to exogenous interventions, feedback from the economic module and fear of infection (as explained in the sections below).

We use a stochastic, discrete-time infection transmission model coupled to the contact network (Figure \ref{fig:figureMaM}C) that extends the classical Susceptible, Latent, Infected, Removed (SLIR) model. A susceptible individual ($S$) may become infected upon contacting an infectious individual, moving to the latent compartment ($L$). Three states describe individuals who are potentially infectious, each of them with their corresponding transmission rate: pre-symptomatic ($P_S$), with transmission rate $\beta_S$; infectious symptomatic, with rate $\beta$; and infectious asymptomatic ($I_A$), with rate $r\beta$. Contacts between infectious and susceptible individuals depend on the contact network estimated for each day. Therefore, the probability that a susceptible node $i$ gets infected by an infectious node $j$ in infectious compartment $type$ and place $p$ is:
\begin{equation}
    \label{eq:pSI}
    P(S_i + I_j \rightarrow L_i + I_j) = 1 - e^{-\beta_{\text{type}} w_{i,j,p} (t) \Delta t}
\end{equation}
where $\Delta t = 1 $ day and $w_{i,j,p}$ is a weight that modulates the effectiveness of contacts in a given setting in terms of spreading. We assume that all locations within the same layer (schools, workplaces, households and community) have the same weight, except for indoor/outdoor spaces in the community layer (see Supplementary Section \ref{apx:epicalibration}).

Once infected, the individual will enter the incubation compartment ($L$) for $\epsilon$ days, during which she will be infected but not infectious yet. A latent individual will become infectious $\gamma$ days before the end of the incubation period, to account for pre-symptomatic transmission. Whether an individual becomes symptomatic or not depends on the specific age-specific symptomatic probability, $p$. Lastly, the individual will be removed (R) from the infectious pool according to an exponential process with rate $\mu^{-1}$, where $\mu$ is the average length of the infectious period in days. Note that the removed compartment does not imply recovery, only that the individual is no longer able to infect. After $\delta$ days removed individuals might transition to the death compartment according to the empirical age-dependent Infection Fatality Ratio (IFR). A new death is reported $T_n$ days after the actual event to account for notification delays.

\paragraph{Economic module.} We introduce a dynamic macroeconomic model that is specifically suited to study the economic effects of COVID\nobreakdash-19, both at the macro-level of industries and at the micro-level of individuals and households. Figure \ref{fig:figureMaM}D shows the causal relations between the variables of the economic module. It distinguishes between variables that are exogenous to the economic module (blue rectangles), such as the epidemic trajectory or the shelter-in-place policies that lead to supply shocks, and endogenous variables. These are further distinguished into industry-level endogenous variables (black rectangles) and individual-level endogenous variables (red diamonds). Examples of industry-level variables include employment, output, consumption and total demand. The only individual-level variable that we consider is employment status, although we do consider other agent attributes (such as age) that are fixed within our simulation period. At every time step $t$:
\begin{enumerate}
    \item Industries decide the size of the workforce they need based on their past employment, past demand, and the current levels of restrictions (we do not consider labor shortages due to illness or quarantine, as they would be difficult to model and preliminary simulations showed that they were a second-order effect).  Conditional on the restrictions and on the previous employment status of individuals, industries decide which specific workers they hire or fire uniformly at random.
    \item Individuals, grouped into households, decide their consumption demand based both on fixed attributes such as age or income and on variable outcomes such as the situation of the epidemic, and on their employment status (workers who lose their job may cut back on spending). Aggregating over agents produces a total consumption demand for each industry. 
    \item Total final demand for each industry is obtained by summing up consumption demand, orders of intermediate goods from other industries, and other components of final demand (including government expenditures, investments, imports and exports).
    \item Finally, industries produce goods and services. Industries aim to produce as much as demanded, but their production can be limited by labor shortages (for simplicity, in this paper we do not consider intermediate inputs shortages, which were not a first-order effect in the first few months of the pandemic \cite{pichler2021and}). In the case of shortages, demand is rationed on a pro-rata basis across intermediate and final consumers.  
\end{enumerate}

\paragraph{Coupling the epidemic and economic modules.} The epidemic and economic modules are strongly coupled, in the sense that  at time $t$ both modules take as input the output that the other module generated at $t-1$. More specifically, the economic module takes the number of deaths reported in the epidemic module, $D_{t-1}$, as an input to compute reduction in consumption demand, as explained in the ``Fear of infection'' section below. At the same time, the epidemic module takes the employment status of each individual in the synthetic population as an input from the economic module. This information is used in the epidemic module as previously employed individuals who get fired can no longer get infected in the workplace, while previously unemployed individuals who are hired can get infected. From a technical point of view, the epidemic module is written in \texttt{C}, while the economic module is written in \texttt{Python}. To implement coupling between the two models, we use a \texttt{Python-C API} (\url{https://docs.python.org/3/c-api/}) that makes it possible to initialize a Python interpreter from within a \texttt{C} run.

\paragraph{Timeline.} A time step in the epidemic-economic model corresponds to one calendar day. Time effectively begins on February 12, 2020. On this date, the epidemic module starts running in calendar time (Supplementary Section \ref{apx:epicalibration}) and producing epidemic outcomes at a daily scale. The economic module starts in a steady state that represents the economic situation at the beginning of 2020. All our simulations finish at the end of June 2020, giving a total of 140 time steps. In the empirical scenario, where we use our model to reproduce what happened during the first wave of COVID\nobreakdash-19 in the NY MSA, we impose a number of exogenous interventions  (see below) to both the epidemic and economic modules on March 16, 2020. We choose that date for simplicity, as several interventions were imposed at different times by the States of New York and New Jersey from March 9 to March 23, and we see an abrupt change in the social dynamics in our data around that date \cite{aleta2022quantifying}. We remove the economic interventions on May 15, 2020, again as an approximation to the actual relaxation of protective measures that took place in the NY MSA during spring 2020. Other measures, such as the closure of schools and work from home, are kept in place until the end of the simulation. Similarly, for the modeling of counterfactuals we start protective measures at various times, ranging from February 17 to March 30, but still relax them on May 15, 2020.

\paragraph{Exogenous interventions.} During the COVID\nobreakdash-19 pandemic, governments and local authorities imposed a number of non-pharmaceutical interventions with the goal of reducing epidemic spreading. In this paper, we consider three types of interventions:
\begin{enumerate}
    \item Closure of economic activities. We assume that a certain fraction $s_{k,t}$ of industry $k$ at time $t$ can be exogenously shut down. This implies that a fraction $s_{k,t}$ of the in-person workers of industry $k$ cannot work at $t$, reducing economic output of industry $k$. In the epidemic module, we assume the same reduction $s_{k,t}$ of contacts in industry $k$. In the empirical scenario we reproduce the closures that occurred in the NY MSA in spring 2020 \cite{del2020supply}, and we name this set of closures ``Non-essential''. We consider alternative closing strategies when studying counterfactuals, such as closure of customer-facing industries only.
    \item Imposition of work from home. All workers who can work from home must do so. We assume that this has no impact on the economic module --a worker who can work from home is as productive at home as in the workplace-- but it reduces contacts and so infections in the workplace in the epidemic module.
    \item School closures. All contacts between children going to school are removed, effectively cutting off schools from disease transmission and thus reducing overall infections. We do not consider the impact of this policy on economic outcomes, due to the difficulty of calibrating the productivity loss related to childcare.
\end{enumerate}

\paragraph{Fear of infection.} Further to government interventions, a distinctive hallmark of the COVID\nobreakdash-19 pandemic has been behavior change: Due to fear of getting infected, several individuals reduced their in-person consumption and contacts in the community as the epidemic situation worsened. We follow the approach introduced in Ref. \cite{perra2011towards} and let behaviour change follow the functional form
\begin{equation}
    \Lambda_t (\phi, D_{t-1}) = 1-\exp\left(-\phi D_{t-1}\right), 
    \label{eq:behav_change_gen}
\end{equation} 
where $D_{t-1}$ is the number of daily reported deaths in the NY MSA on day $t-1$, and $\phi$ is a sensitivity parameter that we name \textit{fear of infection}. When $D_{t-1}=0$, $\Lambda_t=0$, so there is no behavior change. In contrast, when $D_{t-1}$ grows large there is strong behavior change, as $\Lambda_t \rightarrow 1$. Since $D_{t-1} \geq 0, \hspace{5pt} \forall t$, behaviour change increases with fear of infection. The exponential functional form in Eq. \eqref{eq:behav_change_gen} corresponds to a non-linearity in behavioral response that gives larger weight to an initial increase in $D_{t-1}$, with saturation afterwards \cite{perra2011towards}, in line with the behavioral response to the first wave. (See Supplementary Section \ref{sec:consumption} for a comparison between our approach and other modeling efforts for fear of infection during the COVID\nobreakdash-19 pandemic). 

For parsimony, we assume that fear of infection is constant. This implies that the dynamics of the behaviour change are driven by the evolution of the death rate. We also assume that fear of infection is homogeneous across individuals. While it could be expected that old, at-risk individuals intend to reduce their consumption and contacts the most, survey evidence suggests that this is unlikely to be a first-order effect, and in some cases it may be youngest individuals that intend to reduce their consumption the most \cite{hodbod2021covid} (of course, whatever one's intentions, socio-economic groups differ in their ability to actually avoid contacts, e.g. depending on whether individuals can work from home).

While behaviour change in both the epidemic and economic module is based on the functional form in Eq. \eqref{eq:behav_change_gen}, there are slight differences in how behaviour change affects consumption and workplace and community contacts.
\begin{itemize}
    \item Reduction in consumption (effect in the economic module). During the COVID\nobreakdash-19 pandemic, individuals reduced ``risky'' consumption of customer-facing services such as restaurants, cinemas, hairdressers, etc. However, they did not reduce consumption of financial and real estate services due to fear of infections. Most people kept paying rent, and even increased consumption of some manufacturing goods such as houseware. Therefore, we consider that fear of infection only decreased consumption in consumer facing industries. using the following functional form
    \begin{equation}
        \Lambda^{\text{ECO}}_{t, k}=\Lambda\left(\phi^{\text{ECO}}, D_{t-1}\right)\tau_k,
        \label{eq:fear_inf_ec}
    \end{equation}
    where $\tau_k$ is an indicator that takes value $\tau_k=1$ if industry $k$ is customer-facing, and $\tau_k=0$ if it is not. Section \ref{sec:apx_def_customer_contact} lists which industries are customer-facing, and explains our classification. The parameter $\phi^{\text{ECO}}$ is a fear of infection parameter specific to the economic module, as we detail below.
    
    \item Reduction in community contacts (effect in the epidemic module). As individuals reduce consumption of customer-facing services, they also decrease their contacts in the community, which has an effect on the epidemic module. However, the reduction in consumption is not identical to the reduction in community contacts. For instance, individuals may order take away meals from restaurants, thus reducing contacts but not consumption. We assume that reductions in consumption and community contacts are proportional, letting fear of infection in the epidemic module be given by
    \begin{equation}
        \phi^{\text{EPI}}= \phi^{\text{ECO}} / \Tilde{\phi},
    \end{equation}
    with $\Tilde{\phi}$ as a parameter giving the proportion between the two fear of infection parameters (see Section \ref{apx:calibration} for how we calibrate these parameters). The reduction in community contacts is then given by
    \begin{equation}
        \Lambda^{\text{EPI}}_{t,k}=\Lambda \left(\phi^{\text{EPI}}, D_{t-1}\right)\tau_k.
    \end{equation}
    \item Reduction in workplace contacts (effect in the epidemic module). If work from home is not imposed by the government or local authority, individuals may nevertheless decide to work from home if they can. We assume that this is uniform across industries, so that the reduction in workplace contacts among individuals who can work from home is given by
    \begin{equation}
        \Lambda^{\text{EPI}, \text{work}}_{t, k}=\Lambda \left(\phi^{\text{EPI}}, D_{t-1}\right).
    \end{equation}
    We assume no reduction in productivity from working from home --a worker who can work from home is as productive at home as in the workplace--, so this has no effect on the economic module.
\end{itemize}

\paragraph{Calibration and stochasticity.} For both the epidemic and economic results, uncertainty comes from: (i) stochasticity in the simulation runs, namely inherent stochasticity of transmission in the SLIR model and inherent stochasticity in the hiring/firing process of the economic module; (ii) uncertainty over parameter values, as obtained from the Approximate Bayesian Computation (ABC) calibration algorithm that we use to calibrate the seven parameters that we cannot pin down independently (see Section \ref{apx:calibration}). (This means that, in line with a Bayesian approach, we run simulations sampling from all parameter values accepted by ABC.)

}
\newpage

\onecolumn

\renewcommand{\appendixname}{Supplementary Information}
\appendix
\setcounter{section}{0}
\setcounter{figure}{0}
\setcounter{table}{0}

\makeatletter
\renewcommand \thesection{S\@arabic\c@section}
\renewcommand\thetable{S\@arabic\c@table}
\renewcommand \thefigure{S\@arabic\c@figure}
\makeatother

\begin{center}
{\Huge \textbf{Supplementary Information}}
\end{center}

\tableofcontents

\vspace{2cm}
\begin{itemize}
    \item Section \ref{apx:epi_model} gives details on the epidemic module, which is a modified SLIR model that is simulated at the level of individuals who are heterogeneous in their contacts. Given that we have a lot of information to build the contact network on which infections take place, but not detailed enough information to build it exclusively from the data, we explain our probabilistic approach to building a multilayer contact network using all information available as well as reasonable assumptions.
    \item Section \ref{apx:model} explains all details of the economic module, in particular its employment and consumption submodules. It clarifies the mechanisms that connect the macro level of industries to the micro level of individuals, as well as the reverse mechanism.
    \item Section \ref{apx:data} explains how we initialize variables and agent-specific parameters. In this paper, we use the word \textit{initialization} to refer to the setting of initial conditions of model quantities that change over time (variables), but also to setting individual- and industry-specific parameters. 
    \item Section \ref{apx:calibration} shows how we calibrate model-wide epidemic and economic parameters. We select some parameters directly from data or from other studies, while the remaining seven parameters are calibrated using Approximate Bayesian Computation (ABC).
    \item Section \ref{apx:first_wave} presents additional results on the empirical scenario that we cannot report in the main paper due to space constraints. These include more detailed results on infections, employment and consumption across income levels, occupations and industries.
    \item Section \ref{apx:scenarios} presents additional results on counterfactuals. These include partial closures of customer-facing industries, no school closures or imposition of work from home, and more detailed results across scenarios on infections, employment and consumption across income levels, occupations and industries.
\end{itemize}

Table \ref{tab:notation} shows the notation that we use throughout this supplementary material. This list is not exhaustive, as it misses some minor parameters or variables of both the epidemic and economic module, but it includes all the symbols that are used repeatedly.

\begin{table}[!h]
	\centering
	\small
			\caption{Notation. We distinguish between indexes (first block), variables and agent-level parameters (second block) and model-wide parameters (third block). }
	\label{tab:notation}
		\begin{tabular}{|C{3.2cm}|C{12.8cm}|}
			\hline
			Symbol & Name \\
			\hline
          $i,j$ & Individual \\
          $h$ & Household \\
          $k,l$ & Industry \\
          $o$ & Occupation \\
          $g$ & Group of agents with same consumption preference \\
          $p$ & Place \\
          $t$ & Time (day) \\
          \hline
          $\omega_{i,j,t}^C, \omega_{i,j,t}^W, \omega_{i,j}^H, \omega_{i,j}^S$ & Links between agents $i$ and $j$ in the Community, Workplace, Household and School layers  \\
          $D_t$ & Daily reported deaths \\
          $\Lambda^{ECO}_t$ & Reduction in consumption demand of customer-facing products  \\
          $\Lambda^{EPI}_t$ & Reduction in community and workplace contacts due to fear of infection  \\
          $s_{k,t}$ & Fraction of in-person workers of industry $k$ that cannot work due to restrictions \\
          $\mathcal{G}_t,\mathcal{S}_t$ & Shocks to government demand and other (investment and export) final demand \\
          $\tau_k$ & Indicator of whether industry $k$ is customer-facing \\
          $l_{k,t}^\text{P}, l_{k,t}^\text{H}$ & Number of in-person and from-home workers employed in industry $k$\\
          $x_{k,t}$ & Output of industry $k$  \\
          $O_{k,l,t}, Z_{k,l,t}$ & Orders and realized intermediate consumption by industry $l$ of goods and services produced by $k$  \\
          $A_{k,l}=Z_{k,l,0}/x_{l,0}$ & Technical coefficients \\
          $c_{k,t}^{d}$, $G_{k,t}^{d}$, $f_{k,t}^{d}$ & Consumption, government, and other final demand for products of industry $k$  \\
           $c_{k,t}, G_{k,t}, f_{k,t}$ & Realized private consumption, government consumption and other final demand for products of industry $k$ \\
           \hline
           $R_0$ & Reproduction number \\
           $r, \beta, \beta_S,k,\epsilon,p,\gamma,\mu,\delta$, IFR,$\theta$,$T_n$ & Other epidemic parameters \\
            $\phi^{EPI}$ & Fear of infection in the epidemic module \\
            $\phi^{ECO}$ & Fear of infection in the economic module \\
            $\Tilde{\phi}$ & Factor of proportionality between $\phi^{EPI}$ and $\phi^{ECO}$ \\
            $\phi^U$ & Fear of unemployment \\
            $\Delta s$ & Reallocation pararameter \\
            $\gamma_H, \gamma_F$ & Speed of hiring and firing \\
			\hline						
		\end{tabular}
\end{table}

\clearpage

\section{Epidemic module}
\label{apx:epi_model}

We use a standard stochastic, discrete-time infection model on top of a contact network extracted from GPS location data. The infection model is already described in Materials and Methods, so in this section we discuss our theoretical framework to define contact networks between individuals using available data and plausible assumptions (see also Ref. \cite{aleta2022quantifying}).

The primary route for transmission of SARS-CoV-2 is close contacts between individuals \cite{CDCairborne}. Individual-level contacts are very hard to measure at large scale and investigators have proposed the use of aggregated mobility metrics to infer the amount of contacts in the population \cite{Grantz2020Sep}. This, however, is not always a valid approach since it is possible for a person to travel to many distinct points of interest without actually coming into close contact with others \cite{Crawford2022Jan}. In contrast, anonymized mobile device geolocation data has been shown to be a good proxy to estimate the real contact rates in the population \cite{Crawford2022Jan}.

In this work, we model interaction between individuals as a network in which nodes represent the agents and there is a link between them on a given day if they might have been in close contact. Furthermore, this contact can be associated to the specific place (henceforth, point of interest or POI, see Section \ref{apx:subPOI}) in which it took place. We distinguish four different contexts - or layers - for these contacts: households, schools, workplaces and community, which captures all connections that do not belong to the others. This last context is the one that most benefits from the use of geolocation data, since it is the one most subject to random and occasional encounters.

Ideally, one should be able to extract the links between individuals in the community directly from geolocation data. However, there are intrinsic limitations in the technology that limit this approach: information is collected asynchronously, irregularly over time, dependent on the apps being used in the device... \cite{Crawford2022Jan}. As such, even though the mobility dataset we use is large, co-location events cannot be precisely quantified. Instead, we rely on a probabilistic approach to measure co-presence and build the contact network \cite{Crawford2022Jan,aleta2020modelling,aleta2022quantifying}. For more details on the characteristics of the data see Section \ref{apx:mobility}.

In order to explain better our approach let us consider the homogeneous mixing approach in a contact network perspective. We assume to have $N$ individuals who are homogeneously mixed. This implies that each individual is potentially in contact with anybody else. Thus, we have a connection $\omega_{ij} = 1$, among each pair of nodes. This implies that the rate of contacts $c_i$ for the individual $i$ is $c_i = \sum_j m\,\omega_{ij} = m(N-1)$, where $m$ is an appropriate factor ensuring that the number of average effective contacts per individual unit time in the system is equal to $\kappa$. This implies that
\begin{equation}
\label{response:eq1}
    \kappa = N^{-1} \sum_i c_i = N^{-1} \sum_{i,j} m\, \omega_{i,j}
\end{equation}
yielding
\begin{equation}
    m = \frac{\kappa}{N^{-1} \sum_{ij}\omega_{ij}} = \frac{\kappa}{N-1}
\end{equation}
This finally provides the usual expression for the rate of contact $\omega_{ij} = \kappa/(N -1)$, that is multiplied by the transmissibility per contact $\alpha$ to give the rate (or probability) of infection per contact. This finally leads to the force of infection
of a susceptible as
$$
P_{S\rightarrow I} =1-\left(1- \frac{\alpha\kappa}{N-1} \right)^I=1-\left(1- \frac{\beta}{N-1}\right)^I \simeq \frac{\beta I}{N},
$$
where $\beta = \alpha\kappa$ is the transmissibility used in homogeneous model and the last approximations are valid for very large $N$.

In order to go beyond the homogeneous assumption, from our data we can consider that individuals who are never visiting the same places are never in contact. This is additional information of which we are certain. So for each individual we can list each of the places $p$ that they visit and assume that we
can have a link between two individuals if they have the same place in their list
$\omega_{ij}^p = \delta_{i,p}\delta_{j,p}$, where $\delta_{i,p} = 1$ if the place $p$ is on the list of visited places of individual $i$ and zero otherwise. This step improves on the homogeneous assumption as it rules out possible contacts among individuals that can never
meet. Further we can consider that the potential contacts among individuals is
larger for individuals that can meet in more than one place. We can then define
$\omega_{i,j}= \sum_p \omega_{i,j}^p$ , thus considering that some individuals have more potential contacts. It is worth remarking that we are still considering that each potential contact has the same weight as in the homogeneous assumption. In order to define properly the contact rate/probability per unit time we need to use Eq.\ (\ref{response:eq1}) thus defining
\begin{equation}
    m = \frac{\kappa}{N^{-1} \sum_{i,j}\omega_{ij}} = \frac{\kappa}{\langle \omega_{ij}\rangle}
\end{equation}
where we defined $\langle \omega_{i,j}\rangle$ as the average weighted contacts among individuals.
This yields the effective rate of contact among individuals $i$ and $j$ as
\begin{equation}
   \omega'_{ij} = \frac{\kappa \sum_p \delta_{i,p} \delta_{j,p}}{\langle \omega_{ij}\rangle}   
\end{equation}
In order to improve further on this approach we can consider that places are not visited in a deterministic way. This implies that each individual has a probability to visit a specific place that is $1/n_{i,p}$, where $n_{i,p}$ is the number of places visited by the individual $i$ in a given period. We can therefore define
\begin{equation}
    \omega_{ij} = \sum_p \frac{1}{n_{i,p}}\frac{1}{n_{j,p}}.
\end{equation}
This approach still considers potential contacts only among individuals however with a weight that depends on the variability of places of each individual. As before the rate/probability of contact would be:
\begin{equation}
\omega'_{ij} = \frac{\kappa \sum_p n_{i,p}^{-1} n_{j,p}^{-1}}{\langle \omega_{ij}\rangle}     
\end{equation}
So far we did not consider at all the time spent in each location. We can therefore improve on the probability to be in a place by weighting the number of places $n_{i,p}$ by the time spent on average in each place. This finally leads to the expression:
\begin{equation}
    \omega_{ij} = \sum_p \frac{T_{i,p}}{T_i} \frac{T_{j,p}}{T_j}
\end{equation}
where $T_{i,p}$ is the time spent by individual $i$ at location $p$ and $T_i$ is equal to the sum of all time spent in places in the community by individual $i$. In this case the rate of interaction will be:
\begin{equation}
\label{response:eqfin}
      \omega'_{ij} = \frac{\kappa \sum_p \frac{T_{i,p}}{T_i} \frac{T_{j,p}}{T_j}}{\langle \omega_{ij}\rangle}.
\end{equation}
This is the expression we use in our work. It is important to stress that this expression is improving on the homogeneous assumption as it considers that effective contacts can occur only in places visited by both individuals, and considers that each contact is weighted by the probability for each individual to be in that place. The approach however does not account for concurrency of visits. In this respect it is still adopting an homogeneous perspective in that all places visited at any time corresponds in a potential contact.

The next steps to improve on this approach would be indeed to consider concurrency of visits. It is thus tempting to consider that each contact is weighted by $T_{i,p}/T$, where T would be the specific amount of time of the day. One could assume the 8 hours of the working time or the 24 hours cycle of the day. This is a tempting solution but introduces a number of issues. For instance the time that should be considered in the normalization depends on the places. As an example, restaurants have specific bracket of times during the day, and concurrency should be evaluated on specific hours of the day and specific days (e.g. the weekend). The same happens for places like movie theatres, museums etc. Furthermore, during the lockdown the concurrency normalization should all be re-evaluated to be consistent in their definition as the number of hours in the community of the population drastically changed. In other words, we are not sure if the simple normalization by a fixed number of hours although trying to capture the concurrency of contacts is actually introducing unwanted and uncontrolled biases. For this reason we decided to work with the approach of Eq.~(\ref{response:eqfin}), for which all the assumptions can be clearly stated and provides an obvious improvement with respect to the fully homogeneous assumption.

Using our probabilistic approach to detect contacts, we build our contact network in each of the layers:

\begin{itemize}

\item {\bf Community weighted contact network}. In the community layer contacts are built by estimating co-location of two individuals in the same POI. Specifically, the weight, $\omega^C_{ijt}$, of a link between individuals $i$ and $j$ within the community layer at day $t$ is computed according to the expression:
\begin{equation}
\label{sm:eq1}
\omega^C_{ijt} = \sum_p^n\frac{T_{ipt}}{T_{it}}\frac{T_{jpt}}{T_{jt}}, \qquad \forall i,j
\end{equation}
where $T_{ipt}$ is the total time that individual $i$ was observed at place $p$ in day $t$ and $T_{it}$ is the total time that individual $i$ has been observed at any place set within the community layer that day $t$. 

For robustness and computational reasons, we have included only links for which $\omega^{C}_{ijt} > 0.01$, removing 2.88\% of the original links. For other values of the threshold like $\omega^{C}_{ijt} > 0.005$ and $\omega^{C}_{ijt} > 0.02$ we would remove 1.19\% and 6.19\% of the links respectively. Note however that since those links have very small weights, our results for the epidemic spreading do not depend significantly of the threshold chosen provided that it is small.

\item {\bf Workplace weighted contact network}. For privacy reasons, our data is obfuscated around home and workplaces to the level of Census Block Groups (CBGs). To get a proxy of contacts at the workplace, we assume that all workers in the same Census Block Groups have a probability to interact. To account for the potential number of working places in that area, we weight that probability by the number of POIs at the same census block group. Therefore, the contact weight, $\omega^W_{ijt}$, of a link between individuals $i$ and $j$ in the workplace layer at day $t$ is given by:
\begin{equation}
\label{sm:eq2}
\omega^W_{ijt}=\sum_{\alpha\in\mathrm{CBG}}\sum_{\beta \in POI(\alpha)} 
\frac{\delta_{i \alpha t}}{N_{POI}(\alpha)}
\frac{\delta_{j \alpha t}}{N_{POI}(\alpha)} = \sum_{\alpha\in\mathrm{CBG}} \frac{\delta_{i\alpha t}\delta_{j\alpha t}}{N_{POI}(\alpha)}, \qquad \forall i,j
\end{equation}
where $POI(\alpha)$ is the set of POIs in the census block group $\alpha$, $N_{POI}{(\alpha)}$ is the number of POIs in $\alpha$, $\delta_{i\alpha t}$ is the binary variable of observing or not an individual at her workplace within census block group $\alpha$ at day $t$. As before, we have included only links for which $\omega^{W}_{ijt} > 0.01$.

\item {\bf Household weighted contact network}. We first identify individuals' approximate home place as their most likely visited census block group at night. Then we assign a synthetic representative household and demographic traits as documented in Ref. \cite{mistry2021inferring}. To assign weights, we assume that the probability of interaction within a household is proportional to the number of people living in the same household (well-mixing). Therefore, the weight, $\omega^H_{ij}$, of a link between individuals $i$ and $j$ within the same household is given by:
\begin{equation}
\label{sm:eq3}
\omega^H_{ij}=\frac{1}{(n_h-1)}
\end{equation}
where $n_h$ is the number of household members. This fraction is assumed to be the same for all individuals in the population. We assume this layer is static throughout our period.

\item {\bf School weighted contact network}. To calculate the weights of the links at the school layer, we mix together all children that live in the same census tract. Interactions are considered well-mixed, hence, the probability of interaction at a school is proportional to the number of children at the same school. Therefore, the weight, $\omega^S_{ij}$, of a link between children $i$ and $j$ within the same school is given by:
\begin{equation}
\label{sm:eq4}
\omega^S_{ij}=\frac{1}{(n_s-1)}
\end{equation}
where $n_s$ is the number of school members. This layer is removed on March 16 to account for the imposed school closure.
\end{itemize}

To calibrate the relative importance of each layer in the spreading process we further multiply the weights by their corresponding $\kappa$. In particular, with $\kappa = 4.11$ in the household layer, $\kappa = 11.41$ in the education layer, $\kappa = 8.07$ in the workplace layer and $\kappa = 2.79$ in the community layer \cite{mistry2021inferring}, see Eq.\ (\ref{response:eqfin}).

\clearpage

\section{Economic module}
\label{apx:model}

We introduce a dynamic macroeconomic model that is specifically suited to study the economic effects of COVID\nobreakdash-19, both at the macro-level of industries and at the micro-level of individuals and households. The model is particularly focused on employment and consumption, but also considers the input-output network of intermediates that industries use to produce final goods and services. The model is built so that its key variables can be mapped to quantities defined in national and regional accounts.

As a general modeling principle, we keep our model as simple as possible, with a minimal number of free parameters. We focus instead on a very detailed initialization of the state variables of the model, both at the micro- and macro-level, using multiple datasets coming from various sources (see Section  \ref{apx:data}). 

\subsection{Geography}
Our model is set up for detailed modeling of a relatively small ``local region'' and coarse modeling of a ``large area'' that is strongly interconnected with the local region (but does not include it) and sufficiently isolated from the Rest of the World. In the following, we also call the inhabitants of the region the ``locals'' and the inhabitants of the large area the ``outsiders''. As described in Materials and Methods, in our application the local region modelled in detail is the New York-Newark-Jersey City, NY-NJ-PA, Metro Area, and the large area being modeled at a more coarse level is the Rest of the United States. Nonetheless, our model is general and, in principle, the region could represent a country and the large area could be the Rest of the World.

\subsection{Agents}

\subsubsection{Individuals}

Individuals (or persons) play a role in the economic module as workers and consumers. We model individuals in the large area coarsely and in the local area more granularly. In the large area, individuals can be thought of as a ``continuum'', so that the only state variable is the fraction of employed and unemployed individuals for each industry. All individuals in the large area have the same consumption patterns.

In contrast, in the local region, individuals are discrete units, composing a synthetic population that is representative of the real population. (Typically, the synthetic population is a downscaled version of the real population, but it does not need to be.) Individuals are denoted by $i$ or $j$ and grouped in households $h$. As described in more details in Section \ref{apx:synthpop}, individuals are characterized by the following attributes: id $i$, household id $h$, age, income, employment status, industry where they work, occupation, possibility to work from home. Individuals are located spatially, as we also consider the smaller geographical unit where they live. In our application to New York, this is the census tract, but one could consider finer or coarser geographical units.

Economically active individuals (i.e. those who receive an income) play a direct role in the economic module by working and taking consumption decisions for their household. Economically inactive individuals (e.g. individuals younger than 18 years old) may play an indirect role as they participate in the epidemic module which feeds back into the economic module, but do not work nor take household consumption decisions. Note that we do consider retirees, who are out of the labor force but receive an income, and thus participate in the consumption market.

\subsubsection{Industries}
We consider a representative firm for each industry i.e. we do not differentiate between different firms/establishments within the same industry. We denote the representative firm of each industry with subindexes $k$ or $l$. Industries participate in the labor and consumption markets. Additionally, industries are connected by input-output linkages quantifying the flows of intermediates that they use to produce final goods and services.

Industries are modeled at the same level of detail in the region than in the large area. We consider flows of intermediate and final goods and services both within and across the region and the large area. In our application, this means that we consider flows of intermediate goods between pairs of industries that are both located in New York, between an industry in New York and an industry in the Rest of the US, etc. We also consider final demand by New Yorkers for goods and services produced locally, but also final demand of inhabitants from the Rest of the US for goods/services produced in New York, final demand by New Yorkers for goods/services produced in the Rest of US, etc.

Differently from individuals, industries are not assigned a specific geographical location other than belonging to the region or to the large area. In our application, this means that from an economic point of view the location of workers within the New York metro area does not matter. However, as we detail in Section \ref{apx:epi_model}, from an epidemiological point of view we do consider the specific location where individuals work.

\subsection{Hiring and firing}

In the labor module, industries decide the size of their workforce depending on industry-level variables from the previous time step, $t-1$, and on the level of restrictions in place at $t$. Once an industry makes this decision, it hires or fires individual agents by drawing at random from the employed and unemployed lists of workers associated to that industry. Thus, this module uses macro-level variables (i.e. industry variables) to determine micro-level variables (i.e. each agent's employment status).

In the economic module, we only focus on workers that were employed before the pandemic began (i.e., we ignore the ones that were previously unemployed or out of the labor force). We assign each worker to an industry from the start (Section \ref{apx:emp}). Within an industry, workers can be of two types: ``from-home'', who can work from home when required, and ``in-person'', who cannot. We assume that workers' industry and type are fixed. In other words, a worker fired from a given industry cannot find a job in another industry, and we do not allow workers to switch from ``in-person'' to ``from-home'' or vice-versa. 

\begin{figure}[!h]
    \centering
\includegraphics[width = 1\textwidth]{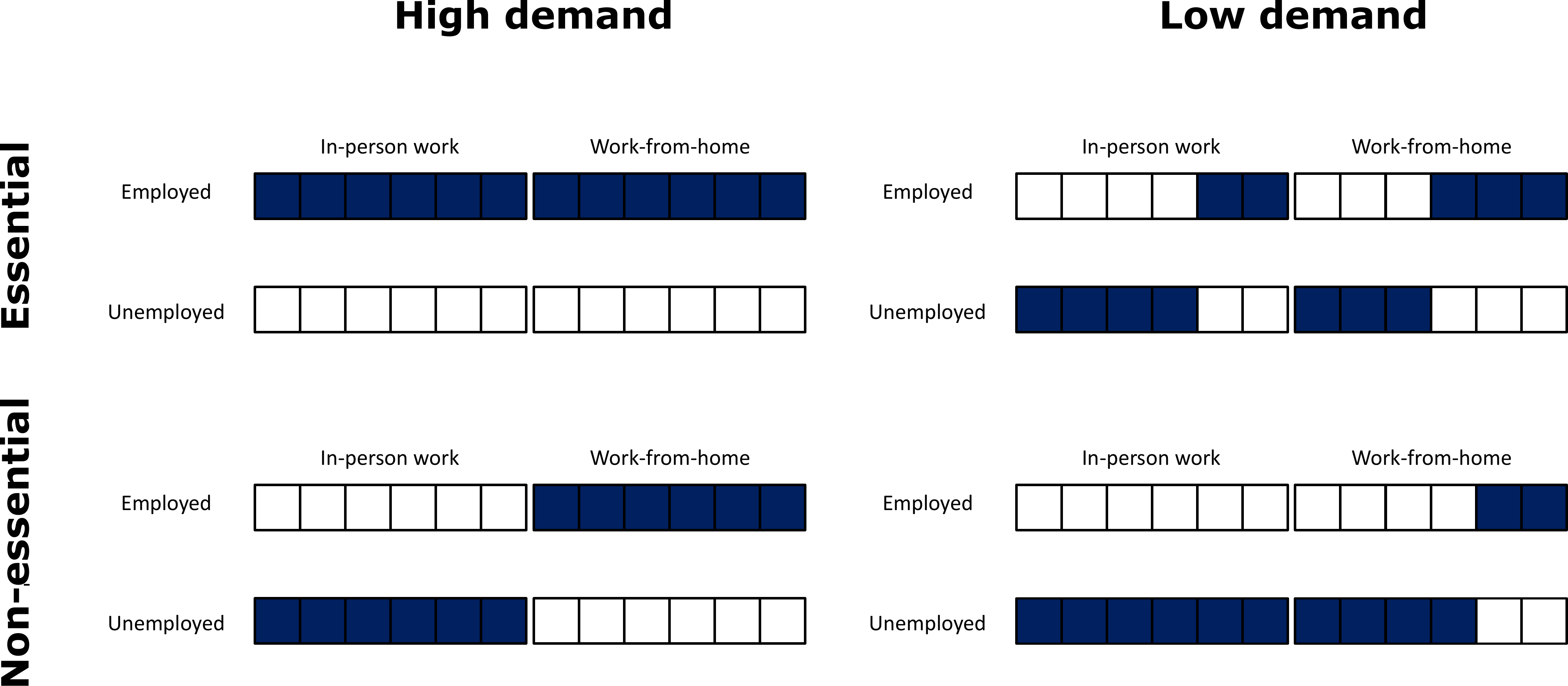}
   	\caption{\textbf{Schematic of the labor module.} Here we represent a single industry: as we do not allow transitions between industries, this case is representative of other industries. We show four combinations of policies and demand: an industry can be essential (i.e. workers can work in person) or non-essential (all workers must work form home), and goods and services produced by that industry can be in high demand or low demand. In each case, both ``in-person'' and ``from-home'' workers can be employed or unemployed. In this example, the industry that we consider has twelve workers, of which six must work in person and six can work from home. The color of each square indicates the state of each worker. When an industry is essential and in high demand, all squares in the ``employed'' row are colored blue and all squares in the ``unemployed'' row are colored white: this means that all workers are employed. Conversely, when the industry is completely non-essential and in high-demand, all squares corresponding to in-person workers are white in the ``employed'' row and blue in the ``unemployed'' row: this means that all in-person workers have been fired. The other cases have the same interpretation.  }\label{fig:schematic_labor}
\end{figure}

Figure~\ref{fig:schematic_labor} explains what the employment status of workers is depending on industry demand conditions and on their ability to work from home. For instance, the top-left panel represents the case of an industry that is essential and faces high demand. In this case, all workers are employed. We take this case as corresponding to the situation before the pandemic.\footnote{
As we focus our analysis on changes with respect to the situation in 2019, it makes sense to consider an initial state of full employment. At any point in time, the level of unemployment predicted by the model is the unemployment generated endogenously by the model plus the level of unemployment in 2019.
} 

\paragraph{Industries' personnel decisions during the pandemic.}
Recall that industries are mandated to close down their business to the extent they are not considered essential. Industries' decisions to hire or lay off employees thus depend on whether they are considered essential, but also on the demand they receive. In their personnel decisions they further consider whether an employee is able to work from.  For example, if an industry is not essential but receives high demand (e.g., certain professional services), it fires all in-person workers and retains from-home workers. Instead, if an industry is essential but faces low demand (e.g., transportation), it fires in-person and from-home workers alike. It is also possible that the industry is both non-essential and faces low demand (e.g., restaurants), so it fires all in-person workers and most from-home workers. Note that while in these examples industries are completely non-essential or essential, in practice our estimates of the shocks typically account for a share of essentialness (see Section \ref{apx:shocks}). 

We capture the intuition portrayed by this example with the following modeling approach. We denote the number of in-person workers employed in an industry $k$ at time $t$ by $l_{k,t}^\text{P}$ and the number of from-home workers by $l_{k,t}^\text{H}$. The total number of workers is $l_{k,t}=l_{k,t}^\text{P}+l_{k,t}^\text{H}$ and is determined by demand and supply constraints as follows.

From a labor demand perspective, an industry $k$ decides to lay off or hire workers depending on the demand for its products, $d_{k,t-1}$, and its productive capacity, $x_{k,t-1}^\text{cap}$. (See Sections \ref{sec:othdemand} and \ref{sec:prod} for a definition of these variables.) When demand is larger than productive capacity, i.e., $d_{k,t-1}>x_{k,t-1}^\text{cap}$, industry $k$ wants to hire workers. Vice versa, when an industry $k$ is not receiving enough demand for its products (i.e. when $d_{k,t-1}<x_{k,t-1}^\text{cap}$), it lays off workers. 

In formula, the labor demanded by industry $k$ at time $t$ is 
\begin{equation}
    l^d_{k,t}=l_{k,t-1} + \Delta l_{k,t},
    \label{eq:labdemand}
\end{equation}
where
\begin{equation}
    \Delta l_{k,t} = \frac{l_{k,0}}{x_{k,0}}\left[ d_{k,t-1}-x_{k,t-1}^\text{cap}\right].
\end{equation} 
The term $l_{k,0}/x_{k,0}$ reflects the assumption that the labor share in production is constant over the considered period. It can alternatively be viewed as a coefficient that translates changes in demand and productive capacity into changes in employment.\footnote{
It makes Equation~\eqref{eq:labdemand} well-defined, in the sense that when demand is equal to initial productive capacity, $d_{k,t-1}=x_{k,0}^\text{cap}$, labor demand is equal to initial employment, $l^d_{k,t}=l_{k,0}$. To see this, note that $x_{k,t-1}^\text{cap}=x_{k,0}\cdot l_{k,t-1}/l_{k,0}$ (Section \ref{sec:prod}), and that initial capacity is equal to initial production, $x_{k,0}^\text{cap}=x_{k,0}$ (we initialize the model in its steady state).
}

The expressions above apply to both labor demand of in-person workers and from-home workers. Labor demand of in-person workers is $l^{\text{P},d}_{k,t}=l^\text{P}_{k,t-1}+\Delta l^\text{P}_{k,t}$, with $\Delta l^\text{P}_{k,t} = l^\text{P}_{k,0}/x_{k,0}\left[ d_{k,t-1}-x_{k,t-1}^\text{cap}\right]$, and an analogous expression holds for labor demand of from-home workers, $l^{\text{H},d}_{k,t}$. 

From a labor supply perspective, we assume that workers are always willing to work, but a lockdown may restrict the number of workers that are allowed to work in person in a given industry. We assume that for an industry $k$ at time $t$ the maximum number of workers allowed to work in person, $l_{k,t}^\text{P,max}$, is a fraction of the initial number of in-person workers $l^\text{P}_{k,0}$, determined by the supply shock $s_{k,t}$ (see Section \ref{apx:supplyshocks}): $l_{k,t}^\text{P,max}=(1 - s_{k,t})l^\text{P}_{k,0}$. 

Given the labor supply constraint, industry $k$ has a target of in-person workers $l_{k,t}^\text{P,T}$ given by
\begin{equation}
    l_{k,t}^\text{P,T}= \min\left(l_{k,t}^\text{P,max},l^{\text{P},d}_{k,t}\right).
\end{equation}
Further, it cannot hire more from-home workers than it initially had, i.e. the target of from-home workers, $l_{k,t}^\text{H,T}$ is given by 
\begin{equation}
    l_{k,t}^\text{H,T}= \min\left(l_{k,0}^\text{H},l^{\text{H},d}_{k,t}\right).
\end{equation}

Finally, we assume that it takes time for industries to adjust their labor inputs. Specifically, we assume that industries can increase their labor force only by a fraction $\gamma_{\text{H}}$ in direction of their target. Similarly, they can decrease their labor force only by a fraction $\gamma_{\text{F}}$ in the direction of their target. These sluggish labor adjustment mechanisms have been used in other economic ABMs \cite{del2021occupational,pichler2021and} and have two main reasons behind them. First, at the microeconomic level, the labor economics literature has established that hiring takes firms time and resources \cite{pissarides2011equilibrium} and found that firing costs (such as severance payments) diminish firms' propensity to fire \cite{bentolila1990firing}. The second reason is aggregation. In these models, one does not consider individual firms, but rather aggregate industries. Thus, an alternative explanation for industry-level sluggishness is that individual firms asynchronously hire or fire groups of workers, but aggregation smooths the dynamics.

The resulting changes in the number of in-person workers are
\begin{equation} 
    l_{k,t}^\text{P} = 
    \begin{cases} 
    l_{k,t-1}^\text{P} + \gamma_{\text{H}} \left( l_{k,t}^\text{P,T} -l_{k,t-1}^\text{P} \right) &\mbox{if } \; l_{k,t}^\text{P,T} -l_{k,t-1}^\text{P} \ge 0, \\
    l_{k,t-1}^\text{P} + \gamma_{\text{F}} \left( l_{k,t}^\text{P,T} -l_{k,t-1}^\text{P} \right) &\mbox{if } \; l_{k,t}^\text{P,T} -l_{k,t-1}^\text{P} < 0.
    \end{cases}
    \label{eq:labor_evolution_inperson}
\end{equation}
The number of from-home workers follows
\begin{equation} 
    l_{k,t}^\text{H} = 
    \begin{cases} 
    l_{k,t-1}^\text{H} + \gamma_{\text{H}} \left( l_{k,t}^\text{H,T} -l_{k,t-1}^\text{H} \right) &\mbox{if } \; l_{k,t}^\text{H,T} -l_{k,t-1}^\text{H} \ge 0, \\
    l_{k,t-1}^\text{H} + \gamma_{\text{F}} \left( l_{k,t}^\text{H,T} -l_{k,t-1}^\text{H} \right) &\mbox{if } \; l_{k,t}^\text{H,T} -l_{k,t-1}^\text{H} < 0.
    \end{cases}
    \label{eq:labor_evolution_fromhome}
\end{equation}

Equations~\eqref{eq:labor_evolution_inperson} and \eqref{eq:labor_evolution_fromhome} describe hiring and firing both in the region that we model in detail and in the large area that we model at a more coarse level (in our application, New York and the rest of the US respectively). 

\paragraph{Hiring and firing at the micro level.}
The equations above determine the number of workers that a given industry fires or hires. However, to estimate economic impacts for the region that we model at a detailed level, we must also specify which workers are laid off or hired. This information is important as workers have different characteristics, especially income, and by tracking their individual employment status we are able to track the characteristics of individuals that are more or less likely to be hired or fired. 

To model hiring and firing at an individual level, we use the arrays shown in Figure \ref{fig:schematic_labor}, which correspond to lists filled with the IDs of individuals from the synthetic population of the region. At the beginning of the simulation only the lists of employed agents are filled. When firing, an industry takes as many people as it needs from the currently employed list and places them in the unemployed list, choosing IDs uniformly at random. When hiring, an industry takes as many individuals as it needs from its fired/furloughed list. So each individual at every time step is either employed or not employed. Hiring and firing of in-person and from-home workers happens independently, in the sense that there are no transitions among the two lists. 

\subsection{Consumption demand}
\label{sec:consumption}

Consumption demand is determined differently in the region (New York) and in the large area (Rest of the US). In the region, it is determined aggregating up the preferences of individual households, while in the large area it is determined at the aggregate level. We start by describing consumption demand in the region.

The basic unit in the consumption market is the household, not the individual. We characterize the properties of a household mainly based on the features of one individual in the household, a fictional ``household head'', who is the individual with highest income in the household. So, the age of a household is the age of the household head and the employment status of the household is determined by the employment status of the household head.\footnote{This is a simplification that avoids the need to consider ``fractional employment'' in the household.} Household income is the sum of the incomes of the individuals belonging to the household.

Consumption demand at time $t$ by households living in the region (``locals''), for goods/services produced by industry $k$ (which can be located either in the region or in the large area), $c^{d,\text{local}}_{k,t}$, is given by
\begin{equation}
c^{d,\text{local}}_{k,t}=\sum_{g} c_{g,k,t}^{d,\text{local}}=\sum_{g} \sum_{h\in g} c_{h,k,t}^{d,\text{local}},
\end{equation}
where $g$ indicates a group of households $h$ that have the same consumption demand. In this paper, this corresponds to households that belong to the same age and income categories. 

Consumption demand of group $g$ evolves according to
\begin{equation}
c^{d,\text{local}}_{g,k,t}=c^{d,\text{local}}_{g,k,0} \times \text{PANDEMIC PREFERENCE}_{k,t} \times \text{INCOME EFFECTS}_{g,t}.
\label{eq:consdemandbygroup}
\end{equation}
Let us comment on these terms one by one:
\begin{itemize}
\item $c^{d,\text{local}}_{g,k,0}$ is simply total yearly consumption of locals in group $g$ for goods and services of industry $k$, as initialized on pre-pandemic data (Section \ref{apx:cons}).

\item PANDEMIC PREFERENCE is the shift in consumption patterns due to the COVID\nobreakdash-19 pandemic. In general, we model a reduction in consumption demand of goods and services produced by industries that consumers perceive ``risky'', as they could be infected during consumption. These include, for instance, bars and restaurants, retail and personal services. At the same time, we assume that consumers reallocate part of what they save on these risky industries towards goods and services produced by industries that do not require contact with customers. For instance, consumers may decide to use the money that they are saving on restaurants to buy food produced by the agriculture and food manufacturing industries.

Behavior change has received a lot of attention by economists, who tend to treat behavioral response to economic policy by assuming that agents respond optimally to ``rational'', or model-consistent, expectations. Since the start of the pandemic, economists have produced several models in which individuals optimally choose how much to expose themselves to infections as they know the equations that govern epidemic spreading, and in turn epidemic dynamics is consistent with this optimal response. In this sense, expectations of individuals about risk of exposure induced by the epidemic dynamics are ``model-consistent'' (see e.g. Ref. \cite{farboodi2021internal} for a model on the COVID\nobreakdash-19 pandemic, and Ref. \cite{fenichel2011adaptive} for a classical model in this tradition). 

Here, we simulate epidemic dynamics at the individual level, rather than having an aggregate SLIR model, so it is infeasible from a computational point of view to model behavior change as an optimal response to model-consistent expectations. We also believe that this modeling endeavor does not account for the deep uncertainty of a novel pandemic, where agents cannot immediately learn the epidemic equations and thus form model-consistent expectations. Given these limitations of the rational expectations approach to behavior change, we follow instead a more standard epidemiological literature (see e.g. \cite{perra2011towards}), which models behavior change as a reduction in the contact rate that directly depends on the state of the pandemic. In particular, we assume that individuals change their behavior at the global level, based on information that is publicly available, using the number of daily reported deaths in the NY MSA as an indicator of the state of the pandemic.

In line with this discussion, we use a different functional for $\text{PANDEMIC PREFERENCE}_{k,t}$ %
depending on whether an industry is customer-facing ($\tau_k=1$) or not ($\tau_k=0$). When it is customer-facing, we let $\text{PANDEMIC PREFERENCE}_{k,t}=1-\Lambda_t^{ECO}$, where $\Lambda_t^{ECO}$ is the \textit{reduction} in consumption demand defined in Eq. \eqref{eq:fear_inf_ec}. Thus, $\text{PANDEMIC PREFERENCE}_{k,t}$ is the \textit{level} of consumption demand that would be induced by the reduction $\Lambda_t^{ECO}$. Replacing the expression in Eq. \eqref{eq:fear_inf_ec}, we have\footnote{
In principle, we could make fear of infection dependent on the characteristics of households, e.g. assume a higher fear of infection for older or richer agents. However, we do not do so for two reasons. First, while there is evidence that \textit{realized} consumption decreased more for older and richer agents \cite{eichenbaum2020people}, surveys trying to estimate consumption \textit{demand} do not find a clear difference across these socio-economic characteristics \cite{hodbod2021covid}. Second,  we do not want to introduce too many free parameters that would make it possible to obtain any result. If differences in consumption across socio-economic groups emerge, they would emerge endogenously, not because we assumed them in the first place. 
} 

\begin{equation}
    \text{PANDEMIC PREFERENCE}_{k,t}=\exp\left(-\phi^{ECO} D_t\right) \text{ if } \tau_k=1.
\end{equation}

In contrast, when an industry $k$ is not customer-facing ($\tau_k=0$), consumption demand does not decrease when the epidemic situation worsens. In fact, it may increase, as consumers redirect a fraction $\Delta s$ of the money that they are saving on customer-facing industries. The total amount of money saved on customer-facing industries is given by $\sum_k \Lambda_t^{ECO} c_{g,k,0}^{d,\text{local}}$. Thus, in this case we have
\begin{equation}
    \text{PANDEMIC PREFERENCE}_{k,t}=1+\Delta s \frac{\sum_k \Lambda_t^{ECO} c_{g,k,0}^{d,\text{local}}}{\sum_k (1-\tau_k)c_{g,k,0}^{d,\text{local}}}  \text{ if } \tau_k=0,
\end{equation}
where the denominator $\sum_k (1-\tau_k)c_{g,k,0}^{d,\text{local}}$ is the sum of consumption of no customer-facing industries and it is needed to properly normalize changes in consumption demand.\footnote{Assume that $c^{d,\text{local}}_{g,k,t}=c^{d,\text{local}}_{g,k,0} \times \text{PANDEMIC PREFERENCE}_{k,t}$, i.e. that there are no income effects. In this case, total consumption demand of goods/services produced by no customer-facing industries would correctly be given by the sum of the usual consumption demand for these industries plus a fraction $\Delta s$ of all the money saved on customer-facing industries, i.e. $$\sum_k (1-\tau_k) c^{d,\text{local}}_{g,k,0} \left(1+\Delta s \frac{\sum_k \Lambda_t^{ECO} c_{g,k,0}^{d,\text{local}}}{\sum_k (1-\tau_k)c_{g,k,0}^{d,\text{local}}}\right)=\sum_k (1-\tau_k) c^{d,\text{local}}_{g,k,0}+\Delta s \sum_k \Lambda_t^{ECO} c_{g,k,0}^{d,\text{local}}$$.}

\item INCOME EFFECTS is the shock to aggregate consumption related to precautionary savings induced by uncertainty about future employment and income. We simply assume that households whose household head becomes unemployed reduce consumption by a factor $\phi^U$, and  sum over all households belonging to group $g$ to get reductions in consumption demand due to income effects, i.e.
\begin{equation}
    \text{INCOME EFFECTS}_{g,t}=1-\phi^U\frac{\sum_{h\in g} U_{h,t}}{N_g},
    \label{eq:fear_unemp}
\end{equation}
 where $U_{h,t}$ is an indicator function that takes value 1 if the household head of household $h$ is unemployed at time $t$, and $N_g$ is the total number of agents in group $g$. We normalize everything so that we consider changes in employment with respect to pre-pandemic levels, so we take by convention initial unemployment to be zero, i.e. $U_{h,0}=0$ for all households $h$.
\end{itemize}

This detailed micro-to-macro characterization of consumption demand only applies to the ``locals'' (i.e. individuals living in the New York region), but describes consumption demand by locals of goods and services produced both locally and in the large area (Rest of the US). For instance, due to fear of infections, New Yorkers reduce their demand for restaurant services in New York and in the Rest of the US by the same amount.

In contrast to consumption demand by locals, consumption demand by ``outsiders'' (i.e. from outside of the region, denoted as ``rest'' below) is not aggregated from households, but is an aggregate that evolves depending on other aggregate quantities. Denoting by $l_t^\text{rest}$ aggregate employment in ``rest'' at time $t$, with initial condition $l_0^\text{rest}$, $\text{INCOME EFFECTS}_{k,t}^\text{rest}$ is given by
\begin{equation}
    \text{INCOME EFFECTS}_{k,t}^\text{rest}=1-\phi^U \frac{l_0^\text{rest}-l_t^\text{rest}}{l_0^\text{rest}}.
    \label{eq:rest_cons_shock}
\end{equation}
Additionally, we assume that the consumption demand shock driven by fear of infection by outsiders willing to consume locally-produced goods and services is as strong as that experienced by locals. This assumption is justified by the fact that the industries affected by fear of infection require on-site consumption, and so outsiders would need to travel to the region and experience the same epidemic situation. By contrast, in case industry $k$ is located in the large area, the term $\Lambda_t$ is replaced by an equivalent term $\Lambda_t^\text{rest}$ that considers the state of the pandemic in the large area rather than in the region.  

Total consumption demand faced by industry $k$ at time $t$ is given by the sum of consumption demand by locals and by outsiders, i.e.
\begin{equation}
    c_{k,t}^d=c_{k,t}^{d,\text{local}}+c_{k,t}^{d,\text{rest}}.
\end{equation}

\subsection{Other demand}
\label{sec:othdemand}

The orders of intermediate goods and services from industry $l$ to industry $k$ at time $t$ are given by 
\begin{equation}
O_{k,l,t}=A_{k,l} x_{l,t-1},
\end{equation}
where $A_{k,l}$ is a component of the technical coefficient matrix and $x_{l,t-1}$ is the output that industry $l$ produced at $t-1$. In other words, industry $l$ anticipates that its output at time $t$ will be again $x_{l,t-1}$, knows that in order to produce one unit of output it needs $A_{k,l}$ units of $k$'s output, and as a result places an order $O_{k,l,t}$ to industry $k$. Note that industries $l$ and $k$ can be located either in the region or in the large area and flows of intermediate goods can occur both within and across the two areas.

Total demand faced by industry $k$ at time $t$ is given by
\begin{equation}
d_{k,t}=\sum_l O_{k,l,t}+c_{k,t}^d+G_{k,t}^d+f_{k,t}^d,
\end{equation}
i.e. total demand is the sum of intermediate orders across all industries $l$ (including industry $k$ itself), private consumption demand $c_{k,t}^d$, government consumption demand $G_{k,t}^d$ and other components of final demand $f_{k,t}^d$ (investment and net exports). We take $G_{k,t}^d=(1-\mathcal{G}_t)G_{k,0}^d$, where $\mathcal{G}_t$ is an exogenous shock to government demand (if $\mathcal{G}_t<0$, government demand actually increases). We also let $f_{k,t}^d=(1-\mathcal{S}_t)f_{k,0}^d$, where $\mathcal{S}_t$ is an exogenous shock to other components of final demand. See Section~\ref{apx:shocks} for details on exogenous shocks.

\subsection{Production}
\label{sec:prod}

In this paper, we assume that production is never constrained by lack of intermediate inputs or capital goods.\footnote{While shortages of intermediate goods seemed to be present at least in some industries, such as the car manufacturing industries, here it would be difficult to calibrate input bottlenecks as we generally lack detailed data of trade between the region and the large area.}  So, we specify a production function that only depends on labor. Further, we assume that the production function is linear in labor inputs, i.e. we do not consider complementarity between occupations and just assume that industry production changes proportionally to industry employment. This leads to defining productive capacity $x_{k,t}^\text{cap}$ as
\begin{equation}
    x_{k,t}^\text{cap}=\frac{l_{k,t}}{l_{k,0}}x_{k,0},
\end{equation}
that is, as a fraction $l_{k,t}/l_{k,0}$ of initial production $x_{k,0}$. 

Production is the minimum of demand and capacity , i.e.
\begin{equation}
x_{k,t}=\min \left( d_{k,t},x_{k,t}^\text{cap} \right).
\end{equation}

If $x_{k,t}^\text{cap}<d_{k,t}$, not all demand can be satisfied. As a general principle, we assume that demand is satisfied on a pro-rata basis, i.e. all components of final demand receive a fixed fraction of what they demanded (see \cite{pichler2021simultaneous} for alternative rationing schemes). This means that intermediate consumption by industry $k$ of products by industry $l$, $Z_{l,k,t}$ is defined as
\begin{equation}
    Z_{l,k,t}=O_{l,k,t} \frac{x_{l,t}}{d_{l,t}},
\end{equation}
while realized consumption by local households in group $g$ as 
\begin{equation}
    c^{\text{local}}_{g,k,t}=c_{g,k,t}^{d,\text{local}} \frac{x_{k,t}}{d_{k,t}},
\end{equation}
and realized consumption by outsiders, $c^{\text{rest}}_{k,t}$, realized other demand $f_{k,t}$, realized government demand $G_{k,t}$ are defined in a similar way. We finally define value added as output minus intermediate consumption, i.e.
\begin{equation}
    va_{k,t}=x_{k,t}-\sum_l Z_{l,k,t}.
\end{equation}

\FloatBarrier
\newpage
\section{Initialization}
\label{apx:data}

The epidemic module (Section \ref{apx:epi_model}) and the economic module (Section \ref{apx:model}) feature several variables and agent-level parameters that need to be initialized and model-wide parameters that must be calibrated. (We use the term initialization to mean setting of initial conditions for model variables, and setting of values for agent-level parameters, i.e. parameters that are potentially heterogeneous across each different agent.) In this section, we focus on variable and agent-level parameter initialization, while in Section \ref{apx:calibration} we explain the calibration of model-wide parameters. 

\begin{table}[htbp]
	\centering
		\begin{tabular}{|C{6.5cm}|c|c|C{3cm}|}
			\hline
			Variables & Name & Section & Source \\
			\hline 
           $\omega_{i,j,t}^C, \omega_{i,j,t}^W, \tau_k$ & Mobility and places & Section \ref{apx:mobility} & Cuebiq, Foursquare \\[10pt]
			 $s_{k,t}^\text{local}$, $s_{k,t}^\text{rest}$, $D_t^\text{rest}$, $\mathcal G_t$, $\mathcal S_t$ & Shocks & Section \ref{apx:shocks} & NY State, NY Times, Ref. \cite{barrot2021sectoral} \\[10pt]
			 $l_{k,0}^\text{P}, l_{k,0}^\text{H}, g, \omega_{i,j}^H, \omega_{i,j}^S$ & Synthetic population & Section \ref{apx:synthpop} & ACS, BLS\\[10pt]
			 $x_{k,0}, O_{k,l,0}, Z_{k,l,0}, A_{k,l}$, $c_{k,0}^{d,\text{local}}$, $G_{k,0}^{d,\text{local}}$, $f_{k,0}^{d,\text{local}}$, $c_{k,0}^{d,\text{rest}}, G_{k,0}^{d,\text{rest}}, f_{k,0}^{d,\text{rest}}$, $c_{k,0}^{\text{local}}, G_{k,0}^{\text{local}}, f_{k,0}^{\text{local}}$, $c_{k,0}^{\text{rest}}, G_{k,0}^{\text{rest}}, f_{k,0}^{\text{rest}}$ & Input-output & Section \ref{apx:io} & BEA \\[30pt]
			 $c_{g,k,0}^{d,\text{local}}$ & Consumption & Section \ref{apx:cons} & BLS, BEA \\[10pt]
			\hline						
		\end{tabular}
		\caption{Variable initialization. NY=New York; ACS=American Community Survey (Census); BLS=Bureau of Labor Statistics; BEA=Bureau of Economic Analysis; CDC=Center for Disease Control}
	\label{tab:variable_initialization}
\end{table}

Table \ref{tab:variable_initialization} shows the variables and attributes of the epidemic and economic modules, divided into the datasets that we use for initialization. In particular, we use the following datasets:
\begin{itemize}
    \item Mobility and places. To assess which places are visited by individuals at which times, we use a dataset of places from Foursquare and mobility data from Cuebiq.
    \item Shocks. These data are estimations of supply shocks. We mainly derive them from the NY executive order issued by Governor Andrew Cuomo.
    \item Synthetic population. To build the synthetic population, we obtain most variables from the American Community Survey (ACS) of the Census Bureau, but some variables are obtained from the Bureau of Labor Statistics (BLS).
    \item Input-output. This is a reconstruction of the New York-Rest of the US input-output system, including inter-industry intermediate flows, final demand and value added in and across New York and the rest of the US. All data come from the Bureau of Economic Analysis (BEA).
    \item Consumption. To get the heterogeneous consumption of agents with different ages and incomes for goods and services produced by different industries, we make information from the Consumer Expenditure Survey conducted by the BLS consistent with our synthetic population and input-output system.
\end{itemize}

\subsection{Mobility and places}
\label{apx:mobility}

\subsubsection{Mobility data}
\label{sectionSM:data}
The mobility data was obtained from Cuebiq, a location intelligence and measurement company. The dataset consists of anonymized records of GPS locations from users that opted-in to share the data anonymously in the New York metropolitan area over a period of 5 months, from February 2020 to June 2020. In addition to anonymizing the data, the data provider obfuscates home locations to the census block group level to preserve privacy. Data was shared in 2020 under a strict contract with Cuebiq through their Data for Good program where they provide access to de-identified and privacy-enhanced mobility data for academic research and humanitarian initiatives only. All researchers were contractually obligated to not share data further or to attempt to de-identify data. Mobility data is derived from users who opted in to share their data anonymously through a General Data Protection Regulation (GDPR) and California Consumer Privacy Act (CCPA) compliant framework.

Our sample dataset achieves broad geographic representation of the New York metropolitan area. The complete sample of users is slightly biased towards higher income individuals. Specifically, the penetration ratio (number of mobile phone users to adult population) in each census tract is correlated with the median household income, $\rho = 0.28\pm 0.02$. However the correlation of the penetration ratio with the number of people above 64 years old in each census tracts is small $\rho = 0.17\pm0.04$. To reduce the impact of this bias, we downsampled the original sample of users in the Cuebiq dataset to get a more representative distribution of the different (4) quantiles of income. Specifically, the original sample contained 438,178 users and was biased toward high-income people. The penetration rate of high income people was 4.13\% while we only get 2.3\% of people in low-income areas. Thus we downsample the high-income groups to get a more balanced distribution of users by quantile groups. In particular we selected a random sample of 2.3\% of people in each of the income groups. This led us to a final set of 316,070 users. This does not correspond to the size of the synthetic population, as we need to include other agents as explained in Sections \ref{apx:age} and \ref{apx:matching}.

\subsubsection{Points of Interest}
\label{apx:subPOI}

We used the Fousquare Public API to retrieve a large collection of (Points of Interest) POIs in the New York metro area. Although Foursquare data is a crowd-sourced resource, it exhibits some editorial control. Their database of POI not only comes from users of their Swarm (check-in) platform, but is built by aggregating data over 46k different trusted sources \cite{fsqpoi}. Several studies confirm that although none of the POI databases is complete, Foursquare is one of the best in number of POIs, location accuracy and number of categories represented  \cite{hochmair2018data}. 

We use a dataset of 375k Points of Interest (POI) in the New York metropolitan area collected using the public Foursquare API. Those POIs are categorized using the Foursquare taxonomy of places which has ten main categories. There are also 638 subcategories, see \cite{catsfsq} for a complete list of them. Despite our dataset contains many venues and places which are companies or business, some evidence that our dataset covers most of the public places comes by comparing them to official statistics: for example, we have 2,155 art galleries in the NY metro area compared to the 1,500 estimation for NY City only. On the other hand we have 9,810 groceries in the NY metro area in our POI database which compares quite well with the 11,791 grocery business reported by the U.S. Bureau of Labor Statistics in their Quarterly Census of Employment and Wages in the NY Metro area \cite{qcew}.

To link Foursquare POIs to economic activities, we manually associate each of the 638 Foursquare subcategories to the 2-digit NAICS industry that best describes the economic activity occurring in the corresponding Foursquare POI.\footnote{The full matching is available in the replication files.} In some cases, it is possible to associate unambiguously a Foursquare category to a NAICS industry. For instance, ``Art gallery'' certainly belongs to NAICS industry 71 - Arts, Entertainment, and Recreation. In other cases, multiple NAICS industries could correspond to a Fousquare category, and we may have to select a few that are the most plausible. For example, the Foursquare category ``Office'' can in theory be associated to most NAICS categories, and is likely to be associated with one or multiple business and professional services, corresponding to NAICS industries 51 to 56. In a few other cases, finally, it is not possible to associate an economic activity to a Foursquare place (e.g., ``River'', and most outdoor venues). 

To associate a NAICS industry to each of the 375k POIs in our dataset, we proceed as follows. First, when only one NAICS industry is associated to the Foursquare category of the place, we simply assign that industry. Second, if multiple NAICS industries could be associated to the place, we draw at random from the list of NAICS industries associated to the corresponding Foursquare category. This random draw is not uniform, but is rather weighted by the employment share of the associated industries. Finally, if there is no economic activity occurring at the POI, we do not associate any NAICS industry to it. 

Table \ref{tab:foursquare} (third column) shows the number of Foursquare POIs associated to each of the 20 2-digit NAICS industries.

\begin{table}[]
\resizebox{\textwidth}{!}{\begin{tabular}{ll|C{1.5cm}C{3cm}C{1.5cm}|C{1.5cm}}
\hline
NAICS & Industry name & Number of places & Total weight in community locations & Customer-facing & Essential score \\ \hline
11 & Agriculture, forestry, fishing and hunting & 0 & 0 & 0 & 1.00 \\
21 & Mining, quarrying, and oil and gas extraction & 0 & 0 & 0 & 1.00 \\
22 & Utilities & 0 & 0 & 0 & 1.00 \\
23 & Construction & 192 & 3193 & 0 & 0.49 \\
31-33 & Manufacturing & 1101 & 12824 & 0 & 0.56 \\
42 & Wholesale trade & 0 & 0 & 0 & 0.81 \\
44-45 & Retail trade & 53108 & 14091171 & 1 & 0.65 \\
48-49 & Transportation and warehousing & 27210 & 935324 & 1 & 0.99 \\
51 & Information & 2379 & 1897 & 0 & 0.78 \\
52 & Finance and insurance & 11812 & 501401 & 0 & 1.00 \\
53 & Real estate, rental and leasing & 3285 & 27913 & 0 & 0.91 \\
54 & Professional, scientific, and technical services & 12519 & 171610 & 0 & 0.60 \\
55 & Management of companies and enterprises & 189 & 0 & 0 & 1.00 \\
56 & Administrative and support and waste & 3201 & 10898 & 0 & 0.45 \\
61 & Educational services & 19305 & 2153993 & 1 & 0.40 \\
62 & Health care and social assistance & 28438 & 1433625 & 1 & 1.00 \\
71 & Arts, entertainment, and recreation & 49717 & 4203783 & 1 & 0.00 \\
72 & Accommodation and food services & 67877 & 6006817 & 1 & 0.22 \\
81 & Other services, except public administration & 37573 & 1646087 & 1 & 0.66 \\
92 & Public administration & 6055 & 264724 & 0 & 1.00 \\
NA & Not an economic activity & 43571 & 1742638 & NA & NA \\ \hline
\end{tabular}}
\caption{Industry-level information. The first group of three columns shows the distribution of Foursquare POIs across industries, the total weight in the contact network associated to these locations, and the choice of whether an industry is customer-facing that results from these data.  The last column show the industry essential score, obtained as explained in Section \ref{apx:supplyshocks}.}
\label{tab:foursquare}
\end{table}

\subsubsection{Initialization of community and workplace contacts and definition of customer-facing industries}
\label{sec:apx_def_customer_contact}

To get a sense of which industries lead to most contacts in the community, we sum the weights of community contacts within any given industry, between February and May 2020. For instance, letting $\omega_{ijt}^{C,k}$ denote a community contact occurring in industry $k$, we compute the total weight of industry $k$ by taking $\sum_{i,j,t} \omega_{ijt}^{C,k}$. We then report results in the fourth column of Table \ref{tab:foursquare}. 

As we can see, the vast majority of the contact weights are concentrated in seven ``customer-facing'' industries, namely retail trade (44-45), transportation and warehousing (48-49),  educational services (61), health care and social assistance (62), arts, entertainment and recreation (71), accommodation and food services (72), and other services (81, e.g., personal services like hairdressers). Places that are not associated to an economic activity (mostly outdoors) also have a weight greater than one million. Most categories in business and professional services receive fewer visits, but still much more than manufacturing and construction. Finally, ``upstream'' industries such as utilities and wholesale trade do not have any associated Foursquare category and so do not have any weight; Based on Foursquare places manually matched to NAICS, agriculture and mining appear non-existent in the New York metropolitan area.

Based on this information, we choose $\tau_k=1$ for each of the seven customer-facing industries mentioned above, and select $\tau_k=0$ for all other industries.

Now, even though we can use the contacts extracted from the data covering the period from 02/12/20 to 06/01/20 for this kind of analysis, we cannot use them directly in the simulations. Indeed, note that the data from March to June naturally incorporates all the societal changes that occurred during those days: some people were not present in the workplace layer because they were fired or working from home; others stopped going to certain venues in the community, etc. As such, if we were to use the real data from this period, we would not be able to couple the epidemic module with the economic module that estimates who should be fired/hired.  Similarly, we would not be able to estimate any fear of infection since it would already be contained in the data.

Instead, we initialize the community and workplace contacts using data from the period 2020/02/13 to 2020/03/11, which we take as representative of the pre-pandemic situation. Then, at each step, contacts belonging to the corresponding day of the week are sampled from this set. For instance, on any Monday in the simulation we would select the contacts from 2020/02/17, 2020/02/24, 2020/03/02 or 2020/03/09. In this way, we can run simulations under ``normal'' conditions (i.e. without the effects of the pandemic), incorporating non-pharmaceutical interventions into the model (for instance, by removing all links associated to a certain type of industry), including the effect of the firing process by removing individuals from the workplace layer, etc.

\subsection{Shocks}
\label{apx:shocks}

\subsubsection{Local supply shocks}
\label{apx:supplyshocks}

To stop the spread of COVID\nobreakdash-19, the New York Governor announced in March 2020 that all non-essential businesses statewide must close in-office personnel functions.\footnote{Part of the New York metropolitan area is actually in New Jersey, which had slightly different restrictions from New York State. For simplicity, we just assume that the restrictions imposed by New York State are valid for the entire metropolitan area and were applied at the same time.} This announcement included a document explaining which businesses were considered essential.\footnote{
\url{https://www.governor.ny.gov/news/governor-cuomo-issues-guidance-essential-services-under-new-york-state-pause-executive-order}
} Unlike the lockdown announcements from other governments, the definition of an essential business in New York was not linked to industry classification codes. Nevertheless, by using the description of NAICS industries and carefully examining the essential businesses described in New York's lockdown announcement, we classified industries as essential or non-essential according to New York guidelines.

To build New York's list of essential industries, we used the BLS 2019 industry employment data for the NY-NJ-PA metropolitan area. There are employment data for 218 industries at the 4-digit NAICS level. Additionally, there are employment data for 5 industries\footnote{
These industries are: Fishing, hunting and trapping; Mining, except oil and gas; Construction of buildings; Paper manufacturing; and Truck transportation
} at the 3-digit NAICS level that is not available at the 4-digit level. This gives a total of $223$ industries to classify as essential or non-essential.  By carefully reading the description of these industries and the New York government's mandate, we assigned an essential score of 1 to those industries that met the criteria for essentialness and 0 otherwise\footnote{
All industries were assigned a binary essential score, except ''Restaurants and other eating places". Restaurants were considered essential only for take-out and/or delivery. We assigned this industry an essential score of $0.1$.
}. We then aggregated the essential score of industries to the 2-digit NAICS classification level. We did this aggregation by taking the employment weighted average of each industry subclass. The procedure is fully documented in the replication materials.

We interpret the essential score of industry $k$, $\text{ESS}_k$, as the fraction of workers from industry $k$ that are allowed to work on-site according to the New York government regulations. We then define the supply shock at time $t$ as $s_{k,t}^\text{local}=1-\text{ESS}_k$, if restrictions were in place at $t$ (in our baseline calibration, this is between March 16 and May 15, 2020), and $s_{k,t}^\text{local}=0$ if restrictions were not in place.

The last column in Table \ref{tab:foursquare} shows the essential score of the 20 2-digit NAICS industries. We can see that a few industries are completely essential (agriculture, mining, utilities, finance, management, health and public administration), only one is completely non-essential (arts, entertainment and recreation) and all the others are partly essential.

\subsubsection{Supply shocks in the Rest of the US}
\label{apx:supplyshocksrous}

Closures have been very different across states and imposed at different times. To get an estimate of supply shocks in the rest of the US, we rely on the data that have been made publicly available in Ref. \cite{barrot2021sectoral}. The authors examined executive orders issued in each State and, using a methodology similar to the one outlined above, gave an essential score to each 2-digit NAICS industry in each State. Weighing the data by state-industry employment shares, we compute an average essential score for each industry. Our estimates lead to essential scores that are generally higher than the ones in NY, especially for construction and manufacturing, which are 94\% and 86\% essential respectively (compared to 49\% and 56\%). Given the different times at which restrictions were imposed throughout the U.S., for simplicity we impose restrictions in the rest of the US at the same time as we impose restrictions in NY.

\subsubsection{Rest of US pandemic indicator}

While the number of deaths $D_t$ in the NY metro area is produced endogenously by the epidemic module, the time series of deaths in the Rest of the US, $D_t^\text{rest}$ is given exogenously. We compute it using data from the New York Times dataset.\footnote{\url{https://github.com/nytimes/COVID-19-data}}

\subsubsection{Government and other demand shocks}

Both in New York and in the Rest of the US, we model changes in government demand and demand for investment and net exports (``other final demand'') exogenously. We let changes in government demand $\mathcal G_t$ be given by $\mathcal G_t=-0.05$ for all days $t$ after the time restrictions are imposed. This corresponds to assuming a small 5\% increase in government demand throughout the pandemic period. At the same time, we impose a 30\% shock $\mathcal S_t$ to other final demand, i.e. $\mathcal S_t=0.30$. These shocks to government and other final demand lead to changes in \textit{realized} government consumption, investment and exports that are in line with the data (Figure \ref{fig:validation}A).

\FloatBarrier

\subsection{Synthetic population}
\label{apx:synthpop}

Our synthetic population is composed of 416,442 individuals grouped into 153,547 households.\footnote{The size of the synthetic population is determined by the number of Cuebiq users, see Sections \ref{apx:age} and \ref{apx:matching}.} Each synthetic individual has two unique ids: one id is a unique identifier, the other id corresponds to a Cuebiq user (not all individuals in the synthetic population are Cuebiq users). In addition, each individual is characterized by eight variables and attributes: household id, the census tract where they live, their age, income, employment status, occupation, industry where they work, and whether they can work from home. Of these, only the employment status changes over time, while we assume that the other attributes are fixed.

\begin{figure}[htbp]
    \centering
\includegraphics[width = 1\textwidth]{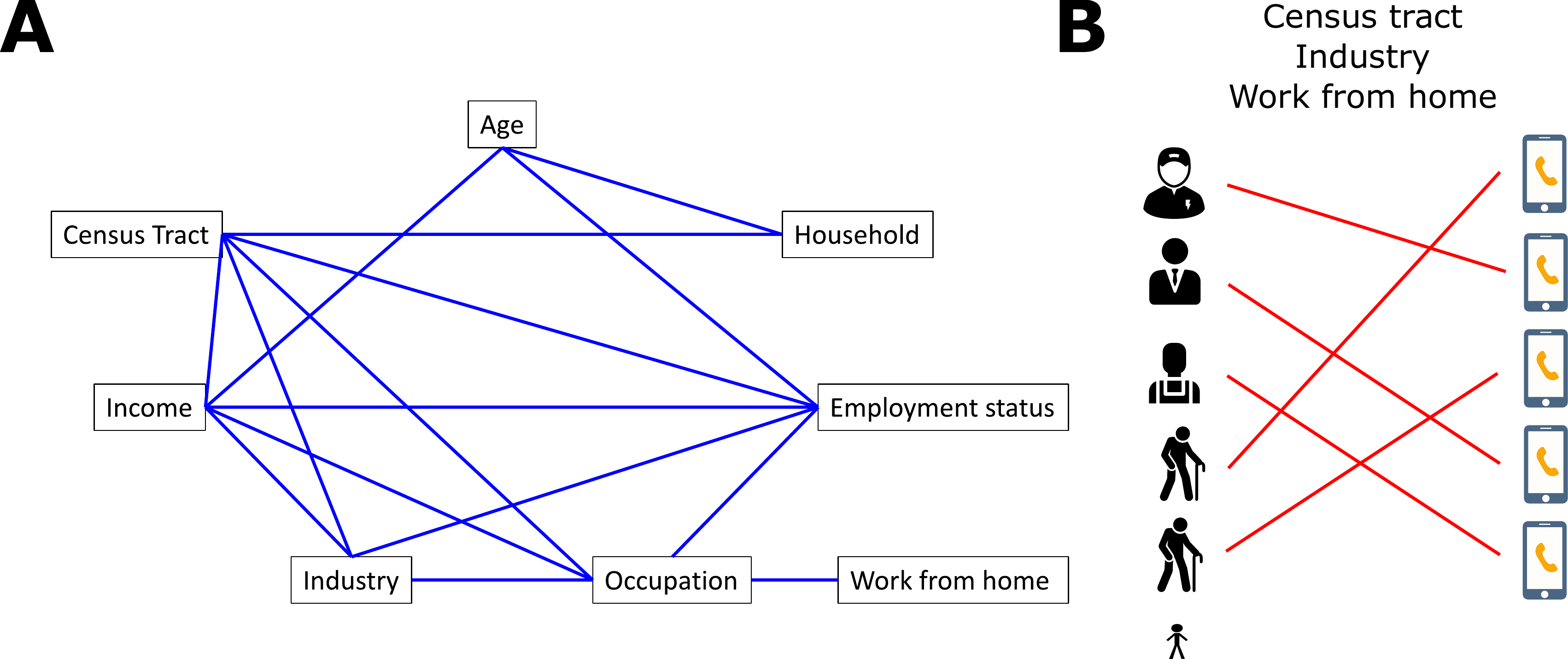}
    \caption{\textbf{Correlation between variables of individuals and matching between synthetic individuals and Cuebiq users.} A: a link between variables indicates that these variables are correlated by construction. When variables form a clique, e.g. income, industry and occupation, it means that they are all determined jointly. B: synthetic individuals are matched to Cuebiq users (mobility traces) mainly using information about the census tract where they live, but also based on the industry they work for and on whether they can work from home. }
        \label{fig:schematic_synth_pop}
\end{figure}

In the following sections, from Section \ref{apx:age} to Section \ref{apx:income}, we explain how we construct these quantities. In Section \ref{apx:summary_synthpop} we show summary statistics and a comparison between the synthetic population and the real New York population. Next, in Section \ref{apx:matching}, we explain how we match Cuebiq users to synthetic individuals. Finally, in Section \ref{apx:initialization_synthpop}, we explain how we initialize model variables based on the characteristics of synthetic individuals. As a reference through this section, Figure \ref{fig:schematic_synth_pop}A summarizes the correlations between variables that are obtained by construction.\footnote{Note that a distance of two in this graph could also imply some correlation. For example, age and industry may be correlated because both are correlated to income. }  

\subsubsection{Age, household, census tract}
\label{apx:age}

We generate as many adults in each census tract as there are Cuebiq users whose ``home'' is in the census tract, plus a few synthetic adults (1\% of all adults) that do not have an associated Cuebiq user but are needed to better match the statistical properties of household size.

We assign an age to synthetic adults, create synthetic children, and group them into households using the procedure documented in Ref. \cite{mistry2021inferring} (Supplementary Information Section 1). For New York, this procedure uses a combination of demographic tables from the ACS, similar to the ones used in the rest of this paper, as well as detailed microdata from IPUMS USA and other data sources. Microdata are used to compute the conditional distribution of a variable given a condition. For instance, microdata can be used to compute the probability that a spouse has a given age conditional on the age of the other spouse.

The approach, documented in detail in Scheme S1 in Ref. \cite{mistry2021inferring}, is to
\begin{enumerate}
    \item Sample household size from ACS demographic tables
    \item Determine the age of a reference adult household member using the conditional distribution of that age conditional on household size
    \item Determine the composition of the household by sampling from the conditional distribution of the existence of a given relation to the reference adult member determined above. Relations include: husband or wife, son, daughter, son-in-law, daughter-in-law, brother, sister, brother-in-law, sister-in-law, mother, father, mother-in-law, father-in-law, niece, nephew, grandchild, and others. 
    \item For each selected household member that has a relation of the type determined above with the reference household member, determine their properties sampling from a conditional distribution that depends on the reference member. For instance, the age of a spouse is sampled conditionally on the age of the reference member, to ensure that the age difference is in line with data.
\end{enumerate}

To ensure a more detailed spatial representation of the household structure, we resample household characteristics across census tracts. For instance, if the age of a household reference person is above median in a given census tract, we sample households with higher age of reference person with slightly larger probability.

As a convention, in this paper we assign to the adult with highest income in the household the title of ``household head''. 

\subsubsection{Employment status and industry.}
\label{apx:emp}

We assign employment status to individuals based on their census tract and their age using ACS Table B23001, reporting the number of individuals by age group\footnote{
While age in the synthetic population is a variable taking integer values between 0 and 103, only the following age groups are available in Table 23001: $[18,19], (19,21],  (21,24],  (24,29],  (29,34],  (34,44],  (44,54],  (54,59],  (59,61], 
(61,64],  (64,69],  (69,74],  (74+]$. We simply give our agents the same age groups as in Table 23001, assuming that all agents within each age group have the same employment rate.
} and census tract in each of the following categories: out of the labor force, unemployed,  employed.\footnote{
The employed category includes armed forces employees.
} We group individuals that are unemployed or out of the labor force into a unique category ``Not employed''. All individuals below 18 years old are, by construction, not employed. We end up, for each census tract and each age group, with a share of employed individuals. We take this share as the parameter of a Bernoulli distribution, and assign employment status to all agents in the same age group and census tract by drawing from such a Bernoulli distribution.
To all agents that are employed, we then assign one of the 20 NAICS 2-digit industries based on the same procedure, using ACS table C24030 to get industry shares of employment for each census tract. We then assign an industry to all employed agents within each census tract drawing from a categorical distribution (i.e., the multivariate version of a Bernoulli distribution) whose vector of probabilities is given by the industry shares of employment.

\subsubsection{Occupation and possibility to work from home}
\label{apx:occ}

The occupation of a worker is strongly linked to the industry they work for.  For instance, it is unlikely to find art performers in agriculture, while production occupations are likely in manufacturing. At the same time, occupations are strongly linked to geography, as, for example, most inhabitants of Manhattan are more likely to be engaged in knowledge-related occupations than inhabitants from other parts of the New York metro area. The ACS does not provide a cross tabulation for industries and occupations at the census tract level. To obtain a distribution of occupations that is jointly correlated with industry and geography, we proceed as follows.

We obtain the joint distribution of occupations and industries from the BLS Occupational Employment Statistics (OES), table natsector\_M2019\_dl.\footnote{
\url{https://www.bls.gov/oes/special.requests/oesm19in4.zip}
} For each of 20 industries, we know the occupational share of 22 occupations, which correspond to the most aggregated occupations in the SOC hierarchy (see Table \ref{tab:rliocc}).\footnote{
Not all occupations exist in any given industry.
} It should be noted that these data are valid for the US as a whole; however, we assume that the occupational composition of industries in NY is not too different from the one in the rest of the US (see Section \ref{apx:summary_synthpop} for a validation of this assumption). To obtain the distribution of occupations for each census tract, we use ACS table C24010, giving the share of employed workers for each of the 22 main occupations in each census tract. 

For all individuals that have been assigned a given industry, we then draw their occupation respecting both the relation between occupation and industry and that between occupation and geography. Focusing on a given industry, we first draw the number of employed workers in a given occupation from a multinomial distribution whose probability vector is given by the occupational share of that industry. We then assign the occupations to specific agents living in a census tract with a probability that is proportional to the number of employed workers in that occupation in their census tract. 

Let us consider an example. Assume that 100 individuals are employed in industry $X$, and that workers in industry $X$ are either performing occupation $Y$ (60 workers) or occupation $Z$ (40 workers). We then draw the number of workers in each occupation from a multinomial distribution with probability vector $[0.6,0.4]$, obtaining, for instance, that 58 workers are employed in occupation $Y$ and 42 in $Z$.\footnote{
In this example, it could also make sense to draw the occupations respecting the exact shares, i.e. 60 and 40 workers. However, this is not possible when certain occupations are very rare but still possible, as rounding to integers would imply that no agents are ever assigned such occupations.
} Assume further that the metro area has two regions, $A$ and $B$, and that the share of inhabitants of A who work in occupation $Y$ is twice that of inhabitants of B. So, to assign the 58 workers of $Y$, we select inhabitants of A with a probability that is twice as much as that of inhabitants of B.

\begin{table}[ht]
\centering
\begin{tabular}{llr}
  \hline
code & occupation\_title & remote\_labor\_index \\ 
  \hline
11-0000 & Management Occupations & 0.71 \\ 
  13-0000 & Business and Financial Operations Occupations & 0.77 \\ 
  15-0000 & Computer and Mathematical Occupations & 0.72 \\ 
  17-0000 & Architecture and Engineering Occupations & 0.63 \\ 
  19-0000 & Life, Physical, and Social Science Occupations & 0.49 \\ 
  21-0000 & Community and Social Service Occupations & 0.59 \\ 
  23-0000 & Legal Occupations & 0.54 \\ 
  25-0000 & Education, Training, and Library Occupations & 0.59 \\ 
  27-0000 & Arts, Design, Entertainment, Sports, and Media Occupations & 0.64 \\ 
  29-0000 & Healthcare Practitioners and Technical Occupations & 0.34 \\ 
  31-0000 & Healthcare Support Occupations & 0.18 \\ 
  33-0000 & Protective Service Occupations & 0.19 \\ 
   35-0000 & Food Preparation and Serving Related Occupations & 0.34 \\ 
  37-0000 & Building and Grounds Cleaning and Maintenance Occupations & 0.18 \\ 
  39-0000 & Personal Care and Service Occupations & 0.28 \\ 
  41-0000 & Sales and Related Occupations & 0.65 \\ 
  43-0000 & Office and Administrative Support Occupations & 0.54 \\ 
  45-0000 & Farming, Fishing, and Forestry Occupations & 0.09 \\ 
  47-0000 & Construction and Extraction Occupations & 0.21 \\ 
  49-0000 & Installation, Maintenance, and Repair Occupations & 0.19 \\ 
  51-0000 & Production Occupations & 0.17 \\ 
  53-0000 & Transportation and Material Moving Occupations & 0.18 \\ 
   \hline
\end{tabular}
\caption{Remote labor index across 22 occupations.}
\label{tab:rliocc}
\end{table}

Finally, we determine whether a worker is able to work from home using the remote labor index ($\text{RLI}$) introduced in Ref. \cite{del2020supply}. This index is bounded between 0 and 1 and captures the proportion of an occupation’s work activities that can be performed from home. When $\text{RLI}_o = 0$ none of occupation $o$'s work activities can be performed at home, while $\text{RLI}_o = 1$ indicates that all of the work activities related to occupation $o$ can be performed from home. Del Rio Chanona et al. \cite{del2020supply} calculated the remote labor index for occupations in the Standard Occupation Classification (SOC) system at the detailed level. We aggregate this index into the SOC system at the major level, which has 22 occupational categories. Table \ref{tab:rliocc} shows the remote labor index for each of the 22 occupations that we consider. As can be seen, workers in the service industries are much more likely to be able to work from home.

We interpret the Remote Labor Index as the probability that a worker can work from home. For each worker in occupation $o$, we decide whether they can work from home or not by drawing from a Bernoulli distribution with success probability $\text{RLI}_o$. We do this at the beginning of the simulation and assume that a worker's capability to work from home remains constant throughout the simulation.

\subsubsection{Income}
\label{apx:income}

We assign income to agents in such a way that income is correlated with industry, occupation, age and geography. As a general problem, different data sources use different definitions and conventions on income, and so it is difficult to compare income levels across datasets. To address this issue, we proceed in steps changing the relative income of different socio-economic groups, and then renormalize income levels at the end to the level indicated in the Consumption Expenditure Survey.\footnote{
https://www.bls.gov/cex/2019/msas/norteast.pdf
} (We normalize to these values for consistency with our initialization of the consumption patterns, see Section \ref{apx:cons}.)

We derive income for each industry-occupation pair from the same BLS OES data that we used to derive the share of occupations within industries. Information on income is presented in terms of quantiles: the BLS provides 10th, 25th, 50th (median), 75th and 90th quantiles.\footnote{A few data points for the 90th quantile, and one for the 75th quantile, are missing for some combinations of industry and occupation. In these cases, we impute the missing quantiles so that the ratio to the median is the same as the average ratio of the quantile to the median for the occupation-industry pairs for which this information is available.} Because we cannot sample incomes from quantiles, we fit a log-normal distribution for each set of quantiles corresponding to an occupation-industry combination.\footnote{There is a debate on whether the best fit to income distributions is with a log-normal or with a Pareto distribution, with some consensus that the Pareto distribution is more appropriate for the tail and the log-normal for the rest of the distribution. Because here we are more interested in the body of the distribution, we choose a log-normal functional form. See e.g. Ref. \cite{pinkovskiy2009parametric} for some pointers to the literature.} In the vast majority of cases, the log-normal provides an excellent fit. (Results available in the replication materials.) We then assign income to all employed individuals drawing from the log-normal distribution whose parameters correspond to the occupation-industry pair of that individual. 

The next step is to adjust income levels by age.\footnote{In principle, differences in income across age could be explained by the different probability to be employed in different industries and occupations by age. Because we could not find data to make the assignment of industry and occupation dependent on age, this correlation is not present in our synthetic population.} We use the Current Population Survey table ``Median usual weekly earnings of full-time wage and salary workers by age, race, Hispanic or Latino ethnicity, and sex, not seasonally adjusted''\footnote{https://www.bls.gov/webapps/legacy/cpswktab3.htm} to derive 2019 income levels in the following age groups: $[18,19],  (19,24],$  $(24,34],  (34,44],  (44,54],  (54,64],  (64+)$. In Q4, mean weekly earnings for these age groups were, respectively, $\$ 494, \$ 616, \$ 876,$ $\$1047, \$1039, \$1053, \$942$. We rescale incomes across age groups so that they are proportional to these values.

After that, we adjust individual incomes so that the resulting household incomes, as averaged over each census tract, match the mean household income in the same census tract as obtained from Census table S1901 (2019, 5-year ACS average). 

So far, we only assigned an income to employed individuals. As a final step, we need to assign an income to individuals who are out of the labor force. This is a difficult step, as it is difficult to find data sources that consistently compare their income to that of workers, and even for workers there are other sources of income than wages and salaries. For instance, focusing on retirement income, Bee and Mitchell \cite{bee2017older} find that using different data sources and definitions of income leads to a 30\% increase in retirement income.  As the majority of non-employed individuals who still receive a substantial income is likely to be composed by retirees, we decide to focus solely on this group. In line with our attempt to be consistent with the Consumption Expenditure Survey, we choose mean retirement income in such a way that the mean income of individuals above 65 years old is about 60\% of that of individuals between 55 and 64 years.\footnote{
https://www.bls.gov/cex/2019/combined/age.pdf
} Given mean income, we assign income values based on a log-normal distribution calibrated over the universe of workers (i.e., not specific industries or occupation). We also make sure that retirement income is rescaled by mean household income across census tracts, obtained from table S1901 as above.

Our final step, as mentioned at the beginning, is to normalize all incomes so that the mean income in the synthetic population corresponds to the 2019 mean income for the New York metro area as indicated in the Consumption Expenditure Survey.

\subsubsection{Summary statistics}
\label{apx:summary_synthpop}

In this section we show summary statistics for the variables in the synthetic population. When possible, we also compare these summary statistics to data, as a way to validate our approach.

\begin{figure}[!h]
    \centering
\includegraphics[width = 1\textwidth]{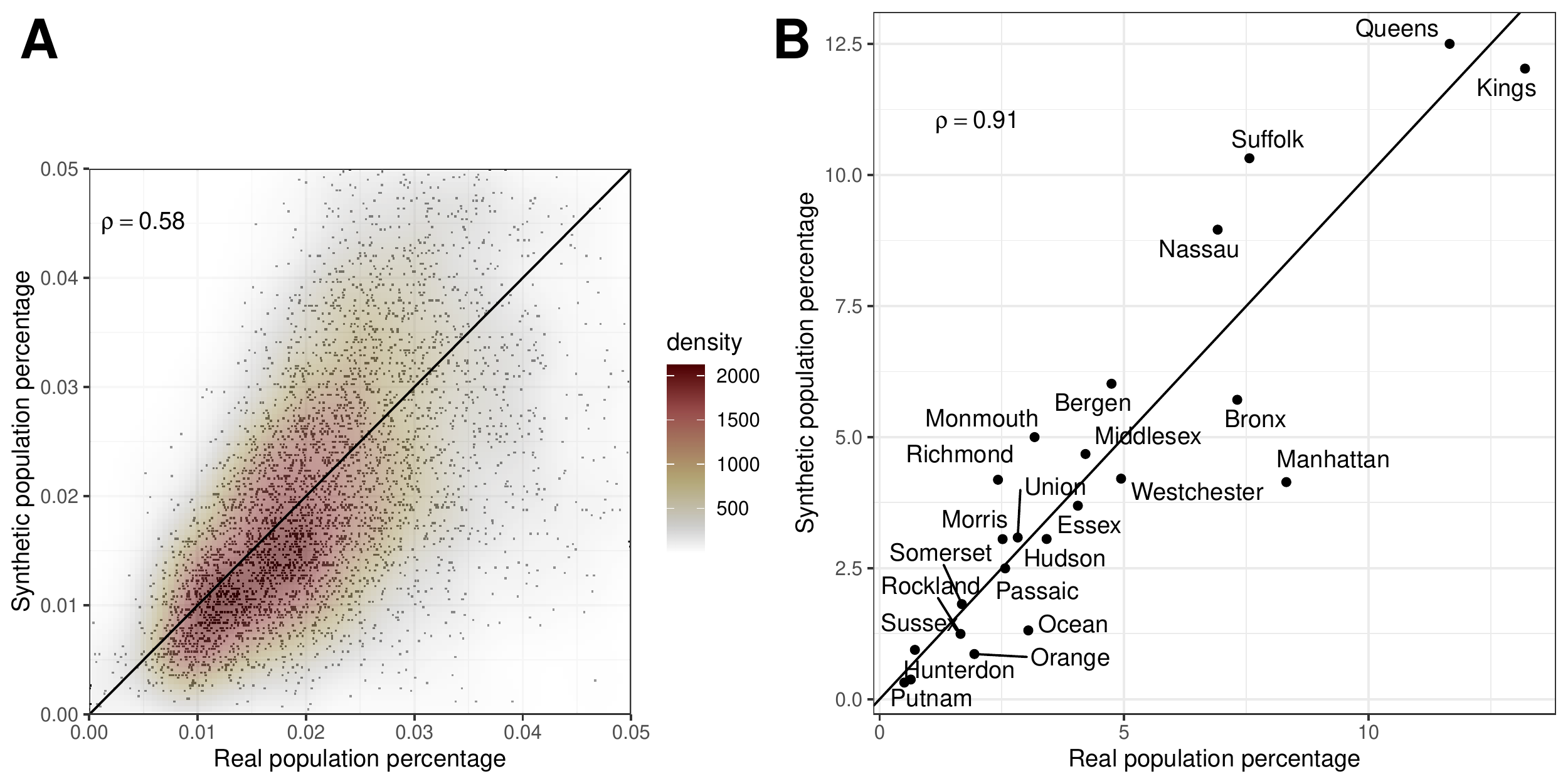}
    \caption{\textbf{Percentage of the real and synthetic populations.} A: percentage at the census tract level. On average, each census tract is 0.02\% of both the real and synthetic population. The Pearson correlation coefficient between the census tract shares in the real and synthetic populations is 0.60. We use a density kernel to show the areas with highest density of points. B: percentage at the county level. Each county makes up on average 4-5\% of both the real and synthetic populations, although some counties such as Queens and Kings are much more populous than others. The Pearson correlation coefficient between the county shares in the real and synthetic populations is 0.92.   }
        \label{fig:percentage_population}
\end{figure}

We begin in Figure \ref{fig:percentage_population}, by showing the shares of both the real and synthetic populations, both across census tracts and counties. There are 23 counties and approximately 4300 census tracts, so, on average, each county makes up 4-5\% of the population, and each tract makes up 0.02\% of the population. However, as it is immediate from the figure, there is a lot of heterogeneity, as some counties and tracts are much more populous than others. For instance, Queens and Kings have many more inhabitants than Putnam or Hunterdon. 

The population shares are not identical in the real and synthetic populations, as can be seen from the fact that not all points are aligned on the identity line. This is because we build our synthetic population based on the number of Cuebiq users in each census tract, and the number of Cuebiq users is not exactly proportional to the population of the census tract. Despite this limitation, there is a good correlation between the real and synthetic population shares, with a Pearson correlation coefficient $\rho=0.60$ in the case of census tracts, and $\rho=0.92$ in the case of counties. The main outlier in the case of counties is Manhattan (New York county), that makes up around 8\% of the real population and only around 5\% of the synthetic population. This is partly due to Cuebiq users leaving New York when the pandemic began (we only retain users that we could identify for the full period, i.e. that were observed in the last two months of the period (May-June 2020) at least once; see Ref. \cite{aleta2022quantifying}).

\begin{figure}[htbp]
    \centering
\includegraphics[width = 1\textwidth]{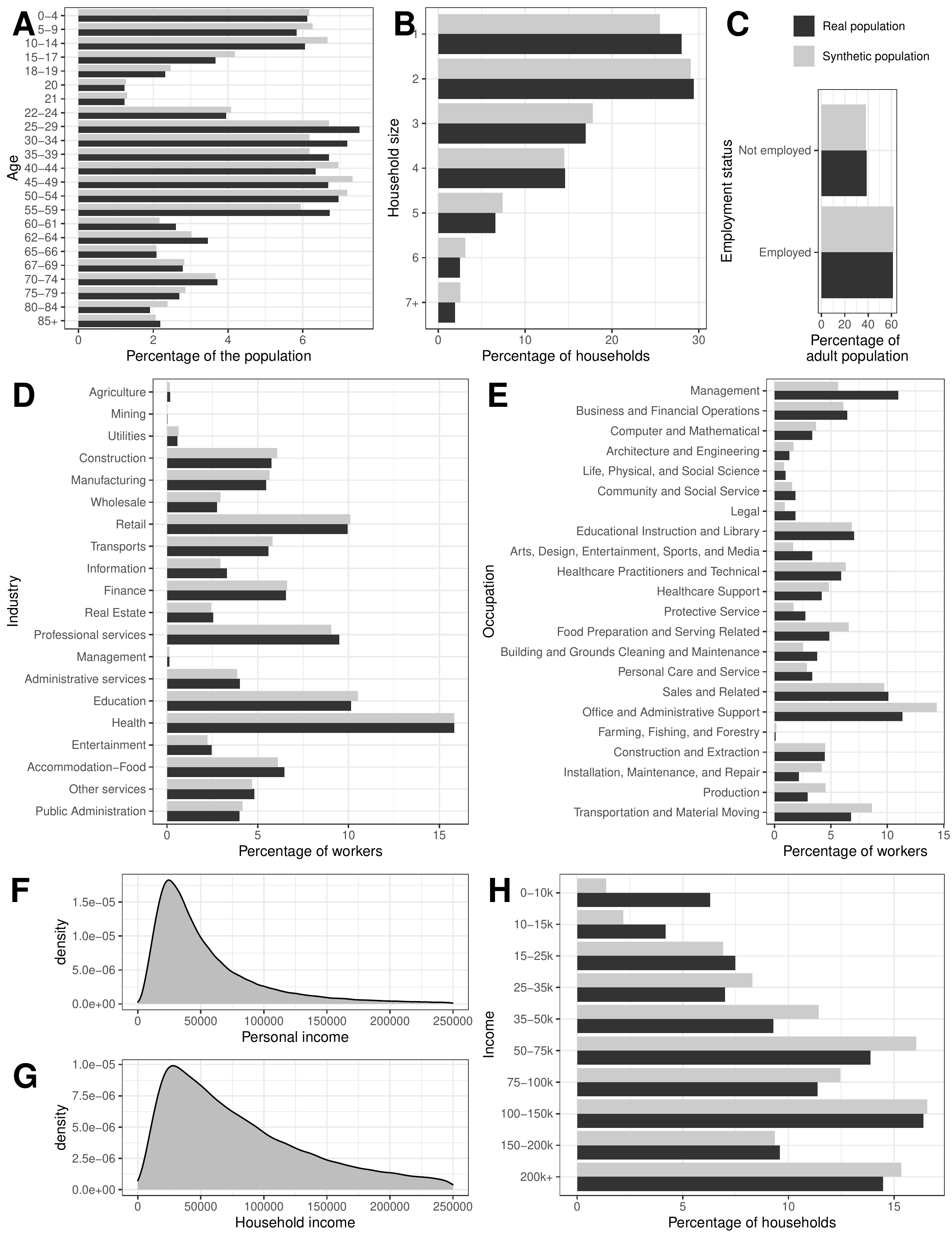}
    \caption{\textbf{Marginal distributions of the main population attributes.} We compare distributions in the real population (dark grey) and in the synthetic population (light grey).   }
        \label{fig:plots_marginals}
\end{figure}

Figure \ref{fig:plots_marginals} shows the marginal distributions of the main attributes of synthetic individuals: age, household size, employment status, industry, occupation, and income. It compares distributions in the real population (dark grey) and in the synthetic population (light grey). 

Starting with age (panel A), we can see that the age distribution in the synthetic population closely resembles that in the real population. The only exceptions are intermediate years, with a slightly larger share of individuals between 25 and 39 in the real population compared to the synthetic population, and a slightly smaller share between 40 and 54. It is difficult to compare the relative size of different age groups as age intervals comprise different numbers of years, but we can see that the bulk of the distribution is concentrated in intermediate years, as in most age distributions in advanced economies.

Panel B shows that household size is very similar in the synthetic and real population. Most households are composed by one or two individuals, but around 2\% of households is composed by seven individuals or more. 

As we can see in panel C, a bit more than 60\% adults are employed, with the remaining share being non-employed. This is true both in the real and in the synthetic population.

Panel D shows the percentage of workers employed in each of the 20 2-digit NAICS industries that we consider. We can see that health is the industry that employs most workers, followed by education, retail and professional services. Unsurprisingly, given the structure of the New York economy, agriculture and mining employ a tiny share of workers. (Management also employs few workers, but this is due to the NAICS industries definition.) The real and synthetic populations correspond almost perfectly.

Panel E shows the percentage of workers employed in the 22 main SOC occupations that we consider. Occupations such as Office and Administrative Support, Sales and Related, Management, Business and Financial Operations, Educational Instruction and Library are among the most frequent occupations, while specialized occupations such as Architecture and Engineering, Life, Physical and Social Science, Legal, are less common. Here, the match between the synthetic and real populations is less good than in the previous panels. This is due to the algorithm that we use to assign occupations respecting both the relation to industry and geography, as explained in Section \ref{apx:occ}. In particular, it is mostly due to the assumption that the occupational composition of industries in New York is the same as in the rest of the US. For instance, we can see in the figure that our algorithm overestimates the fraction of Office and Administrative Support workers and underestimates the fraction of workers in Management occupations, relative to the real population. This is likely because managers of US corporations are likely to reside in New York, while Office and Administrative Support workers are more likely to reside in less expensive areas. Yet, we see that this shortcoming is limited to a few industries, as the percentage of workers is similar in the real and synthetic populations in the majority of cases.

Panels F and G show the distribution of both personal and household income. We see that the peak of the distribution is below \$50,000 in both cases, but the distribution has a long tail, extending above \$250,000. We compare household income in the real and synthetic populations in panel H.\footnote{
We rescale incomes from Table S1901 so that the mean income corresponds to the one in our dataset. This is because mean income in the New York metro area is \$116604 in the ACS and \$103011 in the Consumption Expenditure Survey (our benchmark).
} We see that the correspondence is generally good, except for the very low income categories, which the synthetic population underestimates. This is due to the problem mentioned in Section \ref{apx:income}, that it is difficult to assign an income to individuals out of the labor force, which arguably comprise the majority of individuals in those income categories. As we cannot accurately compare income for part time and occasional work and unemployment benefits to standard labor income, we cannot properly account for these low-income categories.

\begin{figure}[htbp]
    \centering
\includegraphics[width = 1\textwidth]{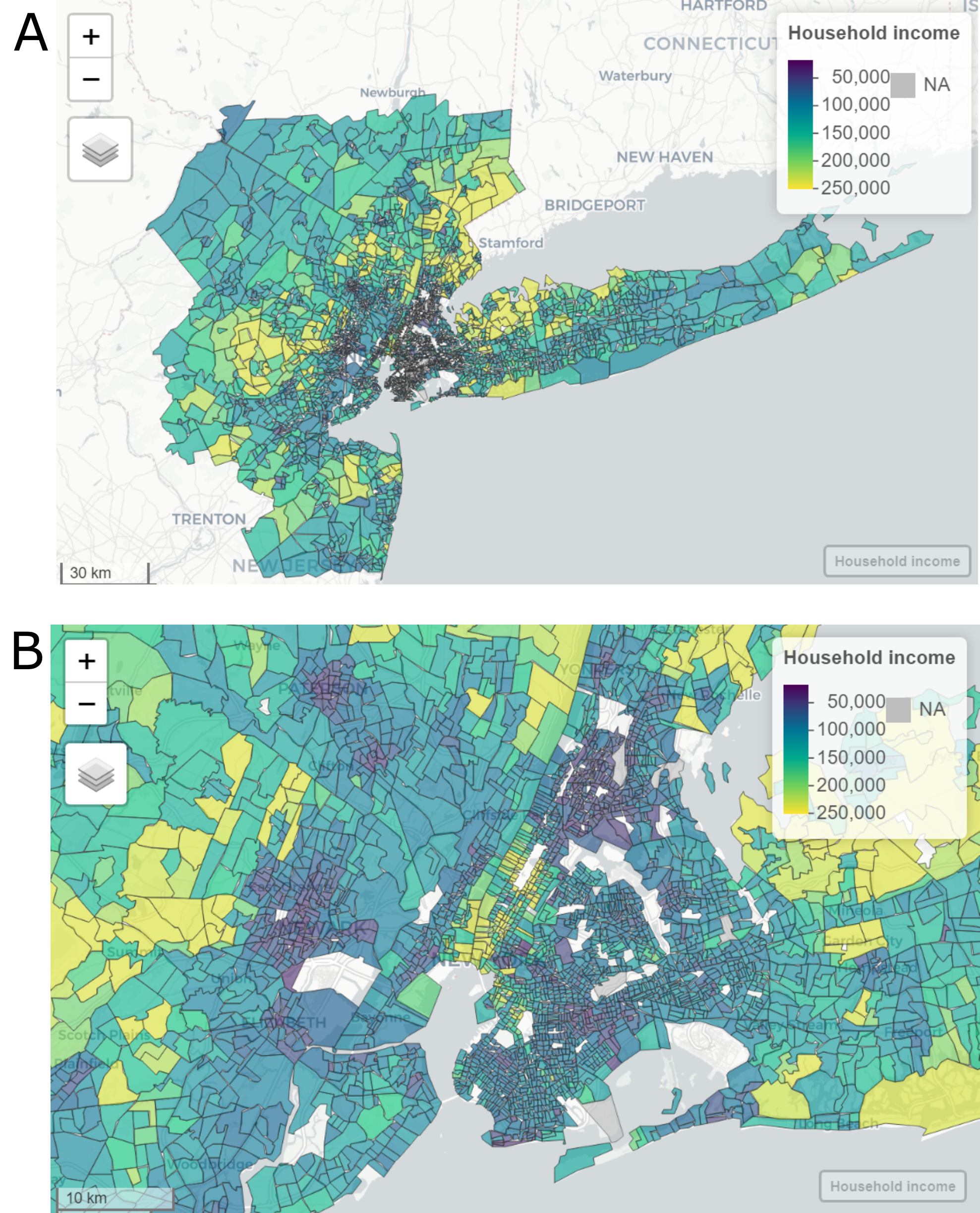}
    \caption{\textbf{Maps of census tracts by mean household income.} Panel A shows the New York metro area as a whole, while panel B zooms in on New York City. To ease visualization, we limit mean census-tract household incomes to \$250,000.}
        \label{fig:tracts_by_income}
\end{figure}

\begin{figure}[!h]
    \centering
\includegraphics[width = 1\textwidth]{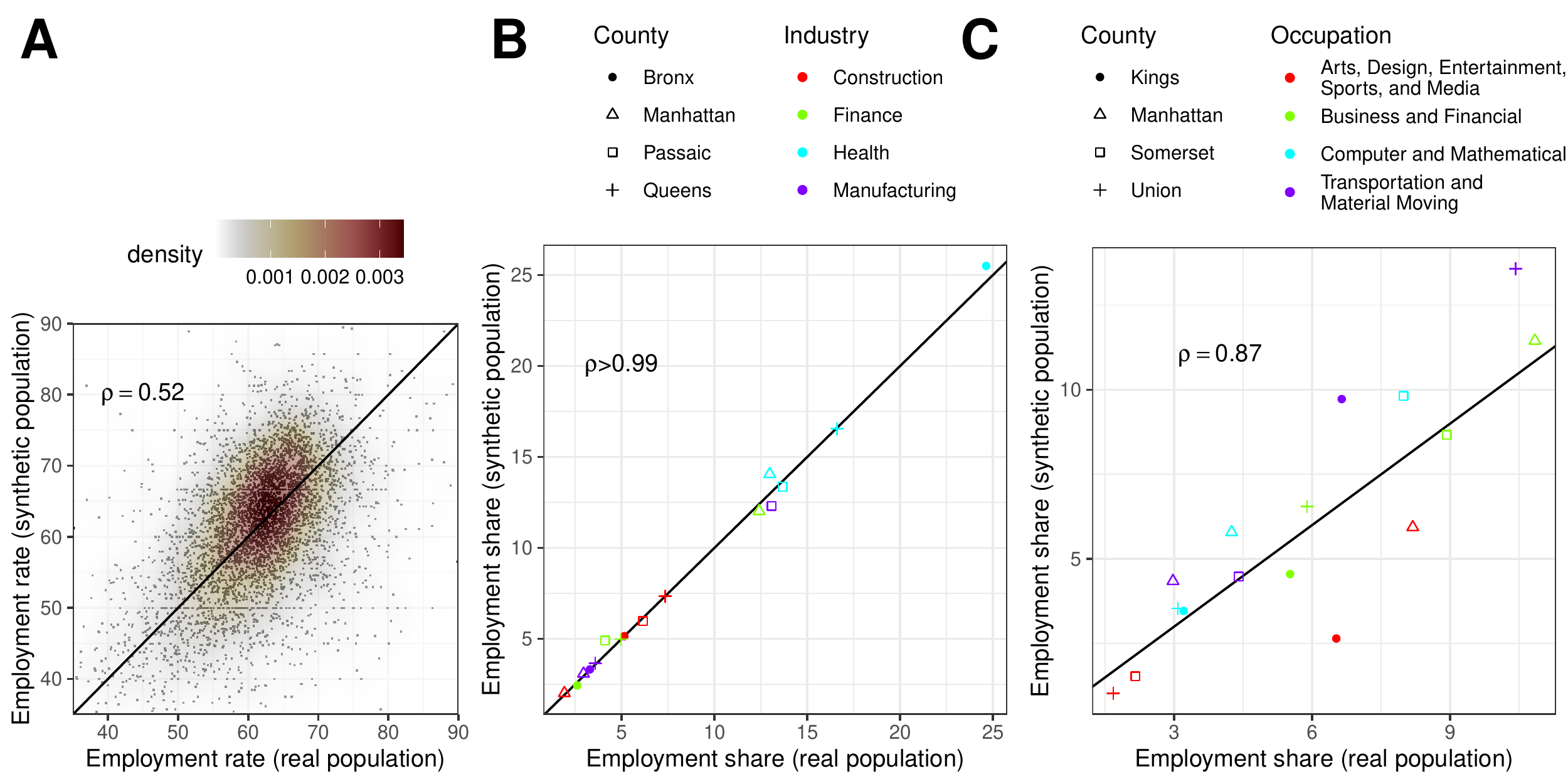}
    \caption{\textbf{Variables by geography.} A: employment rate across census tracts. The Pearson correlation correlation coefficient between the real and synthetic populations is $\rho=0.52$. We use a density kernel to show the areas with highest density of points. B: County share of employment for a few selected counties and industries. There is almost perfect correlation between the model and the data ($\rho=1.00$). C: County share of employment for a few selected counties and occupations. The correlation is still very good ($\rho=0.87$).  }
        \label{fig:plots_joint_geo}
\end{figure}

After looking at marginal distributions in Figure \ref{fig:plots_marginals}, we now show joint distributions. We begin with joint distributions of geography and other attributes in Figures \ref{fig:tracts_by_income} and \ref{fig:plots_joint_geo}. Figure  \ref{fig:tracts_by_income} shows maps of census-tract mean household income. We can see that while high-income areas are clustered in a few zones, they are spread in the metro area, beyond New York City. The Bronx and the Queens are among the lowest-income areas. These data come from the synthetic population but, as we explained in Section \ref{apx:income}, the spatial income distribution is the same as in the real population because we renormalized all incomes to the mean income of each census tract.\footnote{
In principle, this normalization may not be necessary, as long as differences in industry and occupation explain differences in income across census tracts. Indeed, we find that the Pearson correlation coefficient between mean household income in the real and synthetic population, across census tracts and before renormalization, is $\rho=0.58$. To more accurately match geography, we prefer to nonetheless proceed with the renormalization described above, but this correlation suggests that the population building workflow yields consistent results.
}

Figure  \ref{fig:plots_joint_geo} shows the relations between geography and other variables. Panel A shows the employment rate across census tracts. The mean employment rate is a bit above 0.60 in both the real and synthetic population, and the bulk of the distribution is concentrated on employment rates between 55 and 70 in the real population, and 50 and 75 in the synthetic population. The larger dispersion in the synthetic population is likely due to sampling. In any case, there is a good correlation ($\rho=0.52$) between employment rates in the real and synthetic population at this very small spatial level. 

Panel B shows the percentage of employment of a given industry in a given county, for a few selected industries and counties. We can see that there is some specialization, with some counties having a much larger share of employment in a given industry than other counties. This corresponds to the situations where one marker of a given color is very distant from the other markers. For instance, almost 25\% of Bronx workers are employed in Health, compared to 10-17\% in the other selected counties. Likewise, about 13\% of workers in Passaic are employed in manufacturing, against 3\% in the other counties. Unsurprisingly, 13\% of workers in Manhattan are employed by the financial industry, against 3\% in the other counties. The synthetic population captures almost perfectly this variability that exists in the real population ($\rho=1.00$).

Panel C is similar to panel B, but it focuses on occupations. Here, we see for example that Union county is specialized in Transportation and Material Moving occupations, Somerset in Computer and Mathematical occupations, etc. In this case, the correlation between employment shares in the real and synthetic populations is not perfect, but still good ($\rho=0.87$). This is again due to the algorithm explained in Section \ref{apx:occ}.

\begin{figure}[!h]
    \centering
\includegraphics[width = 1\textwidth]{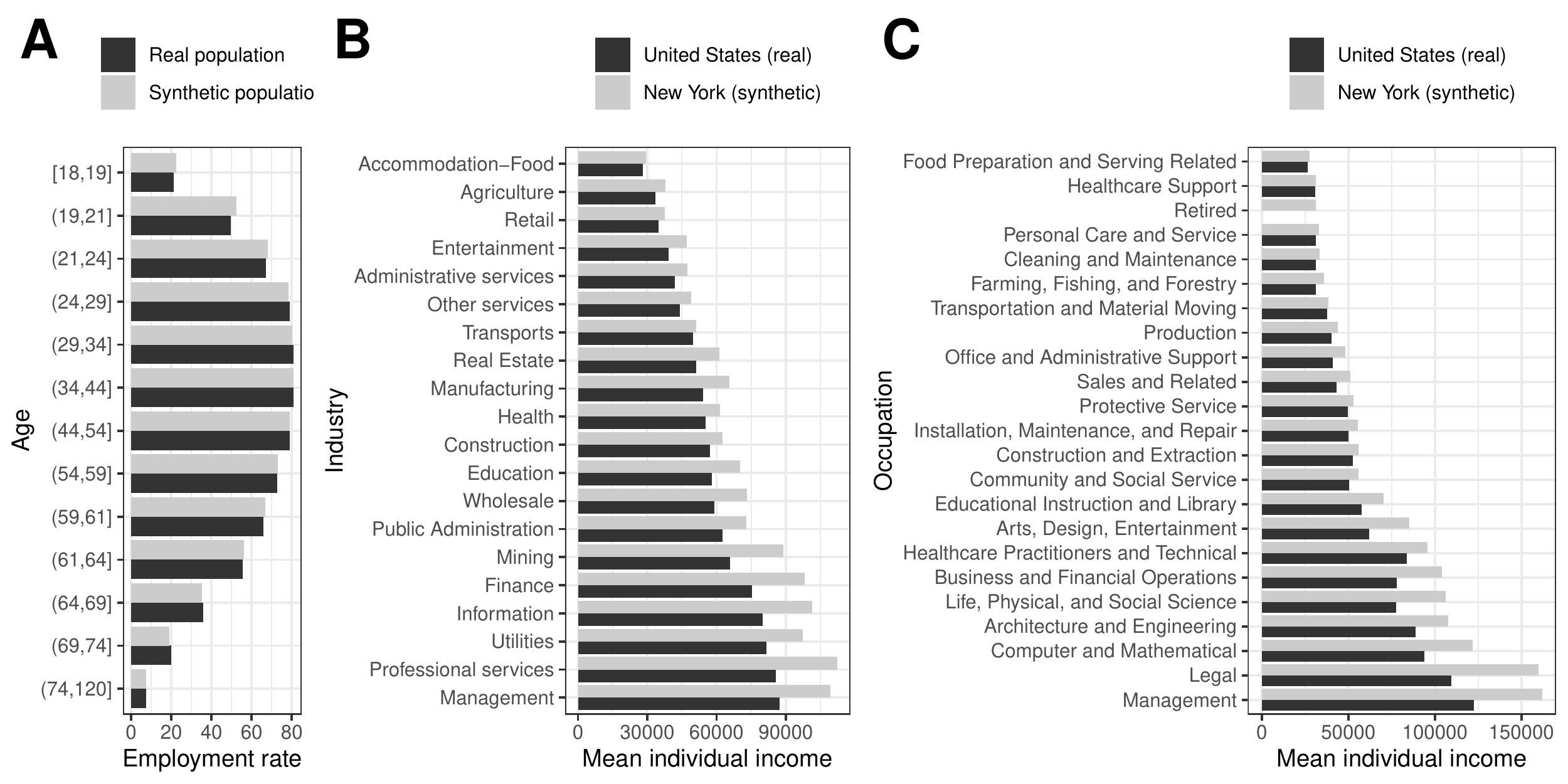}
    \caption{\textbf{Comparisons of joint relations in the real and synthetic populations.} A: employment and age. B: income and industry. C: income and occupation.    }
        \label{fig:plots_joint_bars}
\end{figure}

After showing joint relations between geography and other variables, in Figure \ref{fig:plots_joint_bars} we show other joint relations. In panel A we look at the employment rate across age groups. As we sampled from the joint distribution of employment and age, the employment rate is almost the same in the real and synthetic population. We see that the employment rate is around 80\% between 24 and 54 years old, decreasing at the other ends of the age distribution.

Panel B shows mean individual income across industries. We compare the data for the US as a whole from the BLS dataset that we used in Section \ref{apx:occ} with data for New York in our synthetic population (we do not have real data on this joint relation for New York). We see that the industries that pay their workers the least are accommodation-food, agriculture, retail and entertainment, while industries paying their workers the most are management, professional services, utilities and information. This ranking is the same (except for a few exceptions) both in the real data valid for the US and in the synthetic data valid for New York. As expected, income levels are higher in New York.

Panel C is analogous to panel B, but for occupations. It includes ``Retired'' agents, who are not employed, for comparison. We see that the least paid occupations are Food Preparation and Serving Related, Healthcare Support, Cleaning and Maintenance; Retired individuals also receive very low income. Conversely, the highest paid occupations are Management, Legal, Computer and Mathematical, and Architecture and Engineering.

\begin{figure}[htbp]
    \centering
\includegraphics[width = 1\textwidth]{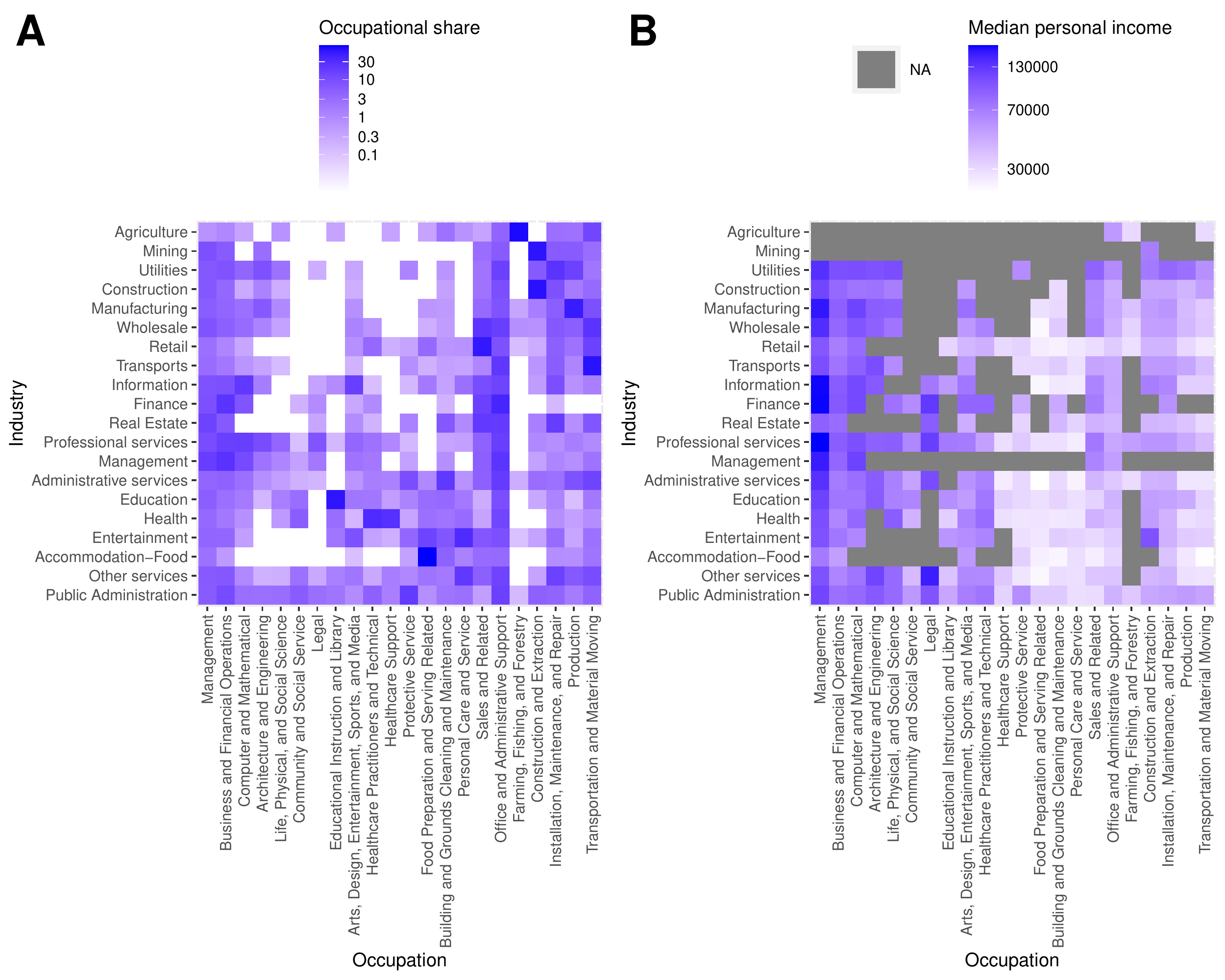}
    \caption{\textbf{Industry-Occupation cross tabulation.} A: heat map showing the occupational share of each industry in the synthetic population. B: heat map showing median personal income for each industry-occupation pair in the synthetic population. Grey squares indicate that there are fewer than 10 individuals working in an industry-occupation pair (so that results would be too dependent on stochastic realizations of the log-normal distribution).   }
        \label{fig:plots_joint_occ_ind}
\end{figure}

As a final illustration of the properties of the synthetic population, Figure \ref{fig:plots_joint_occ_ind} shows a cross tabulation of industries and occupations in the synthetic population. Panel A shows the occupational share of each industry (rows sum to 100). We see that some occupations are spread across all industries more or less equally. For instance, there are some workers in Management, Business and Financial Operations, Sales and Related, Office and Administrative Support occupations in almost all industries, although these occupations are slightly more concentrated in Information, Finance, Real Estate, Professional services and Management. Other occupations are concentrated in a few industries. For instance, Farming, Forestry and Fishing Occupations take up the majority of jobs in Agriculture, Production occupations in Manufacturing, Transportation and Material Moving in Transportation. Certain occupations are almost not represented in certain industries (white squares). 

Panel B shows the same cross tabulation, but the intensity of color indicates median personal income of workers in each industry-occupation pair. We see that income changes more strongly with the occupation than with the industry (there is more variability across columns than across rows within the same column), but certain industries pay their workers more than others, even for the same occupation. 

\subsubsection{Matching to Cuebiq users}
\label{apx:matching}

Our synthetic population features 319,487 adults, and we have 316,070 Cuebiq users. Ideally, we want to match synthetic individuals and Cuebiq users so that each Cuebiq user is matched to a synthetic individual that has the same characteristics as the real user. To do so, we match synthetic individuals and Cuebiq users based on the census tract, the industry, the possibility to work from home and the employment status (see the schematic in Figure \ref{fig:schematic_synth_pop}B). 

As a first step, we restrict the samples of synthetic individuals and Cuebiq users so that synthetic individuals living in a census tract are matched to Cuebiq users whose ``home'' location is in that census tract. As we showed in Section \ref{apx:summary_synthpop}, our synthetic population captures a lot of the variability across census tracts, so already a correct matching based on geography is likely to provide an overall accurate matching.

To further refine this matching, we look at the number of stays of Cuebiq users at their workplace. We rank the Cuebiq users in a census tract by visits to the workplace (which can be zero) and assign to the non-employed synthetic agents the Cuebiq users with minimal number of stays in the workplace. Because workplace data are available since February 2020, when most employed people were normally going to work in person, we are unlikely to be conflating agents who can work from home and unemployed agents.

As another refinement, we assign synthetic agents that cannot work from home and that work in essential industries to Cuebiq users whose number of stays in the workplace decreased the least in April and May compared to February and March. More precisely, we assign to each agent a score of ``essentialness-in-person'', given by the product of the essential score of the industry it works for (from the last column in Table \ref{tab:foursquare}) and an indicator function that is 0 when the agent can work from home and 1 when the agent must work in person. We then match the synthetic individuals with the highest values of this score with the Cuebiq users whose ratio between ``visits to the workplace in April-May'' and ``visits to the workplace in February-March'' is highest. For instance, we are likely to match a synthetic nurse, who works in-person in an essential industry, to a Cuebiq user who did not stop physically going to work in April and May. At the same time, we could match a synthetic financial analyst to a Cuebiq user who stopped going to the office in April and May.

\subsubsection{Initialization of model variables}
\label{apx:initialization_synthpop}

The synthetic population is one of the key building blocks of both the epidemic and economic modules. All our efforts to get the joint relations between socio-economic attributes right is crucial for the results where we marginalize on some variables. For instance, getting the relation between industry and income right is crucial to explain differences in employment across income groups, and getting the relation between age and employment is very important to explain differences in infection rates across ages. Apart from its relevance for the results, information from the synthetic population is used to directly initialize some of the variables of the epidemic and economic module.

In terms of the epidemic module, we use it to initialize $\omega_{ij}^S$ and $\omega_{ij}^H$, as defined in Section \ref{apx:epi_model}. 

For the economic module, we use information on employment, industry, and ability to work from home to create lists of both in-person and from-home workers for each industry, and to initialize $l_{k,0}^\text{P}$ and $l_{k,0}^\text{H}$, for all industries $k$.\footnote{As we do not have a synthetic population for the rest of the US, we obtain industry employment data from the ACS and rescale initial employment levels to be consistent with the size of the NY synthetic population. In other words, given that there are almost 200 thousand workers in the NY synthetic population, compared to 9.6 million in the real population, we rescale the total number of workers in each industry in the rest of the US by a factor equal to 200,000/9,600,000. } As we explain in more detail in Section \ref{apx:cons}, we use information on age and income to group consumers into groups $g$ with the same consumption patterns.

\FloatBarrier

\subsection{Input-output}
\label{apx:io}

We first obtain input-output data for the US as a whole (Section \ref{sec:national_io_tables}), and then follow the literature on regionalization to break it down into a New York-Rest of the US input-output system (Section \ref{apx:regionalization}). Finally, in Section \ref{apx:initialization_inputoutput} we explain how we use this information to initialize model variables.

\subsubsection{Symmetric industry input-output tables}
\label{sec:national_io_tables}

Our input-output data is taken from the US Bureau of Economic Analysis (BEA) and based on the most recent year (2019). Our model is initialized using data on industries, but the BEA provides input-output data as a mix between industries and \textit{commodities}. Roughly, a commodity is defined as the main output of an industry, but a firm that is classified in a main industry may also produce commodities that are classified in another industry. For instance, a hotel that serves dinner to its customers also produces restaurant services; if hotels and restaurants are listed as separate industries, this fact needs to be taken into account.

We derive industry by industry tables from the Make and Use Tables (after redefinitions, at basic prices) provided in the BEA Supplemental Estimate Tables\footnote{
\url{https://www.bea.gov/industry/input-output-accounts-data}
}. To derive the input-output tables, we follow the industry technology assumption outlined in Ref. \cite{bea2009concepts} which we briefly summarize below.

Let $V$ denote the \emph{industry} $\times$ \emph{commodity} make matrix and $U$ the \emph{commodity} $\times$ \emph{industry} use matrix. We define $q$ and $g$ as the vectors of commodity and industry specific outputs, respectively. Vector $g$ can contain an industry's production of scrap, $h$. We let $x=g-h$ denote industry output excluding scrap.
We further use a circumflex to indicate a diagonal matrix, e.g. $\hat{q} = \text{diag}(q)$.

By defining the transformation matrix
\begin{align}
    T &= g \hat{ x }^{-1} V \hat{q}^{-1}, 
\end{align}
we can derive the matrices of intermediate consumption $Z$ and the technical coefficient matrix $A$ as
\begin{align}
    Z &=    T U, \\
    A &= Z x^{-1}.
    \label{eq:techcoef}
\end{align}

Further to trade of intermediate goods, the other key component of input-output tables is final demand. The BEA provides information on final demand as disaggregated in several sub-components: private consumption, government consumption (defense and non-defense), investment in residential and non-residential structures, machinery and other investment goods, exports and imports. We aggregate these sub-components into three final demand categories: private consumption, government consumption, and ``other final demand''. At this point, we have data on final demand by commodity, i.e. we have a matrix $f^c$ whose rows correspond to different commodities and whose columns are the three main categories of final demand listed above. We obtain final demand by industry $f$ applying the same transformation matrix $T$:
$$
f = T f^c.
$$

Our desired number of industries is 20 (2-digit NAICS industries), as this is the finest breakdown of industry information at the census tract level in the ACS. However, the BEA does not provide input-output information at this level of aggregation. Indeed, it either provides information at the level of 15 industries/commodities, or at the level of 3-digit NAICS codes, which comprise 71 industries/commodities. Therefore, we perform the steps above with 71 industries/commodities, and then aggregate gross output $x$, intermediate consumption $Z$ and final demand $f$ to our target 20 industries. We finally re-compute the technical coefficients matrix $A$ using the same formula as above.

\subsubsection{Regionalization}
\label{apx:regionalization}

Since there exists no input-output data on the NY MSA, we use standard non-survey techniques to derive the regional relationships from national tables \cite[ch.8]{miller2009input}.
Specifically, our regional tables are obtained by using the Flegg et al. Location Quotient (FLQ) due to its superior performance over simpler alternative methods \cite{flegg2000regional}.\footnote{We provide a short summary of our approach here but refer to Refs. \cite{bonfiglio2008assessing}, \cite{miller2009input} and the recent literature on industrial ecology virtual laboratories \cite{geschke2017virtual} for a more detailed and critical discussion. } We then obtain the input-output relations for the rest of the US (RoUS) as a residual from the NY MSA regional tables.

Our goal is to split the intermediate consumption matrix $Z$ into matrices of the value of intermediate goods flowing (i) from industries in NY to industries in NY, $Z^{\text{NY}->\text {NY}}$; (ii) from industries in NY to industries in the RoUS, $Z^{\text{NY}->\text {RoUS}}$; (iii) from industries in the RoUS to industries in NY, $Z^{\text{RoUS}->\text {NY}}$; (iv) from industries in the RoUS to industries in RoUS, $Z^{\text{RoUS}->\text {RoUS}}$. Moreover, we want to divide final demand $f$ into final demand of: (i) agents in NY to industries in NY, $f^{\text{NY}->\text {NY}}$; (ii) agents in RoUS to industries in NY, $f^{\text{NY}->\text {RoUS}}$; (iii) agents in NY to industries in RoUS, $f^{\text{RoUS}->\text {NY}}$; (iv) agents in RoUS to industries in RoUS, $f^{\text{RoUS}->\text {RoUS}}$. (In all cases, arrows indicate the direction of the goods and services, from the producer to the consumer.)

Let us start from intermediate consumption. Our starting point is that industries in NY produce goods and services with the same shares of inputs (``recipes'') as in the rest of the US (and as in the US as a whole). Thus, if a certain input is not produced in NY, it must be imported from elsewhere in the US. For instance, a manufacturing firm needing inputs from the mining industry needs to import those inputs from the rest of the US. Defining $A^{\text{NY}->\text {NY}}$ as the technical coefficient matrix for industries located in New York demanding intermediate goods from industries also located in NY, and $A^{\text{RoUS}->\text {NY}}$ as the equivalent technical coefficient matrix for imports from the RoUS, this assumption is translated in the constraint 
\begin{equation}
    A^{\text{NY}->\text {NY}}+A^{\text{RoUS}->\text {NY}}=A.
    \label{eq:idcoef1}
\end{equation}
Additionally, considering industries located in the RoUS, the following constraint must hold:
\begin{equation}
    A^{\text{NY}->\text {RoUS}}+A^{\text{RoUS}->\text {RoUS}}=A.
    \label{eq:idcoef2}
\end{equation}
It is intuitive that coefficients in $A^{\text{RoUS}->\text {NY}}$ are likely to be much larger than coefficients in $ A^{\text{NY}->\text {RoUS}}$. Indeed, a larger share of inputs for firms in NY come from the RoUS than for firms in the RoUS from NY, if only by a matter of size (the NY economy is about 7\% of that of the US a whole). This is likely to be heterogeneous by industries: firms in the RoUS may use information and financial services from NY quite often, while firms in NY may satisfy most of their intermediate demand for information and financial services from the NY industries. To take these effects into account, we proceed as follows.

We let \begin{equation} \label{eq:regionalize}
A_{kl}^{\text{NY}->\text {NY}} = \rho_{kl}^{\text{NY}} A_{kl}.
\end{equation}
Thus, $\rho_{kl}^{\text{NY}}$ indicates by how much intra-regional trade flows differ from national ones. If the (normalized) industry-industry flows within a region are identical to the national flows, we have $\rho_{kl}^{\text{NY}}=1$. As mentioned, typically $\rho_{kl}^{\text{NY}}<1$ since intra-regional flows tend to be smaller than their national equivalents.

To arrive at the estimate of $\rho_{kl}^{\text{NY}}$, we first define the Simple Location Quotient (SLQ) and Cross-Industry Location Quotient (CILQ) for which we write
\begin{equation}
	SLQ_k = \frac{y_k^{\text{r}}/y^{\text{r}}}{y_k^{\text{n}}/y^{\text{n}}}
\end{equation}
\begin{equation}
	CILQ_{kl} = \frac{SLQ_k}{SLQ_l},
\end{equation}
respectively. $y_k^{\text{r}}$ denotes the regional GDP of industry $k$ in a generic region $r$ and $y_k^{\text{n}}$ is the same at the national level. Moreover, $y^\text{r}= \sum_k y_k^\text{r}$ and $y^n = \sum_k y_k^n$.\footnote{We obtain data from the BEA regional accounts: \url{https://apps.bea.gov/itable/iTable.cfm?ReqID=70&step=1&acrdn=5}. As the GDP of some industries is not given for year 2019 due to confidentiality issues, we impute that GDP based on values in previous years. In principle, $y_k^{\text{NY}}$ and $y_k^{\text{US}}$ should be gross output rather than GDP, but using GDP is equivalent as the intermediate share of industry $k$ cancels out between the numerator and the denominator.} 

The SLQ is a measure of regional concentration of industries. If $SLQ_k>1$, sector $k$ is more concentrated at the regional compared to the national level, in the sense that industry $k$'s contribution to GDP is larger in the region than it is in the country. For example, in the case of the information and financial industries in NY, $SLQ_k>1$, while for agriculture and manufacturing we have $SLQ_k<1$. 

The CILQ, in addition, takes into account that the relative importance might be different for sellers and buyers. $CILQ_{kl} >1$ implies that GDP of regional sector $k$ is larger than GDP of regional sector $l$ than it is at the national levels. Thus, the demand of $l$ can be fully met by the supply of $k$. On the other hand, if $CILQ_{kl}<1$, parts of $l$'s demand need to be imported. For example, $CILQ_{kl}<1$ when $l$ is information or finance: as these sectors are really big in NY compared to the RoUS, their intermediate demand for other goods and services must be at least in part satisfied by industries outside of NY. Conversely, $CILQ_{kl}>1$ if $l$ is agriculture or manufacturing: as these industries are small in NY, their intermediate demand can be fully satisfied by NY industries.

We can now define the FLQ equivalently to Ref. \cite{flegg1995appropriate} as
\begin{equation}
	FLQ_{kl} = \begin{cases} 
	\lambda \cdot CILQ_{kl} &\mbox{if } k \neq l, \\
	\lambda \cdot SLQ_{k} & \mbox{if } k = l, 
	\end{cases}
\end{equation} 
where $\lambda = [ \log_2 (1+ y^r/y^n) ]^\delta$ and $\delta \in [0,1)$. The FLQ combines the SLQ and CILQ and also takes into account that the region size affects the modification of technical coefficients. Smaller regions will need to import more from other regions and thus $\lambda$ will be smaller. To what extent region size affects regional coefficients can be controlled by the parameter $\delta$. Since there is no general consensus on the estimation of $\delta$, we choose $\delta=0.15$, which avoids problems with negative demand.

We then substitute 
\begin{equation}
	\rho_{kl}^{\text{r}} = \begin{cases} 
	FLQ_{kl}  & \mbox{if } FLQ_{kl}<1, \\
	1 & \mbox{if } FLQ_{kl} \ge 1, 
	\end{cases}
	\label{eq:regionfinal}
\end{equation}
in Eq.~\eqref{eq:regionalize} to obtain our estimate of regional technological coefficients. By applying these formulas with region $r$ corresponding to NY, $r=NY$, we find the technical coefficient matrix $A^{\text{NY}->\text {NY}}$. We then obtain $A^{\text{RoUS}->\text {NY}}$ from Eq. \eqref{eq:idcoef1}.

Next, we obtain GDP of industry $k$ in the RoUS, $y_k^\text{RoUS}$, as a residual from total GDP and GDP in NY, i.e. $y_k^\text{RoUS}=y_k^\text{n}-y_k^{\text{NY}}$. Identifying region $r$ with the RoUS, we use Eq. \eqref{eq:regionfinal} to obtain coefficients $\rho_{kl}^{\text{RoUS}}$ and $A^{\text{RoUS}->\text {RoUS}}$,\footnote{In this case, we assume $\lambda=1$ to avoid negative final demand for certain NY industries.} and Eq. \eqref{eq:idcoef2} to obtain the residual $A^{\text{NY}->\text {RoUS}}$. We finally calculate intermediate consumption $Z^{\text{NY}->\text {NY}}$ from Eq. \eqref{eq:techcoef}, i.e. $Z^{\text{NY}->\text {NY}}=A^{\text{NY}->\text {NY}} x^{\text{NY}}$,\footnote{We obtain gross output in NY, $x^{\text{NY}}$, from GDP in NY, $y^{\text{NY}}$, by assuming that the shares of intermediate goods is the same in NY as in the US as a whole, i.e. $x_k^{\text{NY}}=\frac{y_k^{\text{NY}}}{y_k^{\text{US}}}x_k^{\text{US}}$. } and proceed similarly to obtain the other intermediate consumption matrices.

\begin{figure}[htbp]
    \centering
\includegraphics[width = 1\textwidth]{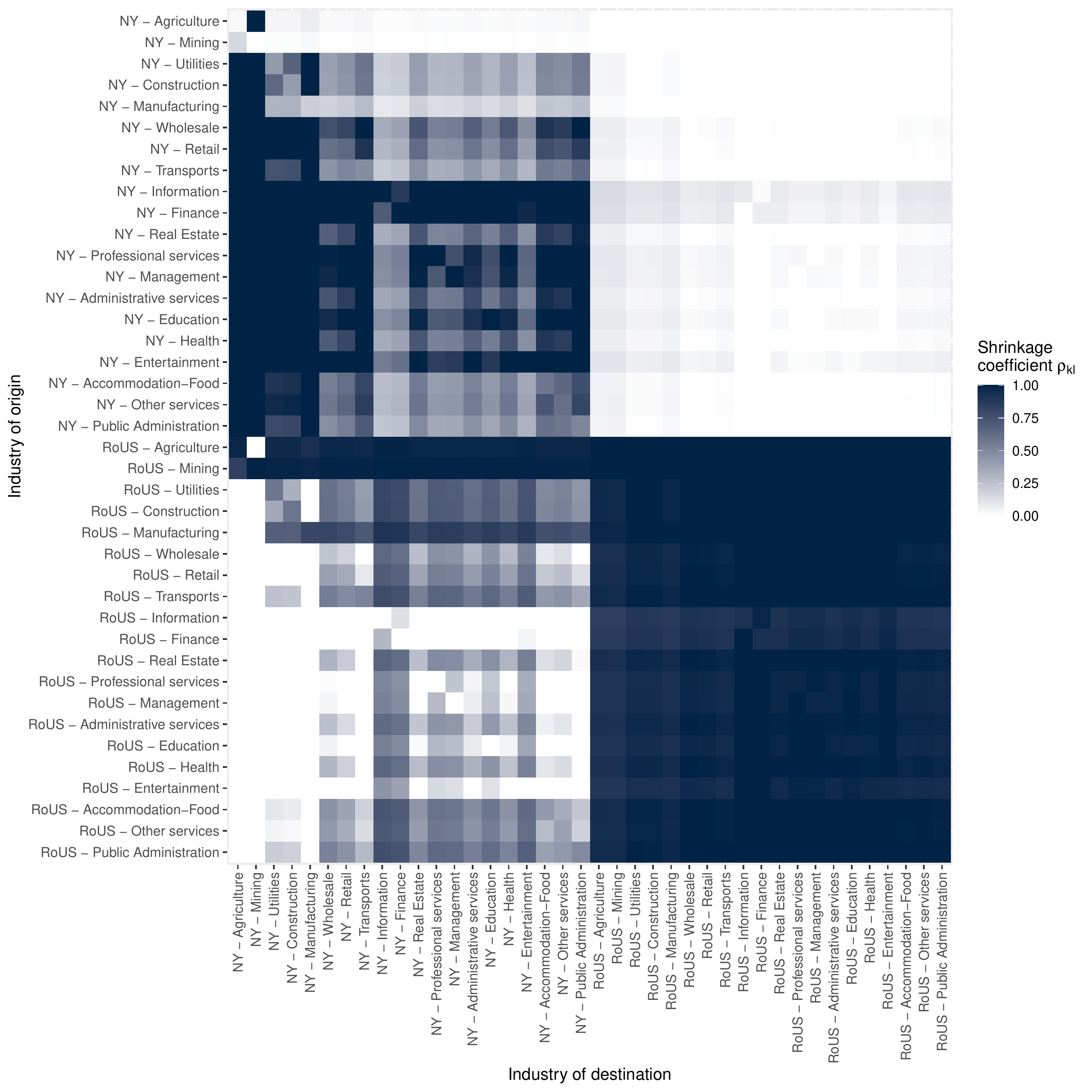}
    \caption{\textbf{Shrinking of technical coefficients.} Each entry represents the value $\rho_{kl}$ that multiplies each technical coefficient from US-wide input-output tables, to account for regionalization. Values $\rho_{kl}=1$ imply no shrinking (dark blue), while smaller values of $\rho_{kl}$ (light blue and white) imply strong shrinking.  }
        \label{fig:FLQ_heatmap_RoUS_CIL}
\end{figure}

Figure \ref{fig:FLQ_heatmap_RoUS_CIL} shows the shrinkage coefficients $\rho_{kl}^\text{NY}$ in the top left block and $\rho_{kl}^\text{RoUS}$  in the bottom right block. It also shows $1-\rho_{kl}^\text{NY}$ in the bottom left, and $1-\rho_{kl}^\text{RoUS}$ in the top right. It is immediately clear that, in relative terms, industries in NY depend much more on industries in the RoUS than industries in the RoUS depend on industries in NY, as is natural given region sizes. In line with the narrative that we kept so far, NY industries such as information or finance need to import a lot of intermediate goods and services from the RoUS because they are so large in NY that local suppliers cannot meet the demand. The only exception is demand for their own products, which they can meet because they are large. Conversely, intermediate demand by agriculture and mining is fully satisfied by industries in NY, because these industries are so small in NY that supply is more than enough; the only exception is again agricultural and mining inputs that need to be imported from the RoUS.

We follow the same procedure to derive regional final demand categories. Let $f_k^r$ and $f_k^n$ be the regional and national final demand values and $f^r = \sum_k f_k^r$ and $f^n = \sum_k f_k^n$. To estimate the regional purchase coefficient, we write
\begin{equation}
	\frac{f_k^r}{f^r} = \rho_k^f \frac{f_k^n}{f^n},
\end{equation}
where $\rho^f$ is obtained from
\begin{equation}
	\rho_k^f = \begin{cases} 
	FLQ_{kk}  & \mbox{if } FLQ_{kk}<1, \\
	1 & \mbox{if } FLQ_{kk} \ge 1. 
	\end{cases}
	\label{eq:iofinaldemand}
\end{equation}

To obtain final demand by New Yorkers to industries in NY, $f^{\text{NY}->\text {NY}}$, we simply apply Eq. \eqref{eq:iofinaldemand} with $r=NY$. This means, for instance, that a good part of NY households' consumption of manufacturing products is satisfied by firms located outside of NY, as manufacturing is small in NY relative to the RoUS. The same applies to other components of final demand (e.g. NY plants will invest in machinery produced elsewhere in the US). We then calculate final demand by agents in NY to industries in RoUS, $f^{\text{RoUS}->\text{NY}}$, as the residual, i.e. $f^{\text{RoUS}->\text {NY}}=f^\text{NY}-f^{\text{NY}->\text {NY}}$. We finally compute demand by agents in RoUS to both industries in NY and in RoUS as a residual from total output, i.e. $f_k^{\text{NY}->\text {RoUS}}=x_k^\text{NY}-\sum_l Z_{kl}^{\text{NY}->\text {NY}}-\sum_l Z_{kl}^{\text{NY}->\text {RoUS}}-f_k^{\text{NY}->\text {NY}}$ (we distribute final demand across private consumption, government consumption and other final demand following the same shares of these components of final demand as in the US as a whole). We proceed similarly to obtain $f^{\text{RoUS}->\text {RoUS}}$.

For consistency, we also compute value added of each industry both in NY and RoUS as the difference between the relevant component of $x$ and the column sum of $Z$, e.g. $va_k^\text{NY}=x_k^\text{NY}-\sum_k Z_{kl}^{\text{NY}->\text {NY}}-\sum_k Z_{kl}^{\text{RoUS}->\text {NY}}$.

\begin{figure}[htbp]
    \centering
\includegraphics[width = 1\textwidth]{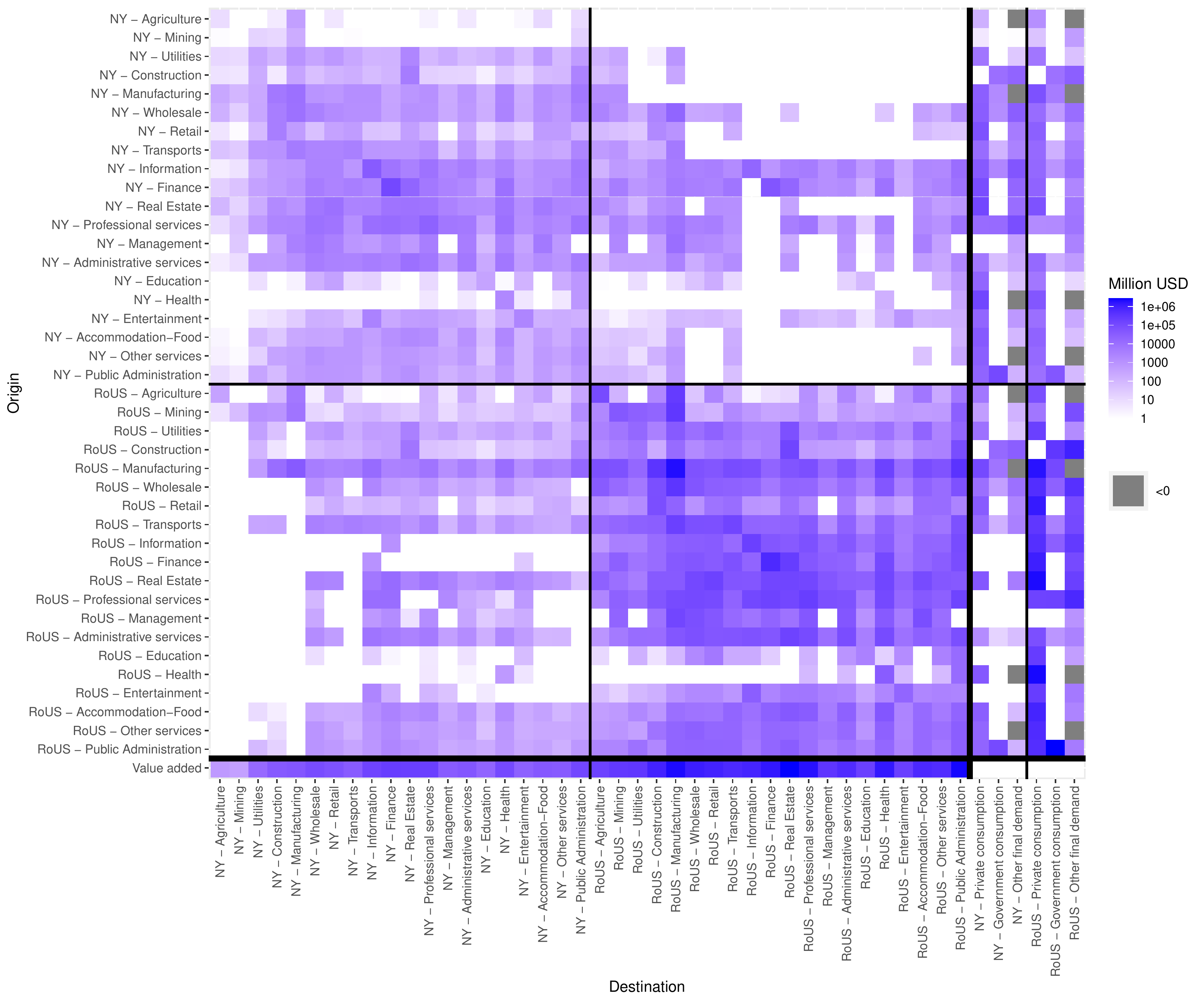}
    \caption{\textbf{New York (NY) - Rest of the US (RoUS) input output system.} Entries represent flows from the row of origin to the column of destination. They are colored based on flow value, in million USD; negative values are colored grey. From top-left to bottom-right, each block delimited by black lines represents: (i) the value of intermediate goods flowing from industries in NY to industries in NY, $Z^{\text{NY}->\text {NY}}$; the value of intermediate goods flowing from industries in NY to industries in RoUS, $Z^{\text{NY}->\text {RoUS}}$; final demand of agents in NY to industries in NY, $f^{\text{NY}->\text {NY}}$; final demand of agents in RoUS to industries in NY, $f^{\text{NY}->\text {RoUS}}$; the same flows as above, but flowing from the RoUS, $Z^{\text{RoUS}->\text {NY}}$, $Z^{\text{RoUS}->\text {RoUS}}$, $f^{\text{RoUS}->\text{NY}}$, $f^{\text{RoUS}->\text{RoUS}}$; value added generated by industries located in NY, $va_{\text{NY}}$  and by industries located in the RoUS,  $va_{\text{RoUS}}$.   }
        \label{fig:NY-RoUS-io-matrix}
\end{figure}

Figure \ref{fig:NY-RoUS-io-matrix} shows the entire NY-RoUS input-output system. The intensity of blue in a cell indicates the magnitude of an intermediate flow or of a component of final demand or value added. (Some components are negative, e.g. because the US import more of a given good from the Rest of the World than they export.) This representation has the advantage of showing compactly the contributions of the various components to the input-output system, and the relative sizes of industries and regions.

\subsubsection{Initialization of model variables}
\label{apx:initialization_inputoutput}

As shown in Table \ref{tab:variable_initialization}, we use the derived input-output system to initialize several model variables:
\begin{itemize}
    \item $Z_{k,l,0}$, intermediate consumption by industry $l$ of goods and services produced by $k$, is obtained from the matrices $Z^{\text{NY}->\text {NY}}$, $Z^{\text{NY}->\text {RoUS}}$, $Z^{\text{RoUS}->\text {NY}}$, $Z^{\text{RoUS}->\text {RoUS}}$, depending on whether $k$ or $l$ are located in NY or in the RoUS. This results in a $40\times 40$ intermediate consumption matrix.
    \item $c_{k,0}^{\text{local}}, G_{k,0}^{\text{local}}, f_{k,0}^{\text{local}}$ are obtained from the private consumption, government consumption and other final demand components of $f^{\text{NY}->\text {NY}}$ and $f^{\text{RoUS}->\text {NY}}$, and $c_{k,0}^{\text{rest}}, G_{k,0}^{\text{rest}}, f_{k,0}^{\text{rest}}$ are obtained from the same components of $f^{\text{NY}->\text {RoUS}}$ and $f^{\text{RoUS}->\text {RoUS}}$. Each of these six variables form a vector with length 40.
    \item $x_{k,0}$, gross output of industry $k$, is the row sum of the input-output system shown in Figure \ref{fig:NY-RoUS-io-matrix}, i.e. it is obtained by summing intermediate and final consumption. This results in a vector of gross output of length 40.
    \item $A_{k,l}$, the $k,l$ coefficient of the technical coefficients matrix, is obtained by $A_{k,l}=Z_{k,l}/x_l$. As $Z$, also $A$ is a $40\times 40$ matrix.
    \item So far, we only listed realized variables. All demand variables, namely intermediate orders $O_{k,l,0}$,  private consumption demand $c_{k,0}^{d,\text{local}}, c_{k,0}^{d,\text{rest}}$, government consumption demand, $G_{k,0}^{d,\text{local}}, G_{k,0}^{d,\text{rest}}$, and other final demand, $f_{k,0}^{d,\text{local}},f_{k,0}^{d,\text{rest}}$, are initially set to their realized counterparts. This corresponds to assuming that before the pandemic all demand was exactly satisfied. Demand matrices and vectors have the same dimensions as their realized counterparts (e.g., orders $O$ are a $40\times 40$ matrix).
\end{itemize}

\subsection{Consumption expenditures}
\label{apx:cons}

The only dataset that has information on consumption disaggregated by age, income and other characteristics is the Consumer Expenditure Survey (from now on, CE). This is a survey, meaning that they ask respondents what they consumed, using a classification system that is just specific to this survey. The CE is produced by the BLS.

The other main dataset on consumption expenditures is produced by the BEA, which also uses it for input-output tables and national accounts. This dataset is called Personal Consumption Expenditures (PCE). It uses a different classification system than the CE, but there exists a crosswalk produced by the BLS and by the BEA to match these two classifications. Differently from CE, PCE is obtained by directly considering receipt data of wholesalers, retailers and other producers and other data sources to estimate actual consumption, so it is not a survey. 

In general, the CE and PCE are different. There is a whole literature comparing them, see e.g. Ref. \cite{passero2014understanding}. Overall, expenditure levels in CE seem to be around 60\% of PCE, and this varies substantially across expenditure items. For instance, purchases of motor vehicles are very similar, but purchases of furniture in CE are 64\% of those in PCE, and purchases of personal care services are 35\% in CE compared to PCE.

In our approach, we consider PCE as our benchmark, and attribute consumption categories in CE to the closest PCE categories to disaggregate consumption by age and income. We finally obtain household consumption of goods and services produced by a given industry by mapping PCE categories to NAICS commodities and finally to NAICS industries.

\subsubsection{Consumer expenditure survey (CE)}

We obtain data for consumption disaggregated by age and income from the consumer expenditure survey (CE). These data are available in several formats, we decide to use tables available on the BLS website.\footnote{https://www.bls.gov/cex/tables.htm} Specifically, we focus on the cross-tabulated tables ``Age of reference person by income before taxes'' for years 2018-2019: these tables provide socio-demographic information as well as income and consumption across many expenditure items for certain joint age-income groups.  (Age refers to the head of the household --simply the respondent to the survey--, while income is the sum of all individual incomes within the household.) Specifically, information is available for all combinations of age brackets $<25$ years old, $25-34$ years old, $35-44$ years old, $45-54$ years old, $55-64$ years old, $>65$ years old, and income brackets $<\$ 15,000$, $\$ 15,000 - \$ 29,999$, $\$ 30,000 - \$ 39,999$, $\$ 40,000 - \$ 49,999$, $\$ 50,000 - \$ 69,999$, $\$ 70,000 - \$ 99,999$, $>\$ 100,000$. This gives a total of $6\times 7 = 42$ age-income combinations.\footnote{The combination $<25$ years old - $>\$ 100,000$ is actually missing in the dataset as not enough households could be found with these characteristics. We thus impute it by simply assuming that, for each consumption category, the ratio between the consumption of $<25$ years old - $>\$ 100,000$ and that of $<25$ years old - $\$ 70,000 - \$ 99,999$ is the same as the ratio between the consumption of $25-34$ years old - $>\$ 100,000$ and that of $25-34$ years old - $\$ 70,000 - \$ 99,999$.  }

Out of all the information provided in these tables, we focus on the number of households for each of the 42 age-income combinations, and on the average yearly consumption of households in each age-income combination for many expenditure items. These items are classified according to a system developed by the BLS named UCC, which comprises several hierarchical levels. For instance, going down the hierarchy from the top, it is possible to know consumption of food, of food at home, of ``meats, poultry, fish and eggs'' and of beef. 

\subsubsection{Personal consumption expenditures (PCE)}

Personal consumption expenditures data are provided by the BEA as part of the U.S. national accounts. These data are used to compute GDP from the expenditure approach, together with information on investment, exports, etc. Differently from the CE, these data are not obtained as part of a survey, but rather estimated from several sources that include receipts from several industries. The BEA has its own hierarchical classification of expenditure items. We focus on the level of aggregation provided in ``Table 2.4.5. Personal Consumption Expenditures by Type of Product'' (year 2019), which comprises 76 expenditure items.

\subsubsection{Mapping CE to PCE}

Sometimes consumption categories in PCE are easily comparable to those in CE. To consider the same example as above, from top to bottom of the PCE hierarchy, it is possible to know consumption of ``Food and beverages purchased for off-premises consumption'', ``Food and nonalcoholic beverages purchased for off-premises consumption'', ``Food purchased for off-premises consumption'', ``Meats and poultry'' and ``Beef and veal''. In other cases, however, items in PCE are not comparable to items in CE. For instance, CE does not provide as detailed information on consumption of financial services as PCE does. Moreover, expenditure in CE underestimates expenditure in PCE: aggregate CE consumption is 60\% of PCE consumption (which is consistent with national accounts).

Thus, we proceed as follows. For each of the 76 expenditure items in PCE, we assign one or more expenditure items in CE. A CE expenditure item can be assigned to multiple PCE items. For instance, PCE categories ``Sporting equipment, supplies, guns, and ammunition'' and ``Sports and recreational vehicles'' do not exist independently in CE, and so we assign the CE category ``Other entertainment supplies, equipment, and services'' to both. This matching is based on the CE-PCE concordance file available on the BLS website.\footnote{https://www.bls.gov/cex/pce-concordance-2017.xlsx} 

We then normalize consumption for each age-income category and each PCE expenditure item so that total consumption for that item corresponds to the value provided in the BEA table 2.4.5 for year 2019. Therefore, information coming from CE is used to tell the consumption shares of PCE expenditure items across age-income categories, while the total consumption values are consistent with those provided by the BEA. 

This process does not preserve age-income shares of total consumption, in the sense that dividing total consumption of any given age-income category by total consumption across all categories after this mapping will in general give different results than in PCE. This is unavoidable: as mentioned above, certain consumption items are underrepresented in CE, so actual shares of total consumption cannot be the same. However, the Pearson correlation coefficient between age-income shares of total consumption in CE and in PCE (after we perform our mapping) is 0.98, suggesting that this is a minor issue.

\subsubsection{Mapping PCE categories to NAICS industries}
\label{apx:mapping_pce_to_naics}

At this point, we have a table with average yearly consumption for all combinations of 76 PCE expenditure items and 42 age-income categories. What we need, however, is a table with average yearly consumption for good and services produced by all 20 2-digit NAICS industries, again for the 42 age-income categories that we consider. This requires performing three steps: mapping the 76 PCE expenditure items to the 71 3-digit NAICS commodities; mapping the NAICS commodities to the corresponding 71 NAICS industries; aggregating the 71 3-digit NAICS industries to the 20 2-digit NAICS industries.

To perform the first step, we use the bridge provided by the BEA, again for year 2019.\footnote{https://apps.bea.gov/industry/xls/underlying-estimates/PCEBridge\_1997-2019\_SUM.xlsx} This is rather straightforward, except for the need to attribute trade and transport margins to the relevant NAICS commodities. To do so, we attribute margins so that total consumption of each of the 71 NAICS commodities matches the data provided by the BEA.

For the second step, we derive the transformation matrix $T$ as detailed in Section \ref{sec:national_io_tables}. This transformation attributes consumption of a commodity to the industry that produced it. The most relevant example is education and health: a significant portion of consumption of these commodities is attributed to Public Administration, reflecting public schools and hospitals.

Finally, we aggregate all 71 3-digit NAICS industries to the 20 2-digit NAICS industries using the official concordance. Our final dataset is a table with 20 rows and 42 columns, reflecting 2-digit NAICS industries and age-income categories. This table is supplemented by a row which gives the total number of households in each age-income category.

\subsubsection{Regional table}

The approach as described so far applies to the U.S. as a whole. Indeed, the data that we used to convert CE consumption into expenditure on NAICS industries are only available at the country level.

To adapt consumption by age and income to the NY MSA, we use both data from the regional input-output tables that we constructed (Section \ref{apx:io}) and from the synthetic population that we built (Section \ref{apx:synthpop}).  

First, we change the number of households in each age-income category to reflect the number of households in the NY MSA (as modeled by our synthetic population), as opposed to the wider US. To do so, recall that our synthetic population only comprises a bit more than 400 thousand individuals, a representative sample of the real population that comprises more than 19 million individuals. Moreover, the synthetic population features 160 thousand households, which again are smaller than the real 7.1 million households.\footnote{See, e.g. \url{https://censusreporter.org/profiles/31000US35620-new-york-newark-jersey-city-ny-nj-pa-metro-area/}.} To address this issue, we rescale the number of households in the synthetic population for each age-income category by a factor of approximately 7100/160. The population share of each age-income category, that is, the number of households in that age-income category divided by the total number of households, is quite similar in our NY synthetic population than in the CE data, in the sense that the Pearson correlation coefficient of population shares across age-income groups is 0.88. Of course, as income in the NY MSA is higher than in the rest of the US, the highest income categories are over-represented. Moreover, as discussed in Section \ref{apx:summary_synthpop} and shown in Figure \ref{fig:plots_marginals}H the lowest income categories are under-represented.

The second step is to adapt spending so that overall consumption on each industry matches consumption by New Yorkers as derived in the regional input-output tables. In principle, changing the population structure and keeping the average spending per household constant should already allow recovering a total consumption that is consistent with input-output tables. Indeed, spending for most industries as computed directly from the consumption table is about 82\% as spending obtained from the regional input-output tables. The remaining discrepancy can be explained by many factors, including different spending patterns in NY than in the rest of the US, different real population structure than in our synthetic population, etc. To be consistent with regional input-output tables, we simply rescale consumption on each industry $k$ by each age-income category $g$ so that it matches total consumption $c_{k,0}^{\text{local}}$, i.e.
\begin{equation}
    \sum_g c_{g,k,0}^{d,\text{local}} = c_{k,0}^{\text{local}}
\end{equation}

\begin{figure}[htbp]
    \centering
\includegraphics[width = 1\textwidth]{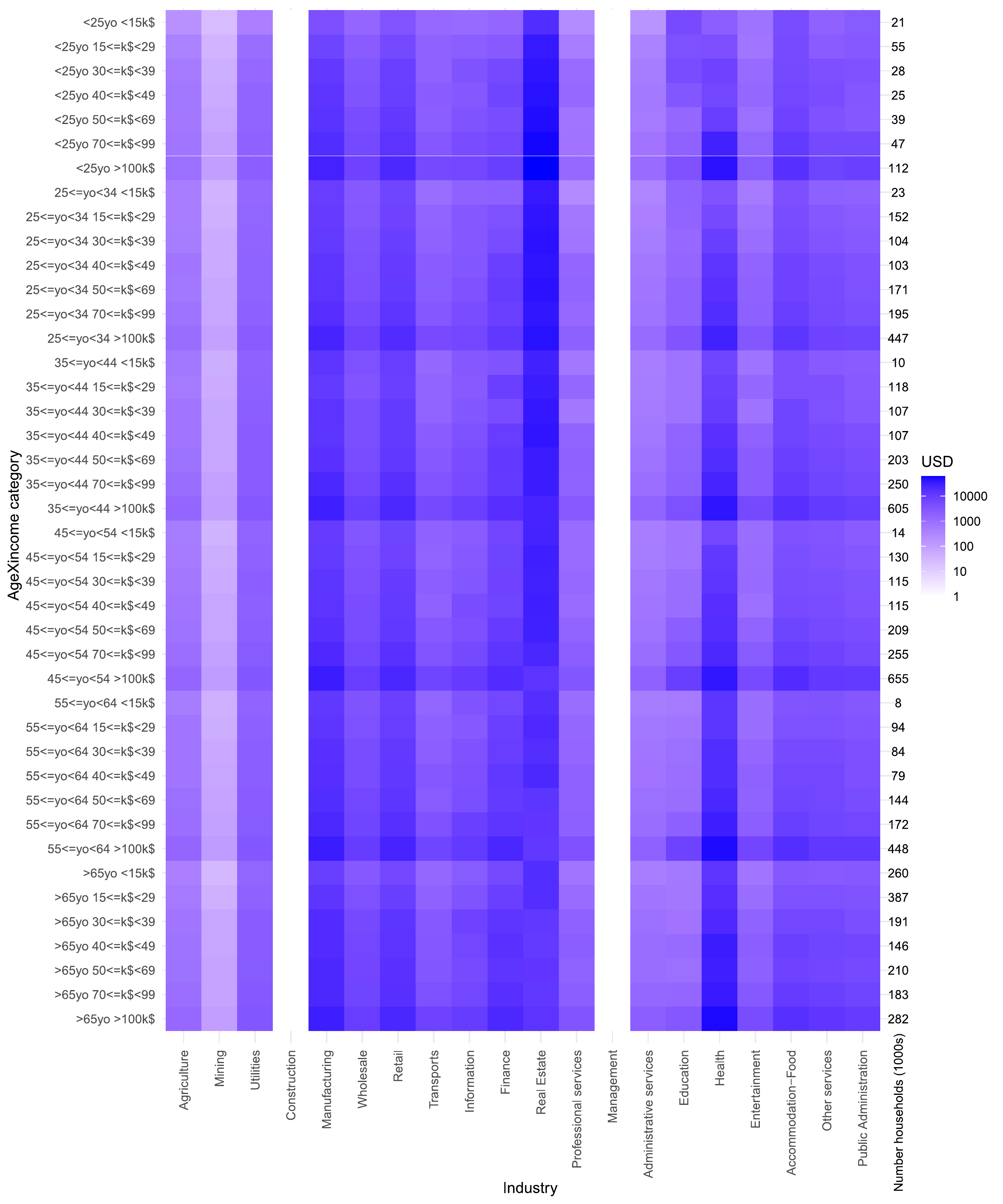}
    \caption{\textbf{Consumption by age and income in the NY MSA.} Each entry is yearly consumption for goods and services produced by any given industry by households within a certain age and income range (yo=years old; k\$=thousands dollars). The numbers on the right of the plot show the  number of synthetic households in the NY MSA (in thousands, normalized to the real population) that belong to the age-income category on the same row.   }
        \label{fig:cons_by_age_income}
\end{figure}

In Figure \ref{fig:cons_by_age_income}, we show the resulting dataset. Two main things are worth noting. First, overall consumption is larger for certain industries, such as real estate, health, manufacturing, and zero for management and construction.\footnote{This is because, for instance, home renovation is classified as investment, not as consumption.} (This pattern could also be seen on the final demand columns of Figure \ref{fig:NY-RoUS-io-matrix}.) Second, most columns are characterized by a periodic pattern by which the shade of blue becomes darker every seven cells. This is because the total consumption of goods and services increases with income. The only exceptions to this pattern are real estate and health, which tend to have higher consumption among the young (at the top of the heat map) and the elderly (at the bottom of the heat map), respectively.

\begin{figure}[htbp]
    \centering
\includegraphics[width = 1\textwidth]{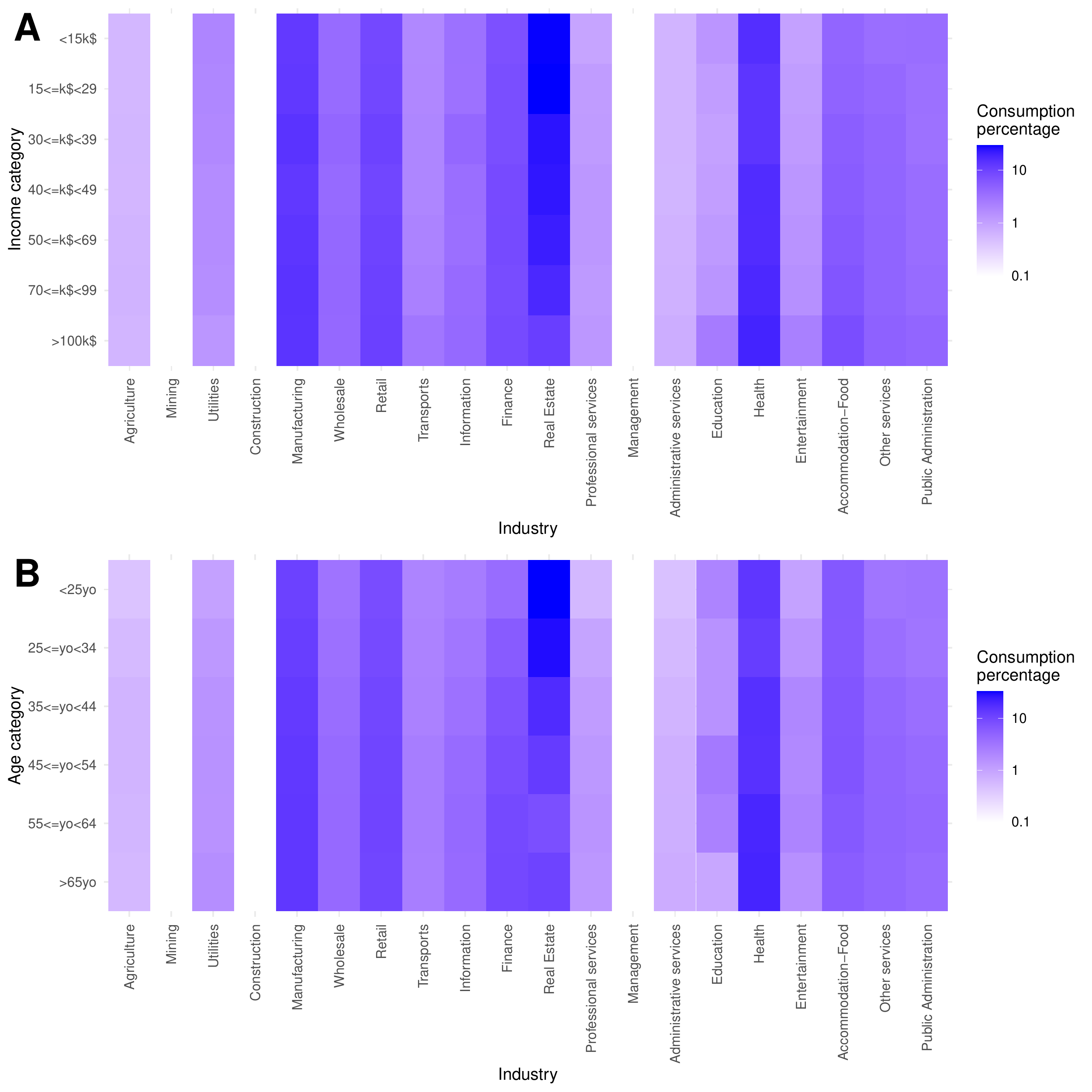}
    \caption{\textbf{Consumption by age and by income in the NY MSA.} Each entry is the percentage share of consumption of goods and services produced by a given industry for a certain income or age group. (In other words, rows sum to 100.)   }
        \label{fig:marginals_age_income}
\end{figure}

To better understand how age and income affect consumption, in Figure \ref{fig:marginals_age_income} we plot industry consumption by income (panel A), aggregating age categories, and by age (panel B), aggregating income categories. In both panels, to focus on the different spending patterns of different income or age groups, we normalize rows to 100. Looking at panel A, we can see a clear pattern that poorer households spend a much larger share of their income on real estate than richer households. Richer households tend to spend more on education, entertainment and accommodation and food, although differences are not enormous. For most industries, there is not much difference across income groups, meaning that households tend to spend the same share of their income on those industries irrespective of their overall income level. Looking at panel B, we see that the only industry for which the younger spend more than the older is real estate. The elder spend more in finance and health.

\FloatBarrier

\section{Calibration}
\label{apx:calibration}
Both the epidemic and the economic modules have some model-wide parameters that we must set to a specific value. For some of these parameters --mostly epidemic-- we are able to find a reference for their value, and simply set them ex-ante. Other parameters need to be calibrated ex-post, using an Approximate Bayesian Computation (ABC) rejection algorithm. In the following Section \ref{apx:epicalibration} we address calibration in the epidemic module, listing which parameters can be directly calibrated and which need to be calibrated using ABC. We then discuss the economic parameters in Section \ref{apx:econcalibration}. Finally, in Section \ref{apx:abc}, we explain the ABC procedure and show our results.

\subsection{Epidemic parameters}
\label{apx:epicalibration}

Many of the parameters of the epidemic module are clinical. These include length of incubation period, proportion of asymptomatic, probability to infect if asymptomatic as compared to a symptomatic case, infection fatality ratio by age group, etc. We list these parameters in Table \ref{table:parameters}, showing their value and the reference from which they have been taken.

\begin{table}[!h]
\begin{tabular}{@{}lllll@{}}
\toprule
\textbf{Parameters} & \textbf{Description}                                      & \textbf{Age group} & \textbf{Value}                         &\textbf{Ref.} \\ \midrule
$r$ & relative infectiousness of asymptomatic individuals & - & 50\% & $\dagger$ \\ \midrule
$k$ & proportion of pre-symptomatic transmission & - & 50\% & \cite{CDCPlanning} \\ \midrule
$\epsilon$ & incubation period & - & 5 days & \cite{Hu2020Nov} \\ \midrule
$p_a$ & symptomatic probability & 0-9 & 0.181 & \cite{PieroPoletti2021Mar} \\ 
    &                          & 10-19 & 0.181 & \\
    &                          & 20-29 & 0.225  & \\
    &                          & 30-39 & 0.225  & \\
    &                          & 40-49 & 0.300   & \\
    &                          & 50-59 & 0.300   & \\
    &                          & 60-69 & 0.360    & \\
    &                          & 70-79 & 0.360    & \\
    &                          & $\geq 80$ & 0.646  & \\ \midrule
$\chi$ & susceptibility & 0-18 & 0.56 & \cite{RussellM.Viner2021Feb} \\
& & $\geq 19$ & 1.00 & \\ \midrule
$\gamma$ & pre-symptomatic period & - & 2 days & \cite{backer2020incubation} \\ \midrule
$\mu$ & time to isolation & - & 2.5 days &  \\ \midrule
$\delta$ & days from isolation to death & - & 12.5 &  \cite{CDCPlanning} \\ \midrule
IFR & infection fatality ratio & 0-9 & 0.00161\% & \cite{verity2020estimates}\ddag \\ 
    &                          & 10-19 & 0.00695\% & \\
    &                          & 20-29 & 0.0309\%  & \\
    &                          & 30-39 & 0.0844\%  & \\
    &                          & 40-49 & 0.161\%   & \\
    &                          & 50-59 & 0.595\%   & \\
    &                          & 60-69 & 1.93\%    & \\
    &                          & 70-79 & 4.28\%    & \\
    &                          & $\geq 80$ & 7.80\%  & \\ \midrule
$T_n$ & Notification of death & - & 7 days & \cite{CDCPlanning} \\ \midrule
$\theta$ & outdoor transmissibility & - & 0.05 &  \cite{Weed2020Sep} \\ \bottomrule
\end{tabular}
\caption{Baseline set of parameters. $\dagger$: assumed ;$*$: calibrated to the generation time $T_g$; $\ddag$ Only applied to symptomatic individuals. As such, a correction factor of 1/p is applied to all age groups.}
\label{table:parameters}
\end{table}

The epidemic module has three remaining free parameters: (1) the number of infected individuals on the first day for which we have data to build the interaction network; (2) the transmissibility, $\beta$; (3) fear of infection $\phi^{EPI}$. 

We begin the simulation on 02/12/20, when it was estimated that there were already several infected individuals in New York. In particular, we use the estimates provided by the GLEAM model \cite{Davis2020Aug}, according to which there were 165 exposed individuals. To initialize the system into such a state, one could select that number of agents randomly from the
simulation and move them into the latent compartment. However, this would not resemble the real evolution of the epidemic,
which does not infect people at random but instead follows the path imposed by the behavior of individuals. For this reason,
we initialize the system with 10 latent individuals and let the simulation evolve. Once the estimated number of individuals is observed, the system starts to run in calendar time
from 02/12/2020 to 06/30/2020 (each step corresponds to 1 day), so that $t = 0$ is 02/12/2020. This allows us to start the simulation on calendar time with the estimated
number of latent individuals without having to select them at random. This also reduces the horizontal drift characteristic of stochastic epidemiological models, although aligning the simulations at such early time still yields a significant temporal uncertainty resulting in large confidence intervals \cite{Kiss2017}.

The other two parameters, $\beta$ and $\phi^{EPI}$, are estimated using ABC, together with the economic parameters. The ABC procedure uses both economic and epidemic data to define the quantities that the model must match. The epidemic data that we use for calibration are the weekly estimated number of deaths as a consequence of COVID\nobreakdash-19 \cite{Dong2020May}.

\subsection{Economic parameters}
\label{apx:econcalibration}

The parameters that we calibrate in the economic module are the fear of unemployment $\phi^U$, the consumer demand reallocation parameter $\Delta s$, the speeds of hiring and firing $\gamma_H$ and $\gamma_F$, and the fear of infection $\phi^{ECO}$. We actually do not directly calibrate $\phi^{ECO}$, because fear of infection in the economic module should be proportional to the same quantity in the epidemic module (see Materials and Methods). We instead calibrate the proportionality factor $\tilde{\phi}$, and then compute  $\phi^{ECO}=\tilde{\phi} \cdot \phi^{EPI}$.

We calibrate these parameters to match six official statistics, mostly coming from national accounts. We choose to consider official sources as they are most likely not biased towards any particular industry, and this makes them most directly comparable to our model that is initialized on national and regional accounts. Additionally, we focus on statistics that are likely to make parameter identification possible, i.e. variables that our model would predict very differently when using different parameter configurations.

These six statistics are: NY employment, NY Gross Domestic Product (GDP), US GDP, US Investments+Net exports (``Other final demand''), US consumption of goods and services produced by customer-facing industries, US consumption of goods and services produced by industries that do not require contact with customers. (When possible, we consider NY statistics, but these are not available for other final demand and for consumption disaggregated by industry. So, in this case we must compare to US data.)  In all cases, we compare the value of these variables in the second quarter of 2020 with the last quarter of 2019, both in the model and in the data. (In the model, this means that we compare to initial steady state values.) In the data, these variables declined, respectively, by 18.9\%, 11.4\%, 10.2\%, 18.8\%, 20.7\% and 3.4\%.\footnote{Sources: QCEW for employment in NY, BEA regional accounts for GDP in NY (GDP in current dollars, table SQGDP2, updated December 23, 2020, NY state -- the data for the metro area are only available annually, so we use NY state as a proxy), BEA national accounts for US GDP and other final demand (Table 1.1.5, last revised on February 25, 2021), BEA underlying detail for US consumption by commodity (Table 2.4.5U, last revised on February 25, 2021), transformed to consumption by industry using the same procedure as in Section \ref{apx:mapping_pce_to_naics}, for US consumption of customer-facing and no-customer-facing industries. }

\subsection{Approximate Bayesian Computation}
\label{apx:abc}

\begin{figure}[htbp]
    \centering
\includegraphics[width = 1\textwidth]{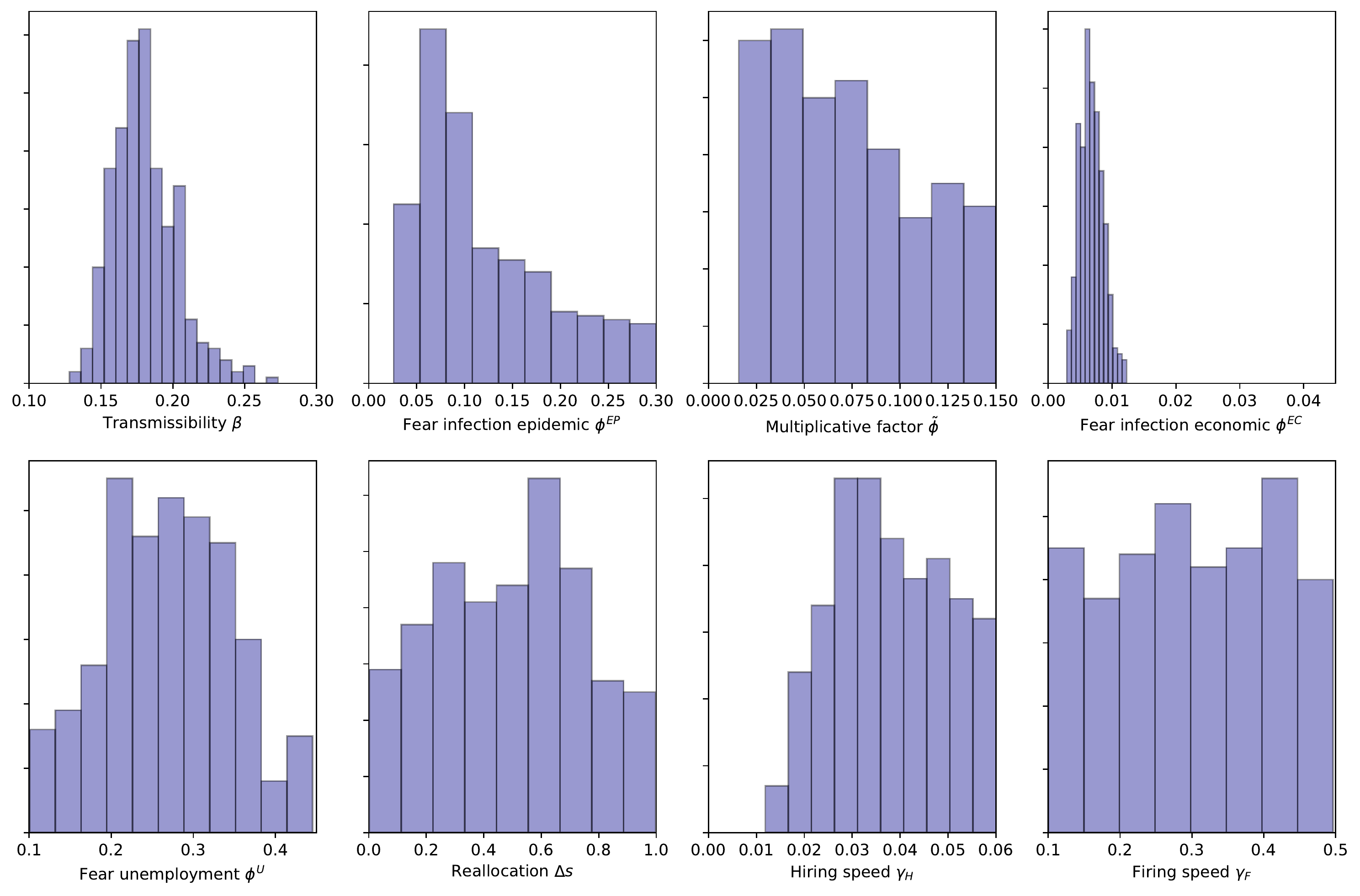}
    \caption{\textbf{Posterior of epidemic and economic parameters.} These histograms show the full range of sampled parameters on the horizontal axis. One can interpret the priors as flat over these ranges.}
        \label{fig:posterior_pars}
\end{figure}

We run 100,000 parameter combinations chosen uniformly at random in the most promising region of the parameter space (which has been identified by running preliminary simulations), shown in Figure \ref{fig:posterior_pars}. We run each parameter combination with a different random seed. We then select only the combinations that yield errors no larger than (in absolute value): 
\begin{itemize}
    \item 0.012 weekly deaths per 1000 synthetic individuals. Compared to the peak of the epidemic wave, this is a 3\% error.
    \item One percentage point for employment and GDP in NY. Here and below, we compare economic statistics produced by the model, as averaged over all days in Q2-2020, with the official statistics listed in Section \ref{apx:econcalibration}. Letting $\mathcal{T}$ be Q2-2020, $T=|\mathcal{T}|=90$ be the number of days in Q2-2020, $x_t$ be an economic quantity at time $t$, and $\hat{x}_\mathcal{T}$ an official economic statistic in Q2-2020, we require $$
    \frac{1}{T}\sum_{t=1}^T x_t - \hat{x}_\mathcal{T} < 0.01.
    $$
    \item Two percentage points for US-wide quantities except other final demand, i.e. GDP and consumption of both customer-facing and no-customer-facing industries.
    \item Four percentage points for other final demand, which is not a primary focus of this study.
\end{itemize}
The choice of thresholds is mostly arbitrary in Approximate Bayesian Computation. We choose these thresholds to stress that, from an epidemic point of view, we want to match the number of deaths very closely, while from an economic point of view we aim at matching NY statistics most strictly, while we allow more slack for US-wide statistics, as we model the rest of the US at a coarser level. 

Based on these thresholds, we select 361 parameter combinations. Figure \ref{fig:posterior_pars} shows the distribution of the eight parameters across the selected combinations. Under Approximate Bayesian Computation, this is interpreted as the posterior distribution of the parameters, obtained without the need to evaluate the likelihood. We use all these accepted parameter combinations to generate uncertainty in the results, according to the Bayesian nature of ABC. In particular, we sample each of these selected parameter combinations uniformly at random and run them with the same random seed that has been used in the calibration. As we perform 200 simulation runs for each counterfactual, this is sufficient.

We see that the distribution of some parameters is clearly peaked. These parameters include the transmissibility $\beta$, the fear of infection in the epidemic and economic modules, $\phi^\text{EPI}$ and $\phi^\text{ECO}$, and the fear of unemployment $\phi^U$. In other cases, such as the proportionality factor $\tilde{\phi}$, reallocation $\Delta s$ and firing speed $\gamma_H$, ABC fails to identify specific parameter values and returns a flat posterior. This is mostly because the economic statistics are averaged over Q2-2020 and so are not much sensitive to parameters such as $\gamma_H$ and $\gamma_F$ that determine the speed of the transient. Moreover, a pairplot (not shown) suggests that for almost all parameter combinations there is virtually no correlation between parameter values, except for two cases. First, we find that $\phi^\text{EPI}$ and $\tilde{\phi}$ are negatively correlated. This is explained by the fact that it is the parameter $\phi^\text{ECO}=\phi^\text{EPI} \cdot \tilde{\phi}$ that determines economic outcomes, and so it is hard to distinguish between combinations of $\phi^\text{EPI}$ and $\tilde{\phi}$ that give the same  $\phi^\text{ECO}$. Second, we find that $\Delta s$ and $\phi^U$ are positively correlated. This is explained by the fact that consumption of goods and services by customer-facing industries declined much more than by non-customer-facing industries (Figure \ref{fig:validation}A). So, because when $\phi^U$ is high consumption demand declines a lot for both types of industries, one needs a high value of reallocation $\Delta s$ to explain the small decline in non-customer-facing consumption. These correlations suggest why the distribution of certain parameters are not peaked, and at the same time elucidate certain mechanisms of the model.

\newpage

\section{Supplementary results on the empirical scenario}
\label{apx:first_wave}

Here we report additional results on the scenario that was used to describe the economic and epidemic effects of the first wave of the COVID\nobreakdash-19 pandemic in the New York metro area. In this scenario we 
\begin{itemize}
    \item Started protective measures on 2020-03-16.
    \item Closed customer-facing and other non-essential economic activities such as large parts of manufacturing and construction.
    \item Closed schools and mandated work from home.
    \item Calibrated fear of infection to its baseline value.
\end{itemize}

\subsection{Economic results}
\label{apx:empirical_economic_results}

\begin{figure}[h]
    \centering
\includegraphics[width = 0.6\textwidth]{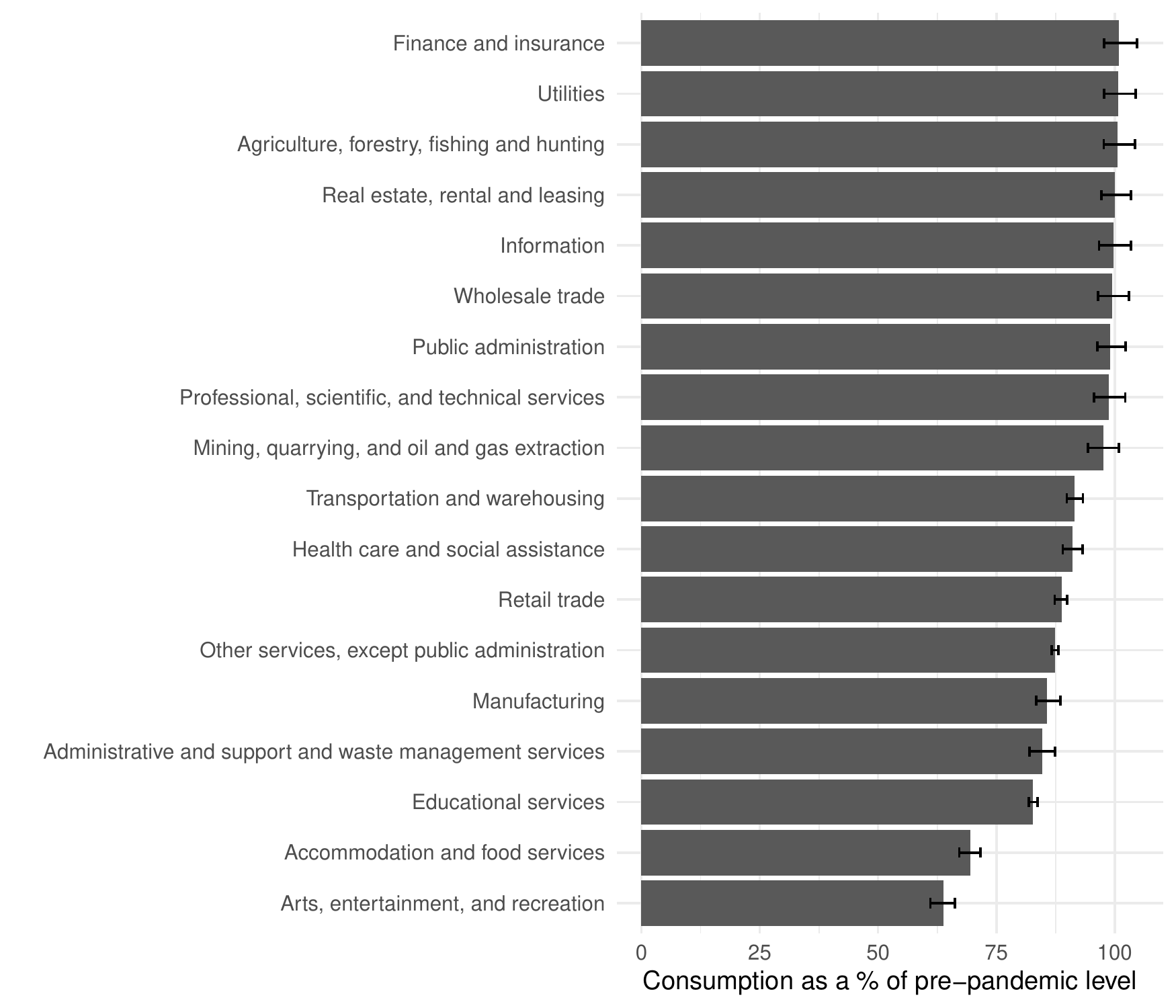}
    \caption{\textbf{Household consumption by industry in the empirical scenario.} For each industry, we show average realized household consumption throughout the simulation period as compared to its steady state level. Industries are ranked from the ones in which consumption was reduced the least (at the top) to the ones in which it was reduced the most (at the bottom). Error bars indicate 2.5-97.5 percentiles.} 
        \label{fig:supp_empirical_consumption_by_industry}
\end{figure}

Figure \ref{fig:supp_empirical_consumption_by_industry} shows that household consumption declines the most among customer-facing industries, although it declines for manufacturing as well. To explain why it also declines for manufacturing (demand for manufacturing goods does not decline in the model since manufacturing is not a customer-facing industry), note that here we are reporting realized consumption, which also depends on supply. As half of the manufacturing industry was considered non-essential in NY (Table \ref{tab:foursquare}), and 14\% of manufacturing was considered non-essential in the Rest of the US, not all demand could be satisfied. Another observation is that consumption of some industries, such as finance and insurance and utilities, in some simulations increased above the pre-pandemic level. This is not in contrast with the assumption that industries cannot produce more than the pre-pandemic level, because household consumption is only part of total demand. Therefore, with a consumption of intermediate goods that is lower than the pre-pandemic level, household consumption can be higher.

\begin{figure}[H]
    \centering
\includegraphics[width = 0.6\textwidth]{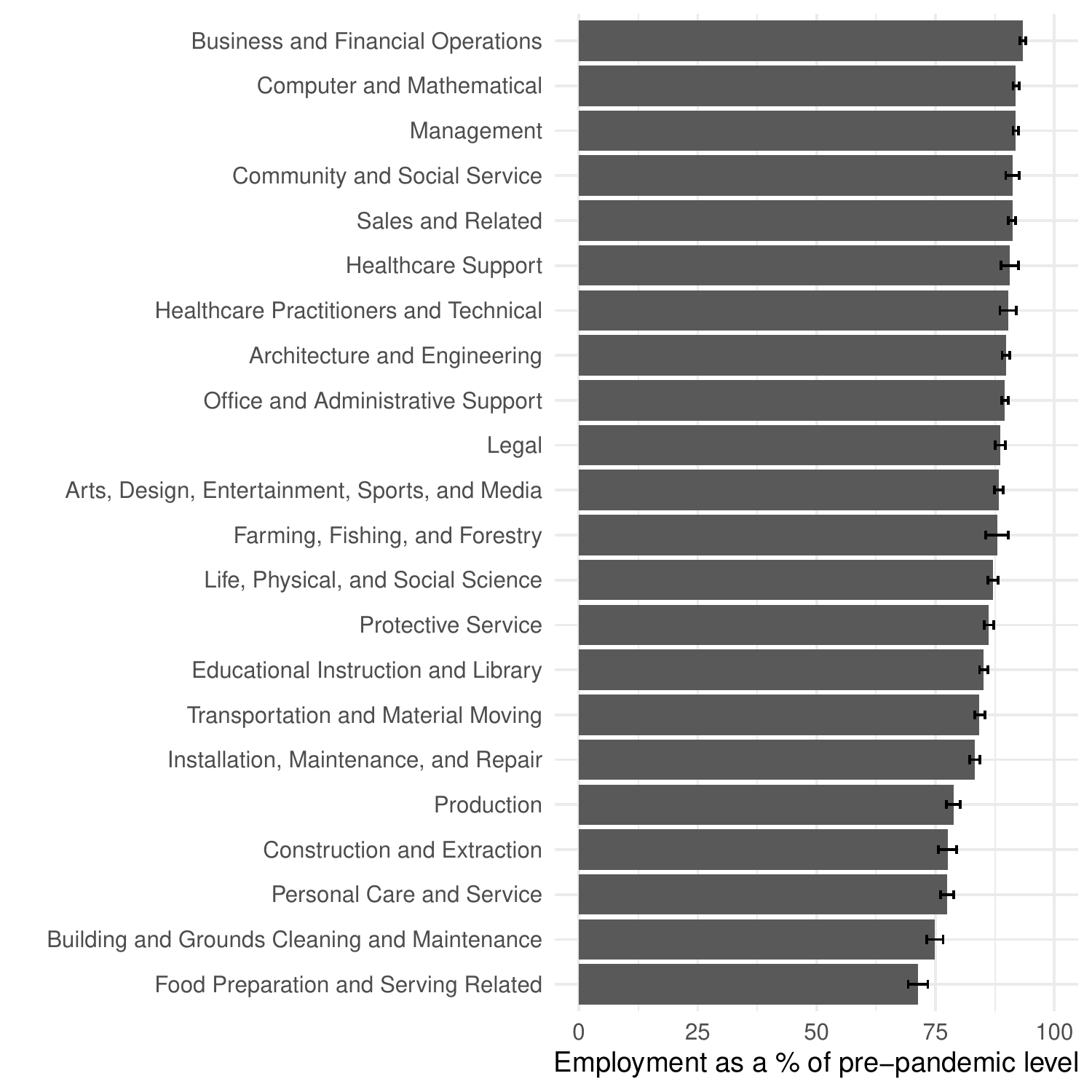}
    \caption{\textbf{Employment by occupation in the empirical scenario.} For each occupation, we show average employment throughout the simulation period as compared to its steady state level. Industries are ranked from the ones in which employment was reduced the least (at the top) to the ones in which it was reduced the most (at the bottom). Error bars indicate 2.5-97.5 percentiles.} 
        \label{fig:supp_empirical_employment_by_occupation}
\end{figure}

Figure \ref{fig:supp_empirical_employment_by_occupation} shows how employment changed across occupations. Employment declined across all occupations, but it declined most strongly among occupations that could not be performed from home and that are particularly common in the industries that were not considered essential (Figure \ref{fig:plots_joint_occ_ind}A). From Figure \ref{fig:plots_joint_bars} we also see that the most affected occupations tend to be the ones with a highest percentage of low-income workers.

\begin{figure}[H]
    \centering
\includegraphics[width = 0.9\textwidth]{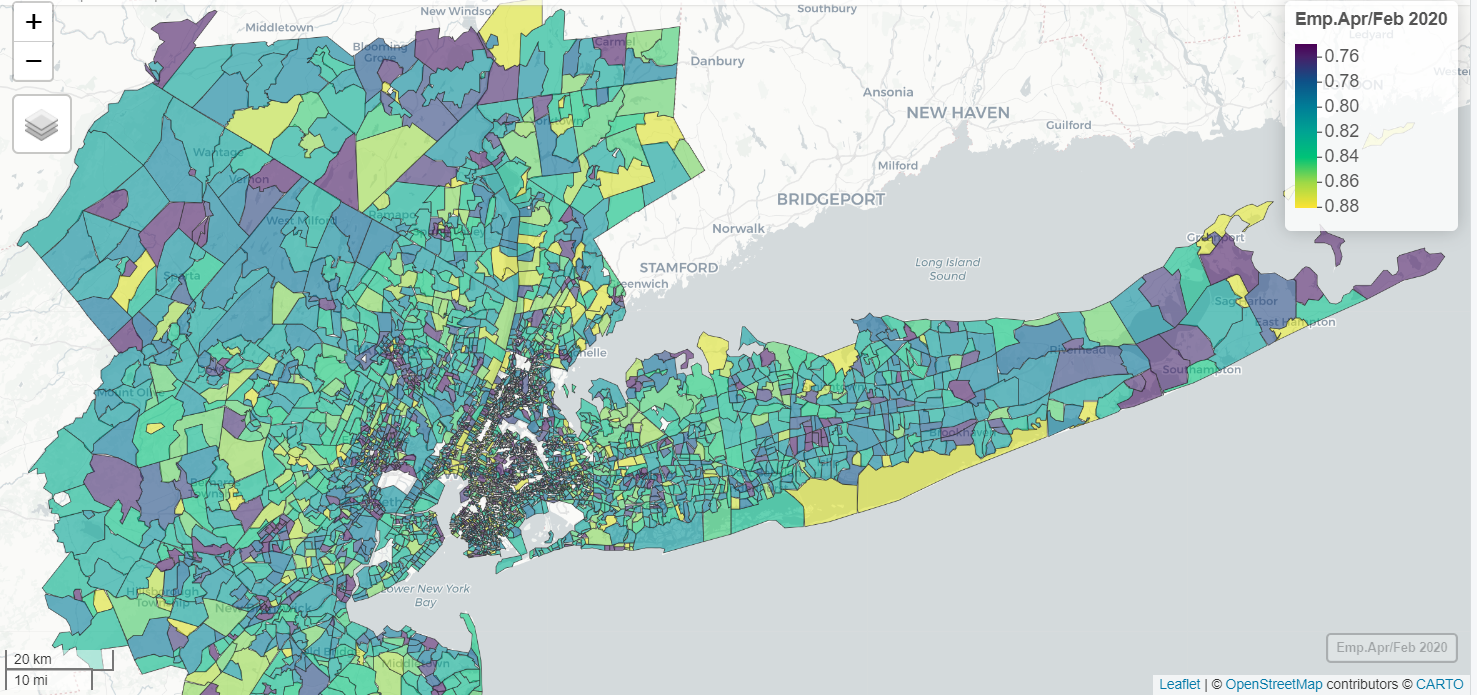}
    \caption{\textbf{Employment by census tract in the empirical scenario throughout the New York metro area.} Census tracts are colored depending on employment in April 2020 compared to steady state values. Census tracts colored purple are the ones in which employment was reduced the most, while census tracts colored yellow are the ones in which it was reduced the least. Note that we crop the color bar so that census tracts in which employment was lower than 76\% of the pre-pandemic level are still colored purple, and census tracts in which it was higher than 88\% of the pre-pandemic level are still colored yellow.} 
        \label{fig:employment_change_by_census_tract_new_york_metro}
\end{figure}

Figure \ref{fig:employment_change_by_census_tract_new_york_metro} shows changes in employment throughout the NY metro area. We see that there is a lot of heterogeneity across census tracts, with a weak pattern of decreasing employment as the distance from Manhattan increases.

\subsubsection{Validation data}
\label{apx:validation_data}

\paragraph{TTR data on employment and consumption by income.} Track the Recovery (TTR, \cite{chetty2020economic}) is a project by Opportunity Insights, a lab based at Harvard University. It uses private-sector data to track several economic statistics in real-time, across demographic groups, and for very granular geographical levels. We downloaded their data\footnote{Repository: \url{https://github.com/OpportunityInsights/EconomicTracker}. Last access: 27/01/2021.} and focused on two key datasets: their combined employment indicator and consumption data by Affinity. For employment, we have data at the county level, which we aggregate at the NY Metropolitan Statistical Area (MSA) level by taking a weighted mean where weights are proportional to county-level employment. For consumption, data are only available at the state level. Because 1/3 of the NY MSA population come from New Jersey State, and 2/3 come from New York State, we aggregate consumption data by using these population shares as weights. For both employment and consumption, raw data are given as percentage changes from the initial day (2020-01-14); for comparison to our simulations we rebase them so that the initial day has value 100 (this transformation does not affect the interpretation of the data at all). Employment and consumption data are further disaggregated across household income levels. For employment data, low-income households have annual income less than 27k\$, while high-income households earn more than 60k\$ per household (these incomes are not ``high'' given the cost of life in NY, but for comparison we need to follow the convention in Ref. \cite{chetty2020economic} that classifies households by income at the US level). For consumption data, low-income households are those who earn less than 46k\$ a year, while high-income households are those that earn more than 78k\$.\footnote{As detailed in Section \ref{apx:cons}, our household income brackets are: $<$15k\$, 15-29k\$, 30-39k\$, 40-49k\$, 50-69k\$, 70-99k\$, $>$100k\$. These do not perfectly overlap with the income brackets in TTR data, so we consider the lowest three income brackets ($<$15k\$, 15-29k\$, 30-39k\$) as corresponding to low-income households, and the last two income brackets (70-99k\$, $>$100k\$) as corresponding to high-income households.} 

\paragraph{QCEW data on industry employment.} The Quarterly Census on Employment and Wages (QCEW) is a data release by the Bureau of Labor Statistics (BLS) that gives monthly data on employment, worker wages and number of establishments, at various levels of spatial aggregation. QCEW data are particularly reliable as they are calculated based on detailed administrative information received by states and counties from all private sector employers. We extract data from January 2020 to June 2020, for the NY MSA, FIPS code C3562. Industry-level data are only available for private employment. Given that we are particularly interested to use these data to compare the predictions of our model regarding differentials in employment across industries, this is unlikely to be a major limitation. A more serious issue is that, due to confidentiality concerns, the QCEW does not provide industry-level data when the number of employees in a given industry is too small. For instance, it does not provide any information for NAICS code 323: Printing and Related Support Activities, presumably because in the NY MSA there are too few printing manufacturing firms and it might be possible to recover firm-level information from industry-level data. This would not be an issue as this paper focuses on quite aggregated 2-digit NAICS industries, but the QCEW extends missing information at more aggregate level. For instance, all 3-digit subsectors of NAICS industry 32 except 323 are available so, if the QCEW was providing information on industry 32, it would be possible to obtain information on industry 323 by taking the difference between employment in 32 and employment in all 3-digit subsectors except 323. This is a problem, as for these reasons the QCEW does not provide any information for several important aggregates, such as manufacturing (31-33), wholesale trade (42), transportation (48-49), administrative services (56). To deal with this issue, we build 2-digit industrial aggregates summing from the subsectors for which data are available, and check using US-level data that the subsectors for which data are not available are a very small fraction (often 1-2\%) of total industry employment.

\subsection{Epidemic results}
\label{apx:empirical_epidemic_results}

\begin{figure}[htbp]
    \centering
\includegraphics[width = 1\textwidth]{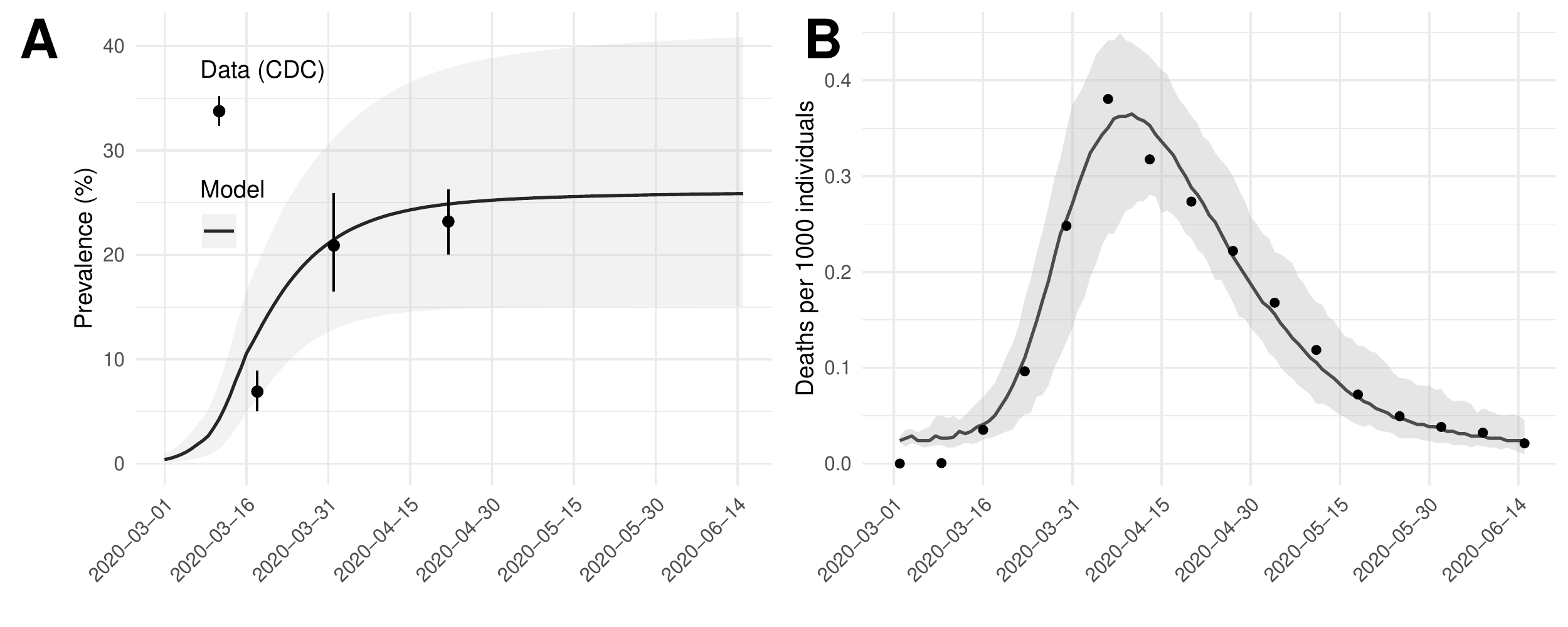}
    \caption{\textbf{Time series of prevalence and deaths.} A: prevalence over time, in the data (seroprevalence surveys) and in the model. B: weekly deaths over time, in the data and in the model. In both panels, error bands indicate the 2.5-97.5 percentiles. } 
        \label{fig:supp_empirical_epidemic_validation_time_series}
\end{figure}

Figure \ref{fig:supp_empirical_epidemic_validation_time_series} shows time series for prevalence and deaths. We see that the model predicts a time evolution for prevalence that is consistent with data, with high uncertainty about the final value of prevalence. The uncertainty over the time series for deaths is much smaller, and we see that the model can successfully track weekly data except in the first two weeks (in Figure \ref{fig:validation} we were only showing monthly aggregation). Note that the prevalence is not directly proportional to the number of deaths since it is possible to have a lot of infections in the younger age groups without significantly increasing the number of deaths. 

\begin{figure}[htbp]
    \centering
\includegraphics[width = 1\textwidth]{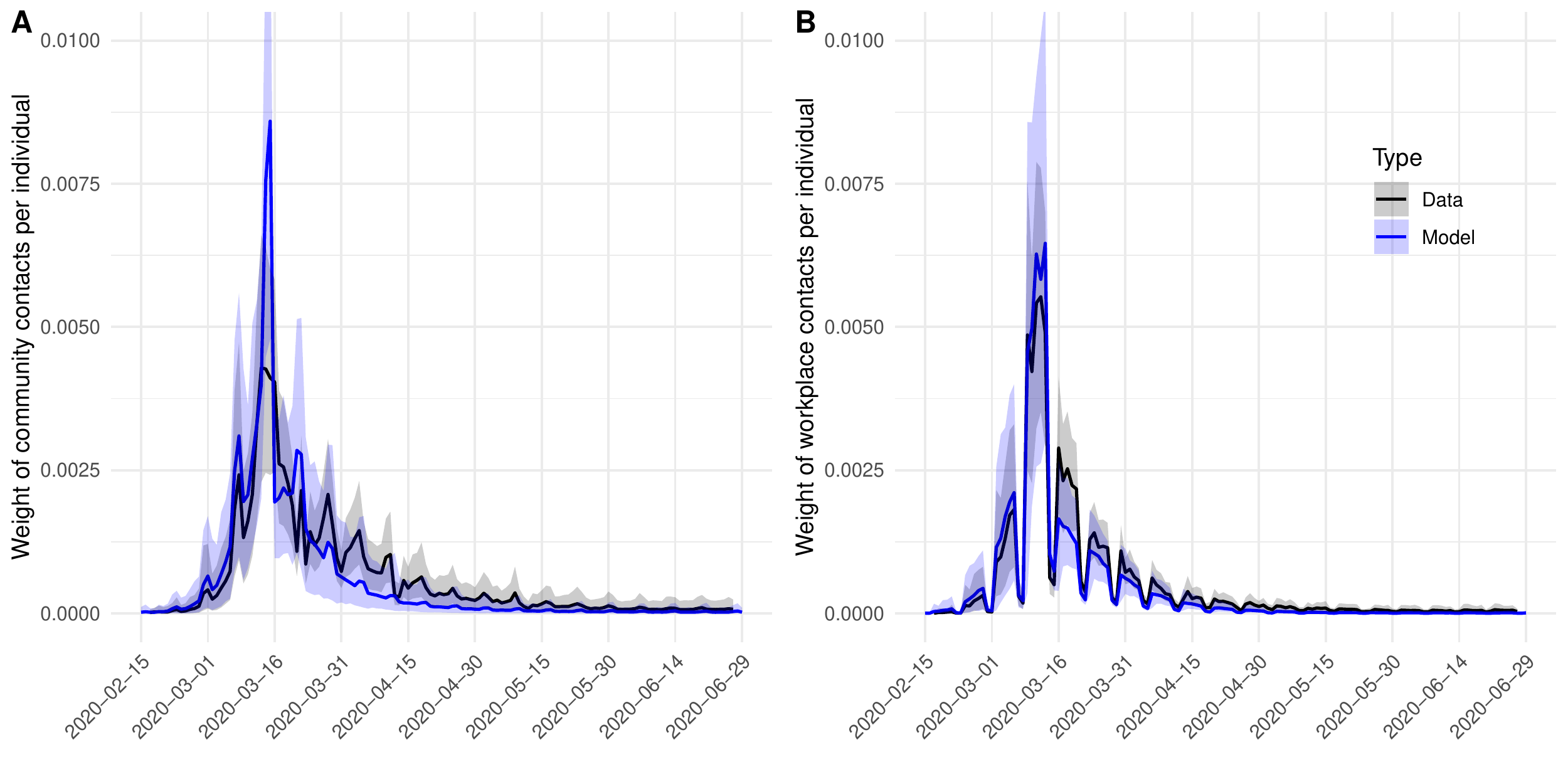}
    \caption{\textbf{Time series of contacts in the epidemic module.} A: average weight of community contacts per infected individual. B: average weight of workplace contacts per infected individual. In both cases, the values are computed by adding up the weight of all the contacts that infected individuals have with susceptible ones, divided by the total size of the population. Error bands indicate the 2.5-97.5 percentiles. } 
        \label{fig:supp_empirical_contacts_validation}
\end{figure}

Figure \ref{fig:supp_empirical_contacts_validation} shows the average weight of contacts in either the community or the workplace layer from infected to susceptible individuals, both in the real data and in our model. Note that in the real data an individual who chose to work from home does not have any connections. Instead, in the data fed to the model (as explained in Section \ref{sec:apx_def_customer_contact}) the link will exist. As such, to properly compare both situations, we only add the weight of those contacts that are materialized (in the model, if either individual works from home the contact cannot take place). As we can see, the model correctly captures the trend of the workplace contacts. For the community, instead, we observe an important peak of contacts in the model yields right before the closure imposed on 03/16 in comparison with the real data. This is probably related to voluntary stay-at-home behavior following the declaration of the National Emergency on 03/13. Indeed, as seen reported in \cite{aleta2022quantifying}, the number of contacts in the community on the weekend from 03/14 to 03/15 is smaller than expected even though most places had not been force to close yet. This behavior cannot be reproduced by the fear of infection that we have introduced since it only depends on the number of reported deaths and not on any exogenous events.

Note also that there is not an important surge of contacts in the real data following the reopening of economic activities. This is probably due to the slow pace at which the riskier venues reopened. For instance, restaurants were only allowed to operate for delivery, and indoor dining came back in September. Similarly, retail stores in New Jersey reopened on May 18 but only for curbside pickup. Thus, even though in the economic module the restrictions to operate are lifted, the epidemic module assumes that most industries are still closed until the end of the simulation period \cite{NJrestrictions,NYrestrictions}.

\begin{figure}[htbp]
    \centering
\includegraphics[width = 0.5\textwidth]{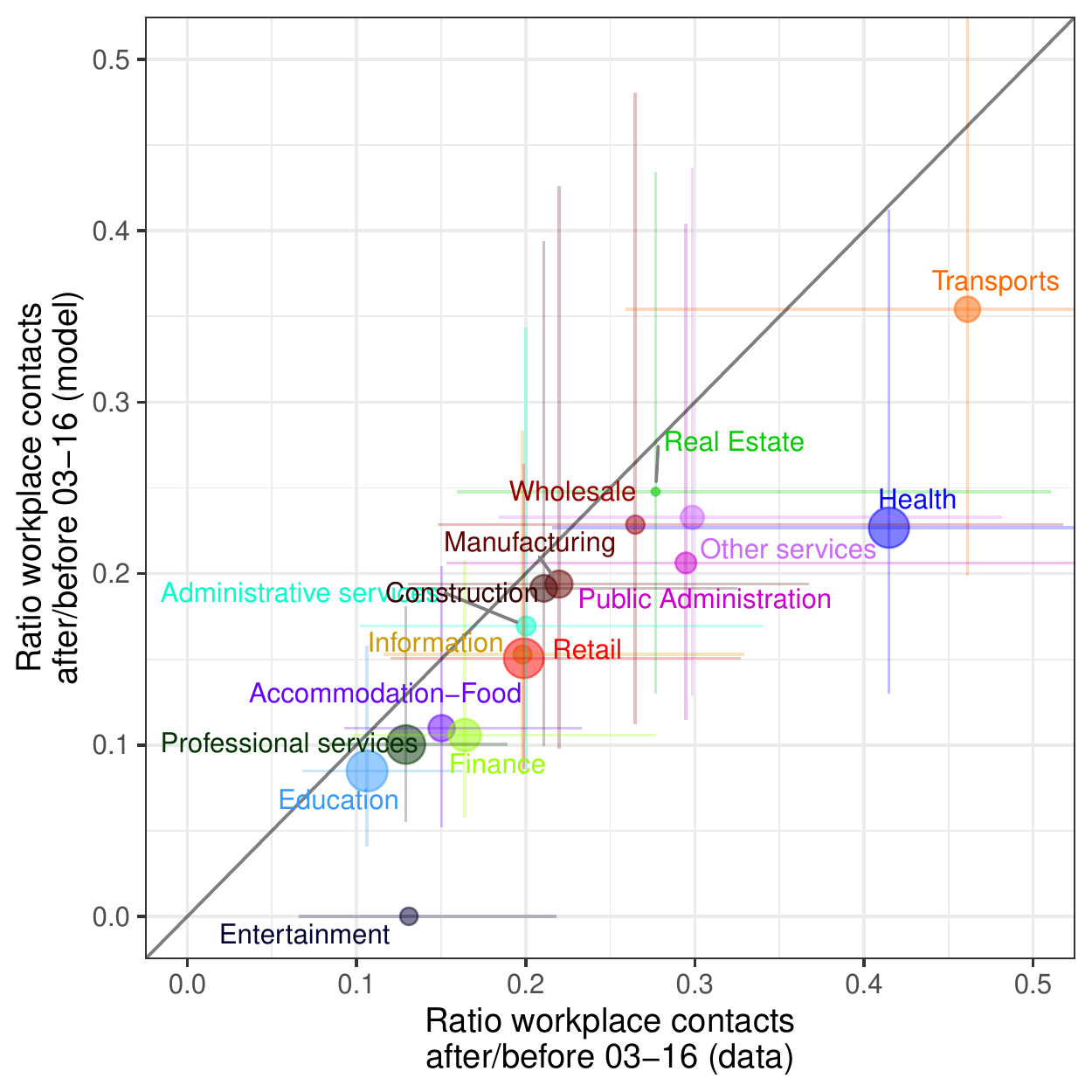}
    \caption{\textbf{Contacts in the workplace.} Ratio between workplace contacts with infectious individuals before and after the imposition of protective measures, in the model and in the data. This figure has the same interpretation as Figure \ref{fig:validation}E. } 
        \label{fig:alt_contacts_plot_workplace}
\end{figure}

Our last results strictly on validation are shown in Figure \ref{fig:alt_contacts_plot_workplace}. Here we show the same plot as Figure \ref{fig:validation}E, but for workplace contacts. We find that these are extremely well correlated between model and data, with a correlation coefficient of 0.88. Workplace contacts declined less in Transportation and Health than Entertainment and Accommodation-Food, in line with the essentialness of industries. Workplace contacts declined a lot also in Professional services, Education and Finance, reflecting the switch to work from home in these industries.

Our next set of results on the empirical scenario cannot strictly be considered as validation, as we will show some results that we could not quantitatively compare to real-world data because such data do not exist. However, as we will discuss in Section \ref{apx:empirical_epidemic_evidence}, we can qualitatively find support for the following results.

\begin{figure}[htbp]
    \centering
\includegraphics[width = 1\textwidth]{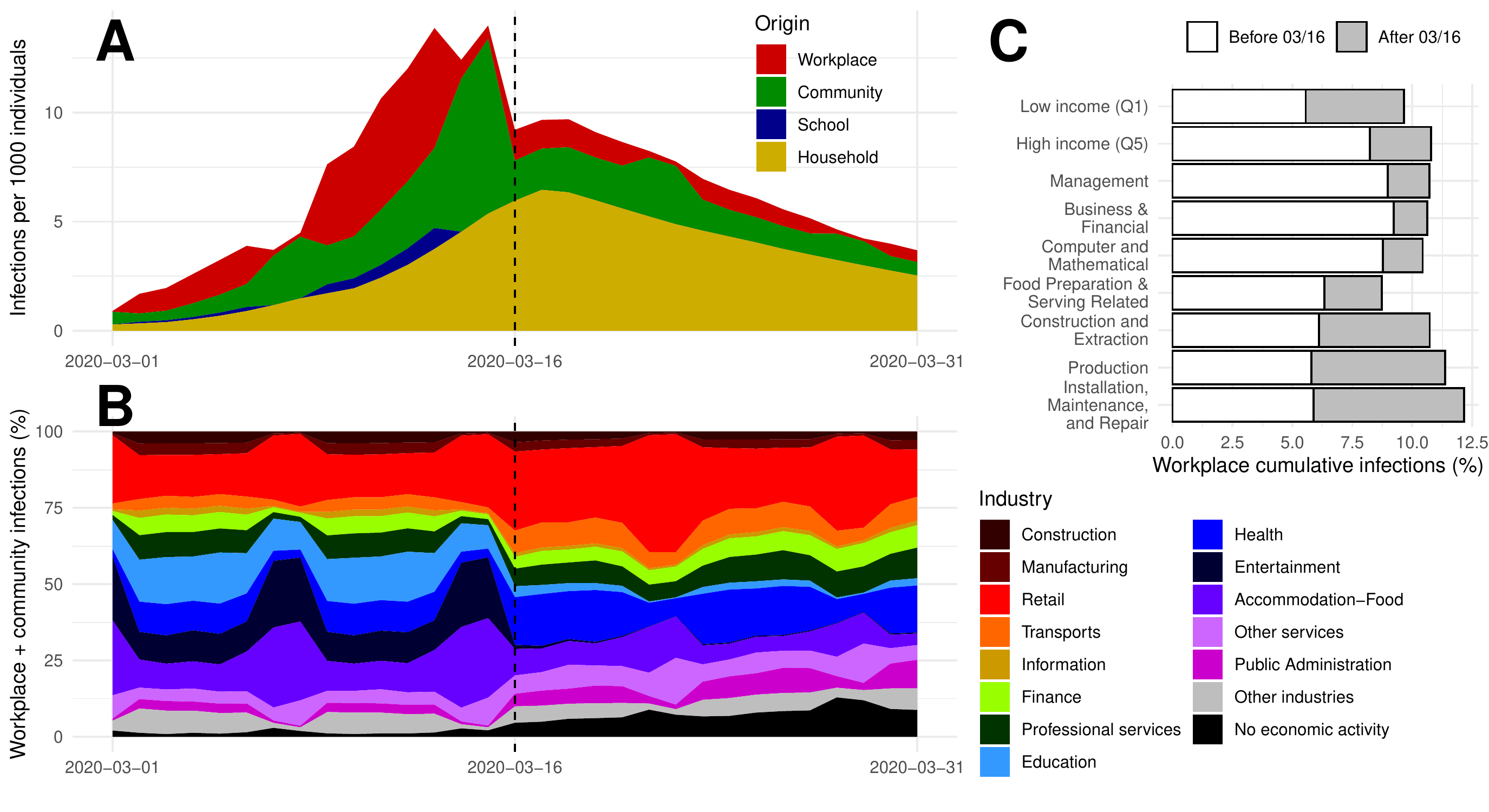}
    \caption{\textbf{Infections by origin and by socio-economic characteristics.} A: Time series of infections disaggregated by origin (workplace, community, school and household layers). The dashed vertical line indicates the day when we impose protective measures (2020-03-16). B: Time series of workplace plus community infections disaggregated by the industry where the infection took place. Time series are normalized by total workplace and community infections. C: Percentage of all workers across some income levels and occupations that has been
infected in the workplace before 2020-03-16 and after 2020-03-16. } 
        \label{fig:supp_empirical_epidemic}
\end{figure}

Figure \ref{fig:supp_empirical_epidemic} shows results on infections by origin and by socio-economic characteristics. Our model predicts that, before protective measures were implemented in mid-March, most infections occurred in the workplace and community layers (Figure \ref{fig:supp_empirical_epidemic}A). After these chains of transmission were severed by social distancing, household
transmission took the highest share of infections. At the industry level (Figure \ref{fig:supp_empirical_epidemic}B), we see a large share of infections in entertainment and accommodation-food before 2020-03-16, especially during weekends, while this share declined after 2020-03-16. In the second period, retail acquired the largest share of infections, as it required in-person interactions and remained partly essential (e.g., consider grocery stores and pharmacies). We also find a general pattern that high-income workers engaged in knowledge-related and business occupations got more infected than low-income workers engaged in physical occupations before 2020-03-16 (Figure \ref{fig:supp_empirical_epidemic}C). This is because they work in denser census tracts that we assume are at higher risk of infection \cite{aleta2022quantifying}. The reverse pattern holds after 2020-03-16, because many high-income workers switch to working from home, while low-income workers in essential occupations must keep working in person.

\begin{figure}[htbp]
    \centering
\includegraphics[width = 0.8\textwidth]{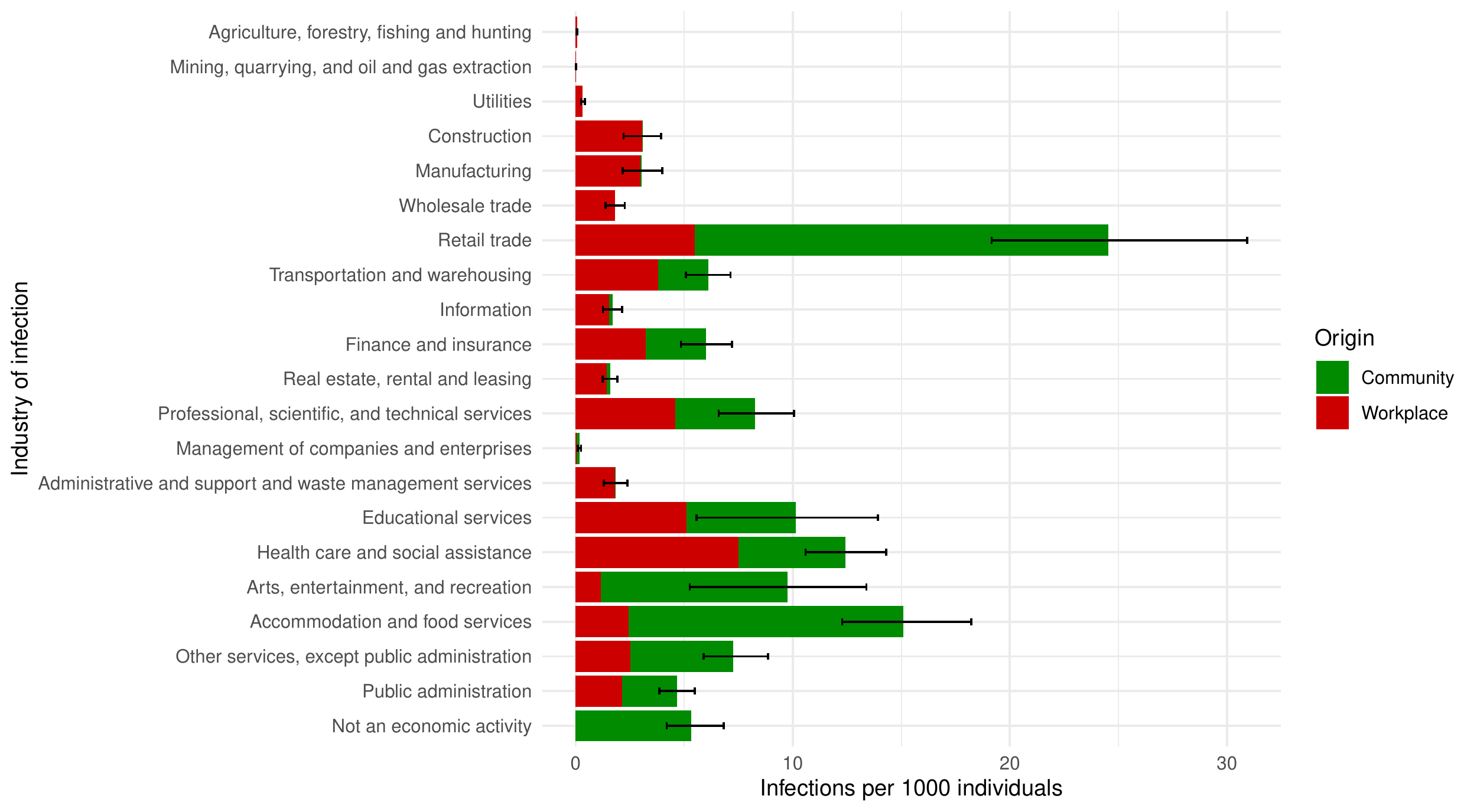}
    \caption{\textbf{Infections by the industry in which the infection happened in the empirical scenario.} We sort industries by NAICS code, and show results distinguishing between the community and workplace layers. Error bars indicate 2.5-97.5 percentiles and refer to the total of infections, not just infections in the community layer. } 
        \label{fig:supp_empirical_infections_by_industry_of_infection_by_origin}
\end{figure}

Figure \ref{fig:supp_empirical_infections_by_industry_of_infection_by_origin} shows infections across industries, distinguishing between infections in the community and workplace layers. It is possible to see that for some industries most infections happen in the community layer (retail trade, arts, entertainment, and recreation, accommodation and food services), while for other industries a majority of infections occur in the workplace layer (health care and social assistance, finance and insurance, professional, scientific and technical services). In a few cases such as construction and manufacturing, infections occur only in the workplace, as these economic activities are not customer-facing.

\begin{figure}[htbp]
    \centering
\includegraphics[width = 0.8\textwidth]{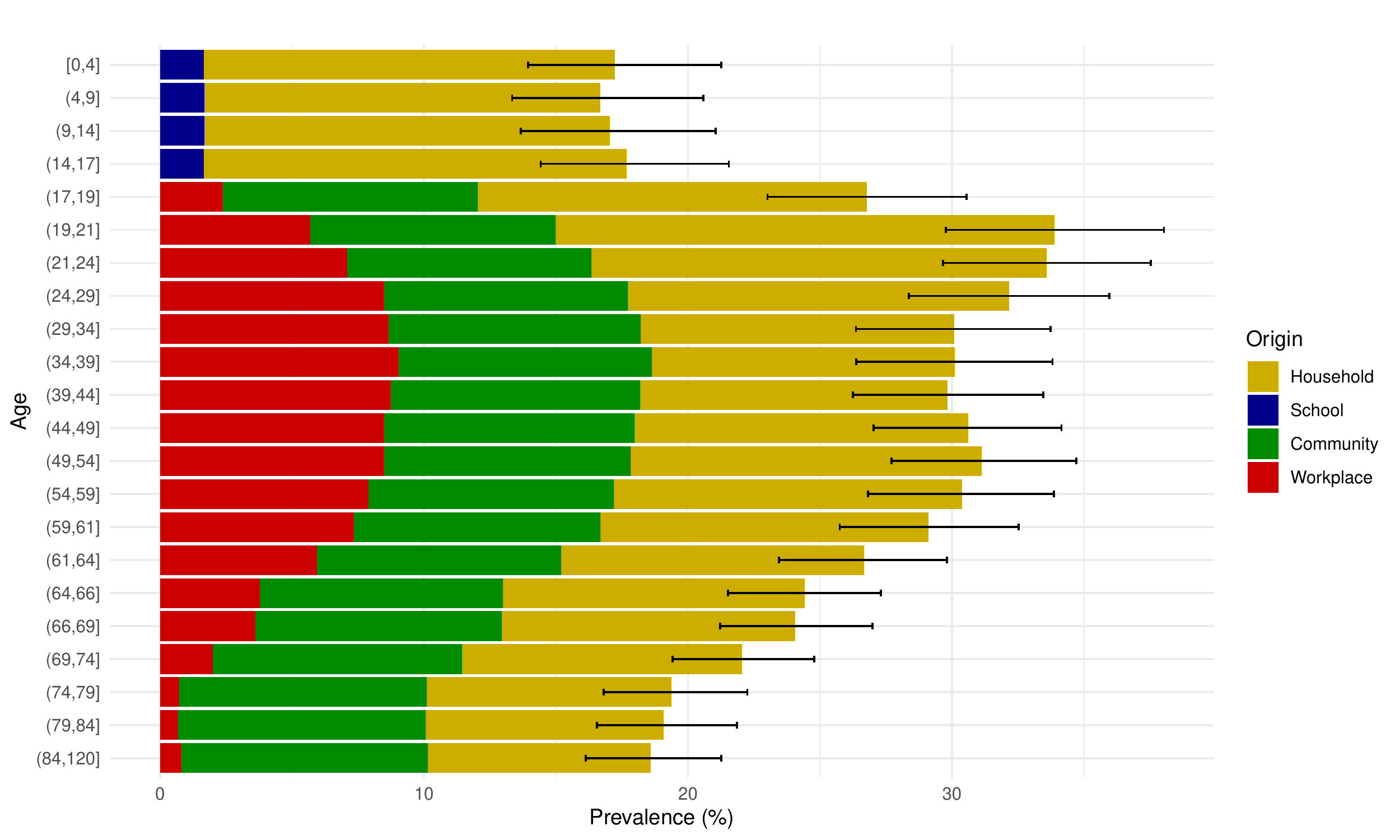}
    \caption{\textbf{Infections by age by origin in the empirical scenario.} For each age band, we show the prevalence at the end of the simulation, distinguishing between the origin of infections. Error bars indicate 2.5-97.5 percentiles and refer to the total of infections, not just infections in the household layer. } 
        \label{fig:supp_empirical_infections_by_age_by_origin}
\end{figure}

Figure \ref{fig:supp_empirical_infections_by_age_by_origin} shows the percentage of people in each age band that got infected, distinguishing by the origin of infections. We see that children (agents up to 17 years old) got some infections in school, but most infections in the household (likely due to the relatively early closure of schools). Across all age groups infections in the household are the majority (likely related to the stay-at-home order which increased time spent in the household while reduced contacts everywhere else), but for intermediate age groups infections are almost equally divided with the workplace and community layers. The fact that intermediate age groups got infected the most is consistent with empirical evidence discussed in Section \ref{apx:empirical_epidemic_evidence}

\FloatBarrier

\subsubsection{Empirical evidence}
\label{apx:empirical_epidemic_evidence}

The empirical evidence on rates of infection across socio-economic characteristics is mixed. Indeed, it is well known that infections confirmed by a PCR test are largely biased, especially in the early period of the pandemic that we are looking at. So, we need to look at serological tests, but the ones available for the New York area are not representative of the whole population. Therefore, we will also consider results for other countries when these are more likely to be representative. 

\paragraph{Sources.} Case counts (confirmed by a PCR test) are available by county using the New York Times dataset. The New York City (NYC) Health Authority (\url{https://github.com/nychealth/coronavirus-data}) provides additional data on case counts, also disaggregated by age. Refs. \cite{moscola2020prevalence} and \cite{venugopal2021sars} report results on serological tests among NYC health care workers, while \cite{rosenberg2020cumulative} offered serological tests to shoppers in several counties in New York state. Ref. \cite{sami2021prevalence} analyzes serological tests to look at prevalence among occupations. Ref. \cite{anand2020prevalence} performed a large-scale serological study among patients on dialysis across all the United States.  Ref. \cite{pollan2020prevalence} studies the outcomes of serological tests in Spain during the first wave: this is the only study that is representative of the population.  Finally, Ref. \cite{magnusson2020occupational} provides a detailed breakdown of infections confirmed by PCR tests across occupations for the whole population of Norway.

\paragraph{Infections by age.} PCR-positives in New York City, as obtained by the NYC Health Authority, show a consistently higher positivity rate as age is increased. By the first week of June, 2.3\% of New Yorkers aged more than 75 tested positive, which compares to only 1.6\% of adults aged 45-54 and 0.7\% of young adults aged 18-24. But these results are biased by the fact that only people with severe symptoms were tested in the first wave of the epidemic. Looking at serological tests, both \cite{moscola2020prevalence} and \cite{anand2020prevalence} found decreasing prevalence with increasing age, while \cite{venugopal2021sars} and \cite{rosenberg2020cumulative} find the highest prevalence in middle ages. The results in \cite{pollan2020prevalence} depend on the type of test used, but generally find a higher prevalence at higher ages. Overall, these results lend some support to our results on age prevalence (Figure \ref{fig:supp_empirical_infections_by_age_by_origin}).

\paragraph{Infections by occupation.} The studies limited to healthcare workers in New York show higher infection rates among service and maintenance workers than among nurses or physicians, but the relative difference varies between studies \cite{moscola2020prevalence,venugopal2021sars}. Ref. \cite{sami2021prevalence} finds the highest prevalence among prison staff and the lowest among laboratory technician, but there is no specific patterns and the set of occupations is not representative of occupations -- the classification is non-standard. Ref. \cite{pollan2020prevalence} finds the highest prevalence among healthcare workers, an intermediate prevalence among workers in transportation, the police, and teleworkers, and the lowest prevalence among retailers and cleaners. This, however, was true for the first wave; in the second wave, cleaners had a much larger prevalence than teachers, and close to the healthcare workers. Ref. \cite{magnusson2020occupational} finds higher prevalence among healthcare workers and bus and tram drivers and lower prevalence among waiters and personal services in the first wave, but higher prevalence among workers in the hospitality sector in the second wave. We find this evidence on occupations to be too mixed to be compared to our results.

\paragraph{Infections by income.} Several studies consider race and, because race is strongly correlated with income, argue about income on the basis of race. For instance, studies showing that blacks and hispanics have higher prevalence than whites also imply that prevalence was high among low-income people. Here, we prefer to look at prevalence by income directly. A NYC Health study shows that ZIP codes with higher poverty rates have higher prevalence according to antibody tests: \url{https://github.com/nychealth/coronavirus-data/blob/master/totals/antibody-by-poverty.csv}. This is in line with the results about race. The Spanish study, which is representative of the full population, does not show a clear trend on infections by income, with higher-income individuals being more hit in the first wave and less in the second wave \cite{pollan2020prevalence}. In this case, these results are supportive of our finding that rich individuals in NY got more infected before the lockdown and less infected after the lockdown (Figure \ref{fig:supp_empirical_epidemic}C).

\section{Supplementary results on the counterfactual scenarios}
\label{apx:scenarios}

Here we report additional results on the counterfactuals. These include:
\begin{itemize}
    \item Starting protective measures on 2020-02-17 (early), 2020-03-16 (baseline) and 2020-03-30 (late). 
    \item Closing all non-essential industries as in the NY lockdown (``Non-essential''), only closing non-essential customer-facing industries (``Cust.-facing 100\%''), only closing 25-50-75\% of these industries (uniformly across industries), and leaving all economic activities open (``All open''). Recall that in all cases the essential part of industries is always left open. For instance, when we say that 100\% of the retail industry is closed, we only refer to non-essential shops, constituting only 35\% of the retail industry (Table \ref{tab:foursquare}).
    \item Closing schools and mandating work from home; opening schools keeping work from home mandatory; and opening schools and making work from home voluntary
    \item Calibrating fear of infection to 0.1 (low), 1 (baseline), and 10 (high) times its baseline value.
\end{itemize}
To show the generality of our results, in this supplementary materials we report more results in one plot than in the main paper. This may be at the cost of legibility -- for instance, while in Figure \ref{fig:main_counterfactuals_aggregate} we only vary fear two factors at a time (with shape and color to reflect all factor combinations), here we will vary three factors at a time (including size). 

\subsection{Aggregate results}
\label{apx:scenarios_aggregate_results}

\begin{figure}[htbp]
    \centering
\includegraphics[width = 0.8\textwidth]{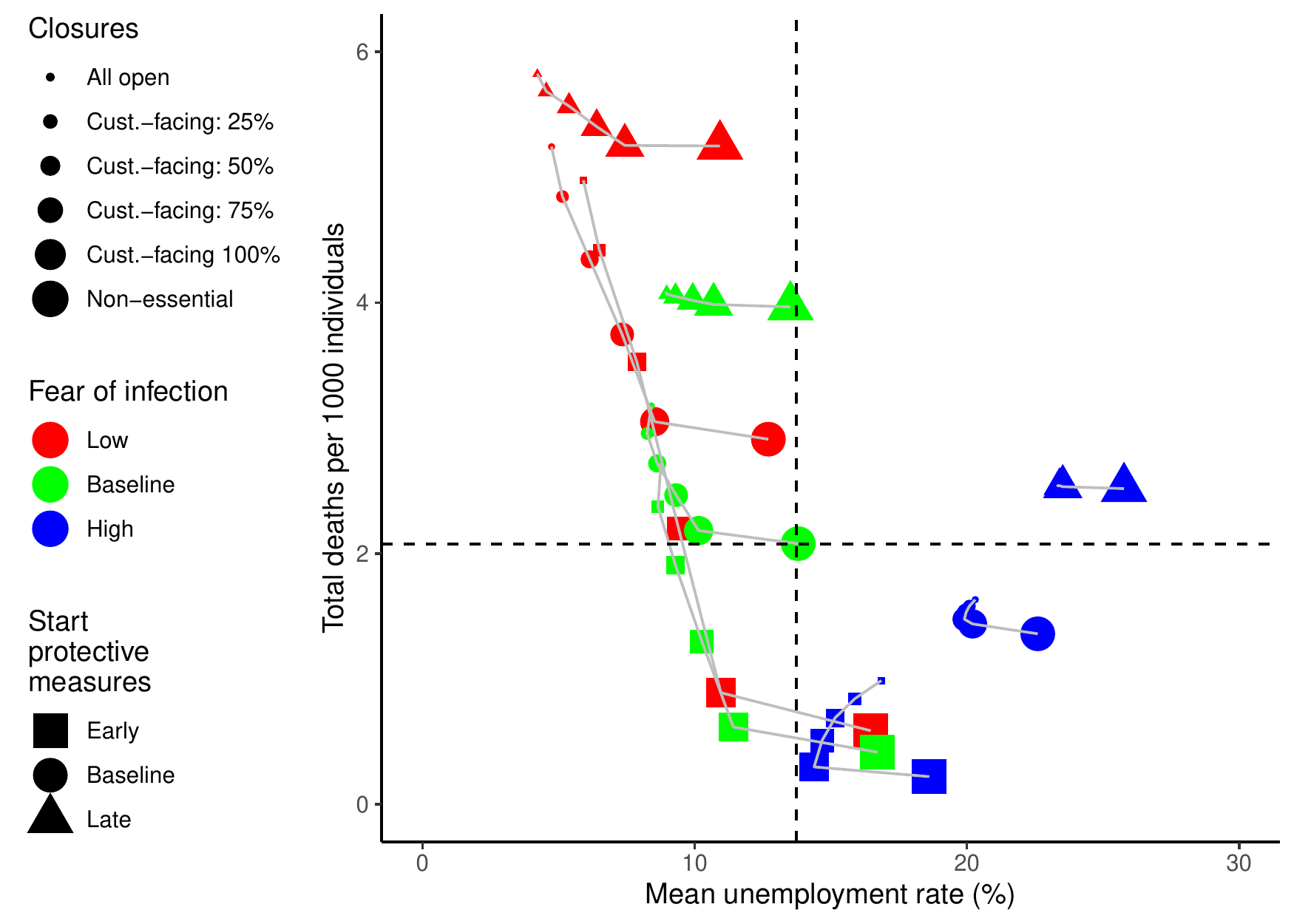}
    \caption{\textbf{Deaths and unemployment across counterfactual scenarios, including partial closures of customer-facing industries.} For each scenario, we show the aggregate unemployment rate as averaged throughout the simulation period, and the cumulative number of deaths. Dashed lines indicate the empirical unemployment and number of deaths. Scenarios are distinguished by the strength of behavior change, as exemplified by the fear of infection parameter (color); by the day when protective measures --closures of economic activities, school closures, imposition of work from home-- were started (symbol); by the specific closure of economic activities (size). Compared to Figure \ref{fig:main_counterfactuals_aggregate}A, we consider a partial closure of customer-facing industries: We allow for these industries to be 25\% closed, 50\% closed, 75\% closed. Error bars indicate 2.5-97.5 percentiles.  } 
        \label{fig:supp_counterfactuals_aggregate_all_closures}
\end{figure}

Figure \ref{fig:supp_counterfactuals_aggregate_all_closures} is similar to Figure \ref{fig:main_counterfactuals_aggregate}A, but contains more information. First, it considers the variation of all three factors -- closures, fear of infection and start of protective measures -- at a time. Second, it includes the scenarios in which non-essential customer-facing economic activities are partly (25-50-75\%) closed. 

The results are in line with Figure \ref{fig:main_counterfactuals_aggregate}A: strong behavior change reduces deaths but increases unemployment as much as strict closures; closing all non-essential rather than just customer-facing industries  greatly increases unemployment while marginally reducing deaths; and waiting to start protective measures is in most cases counterproductive both for epidemic and economic outcomes. This figure also shows other combinations of scenarios that were not shown in the main text. For instance, it is interesting to look at the combination of high fear of infection and an early start of protective measures (blue squares). In this case, leaving all economic activities open (even partially) leads to both more deaths and more unemployment than closing only customer-facing industries. The results for the scenarios that involve partial closure of customer-facing industries interpolate almost linearly between the cases in which these industries are completely open or completely closed.

\begin{figure}[!h]
    \centering
\includegraphics[width = 0.8\textwidth]{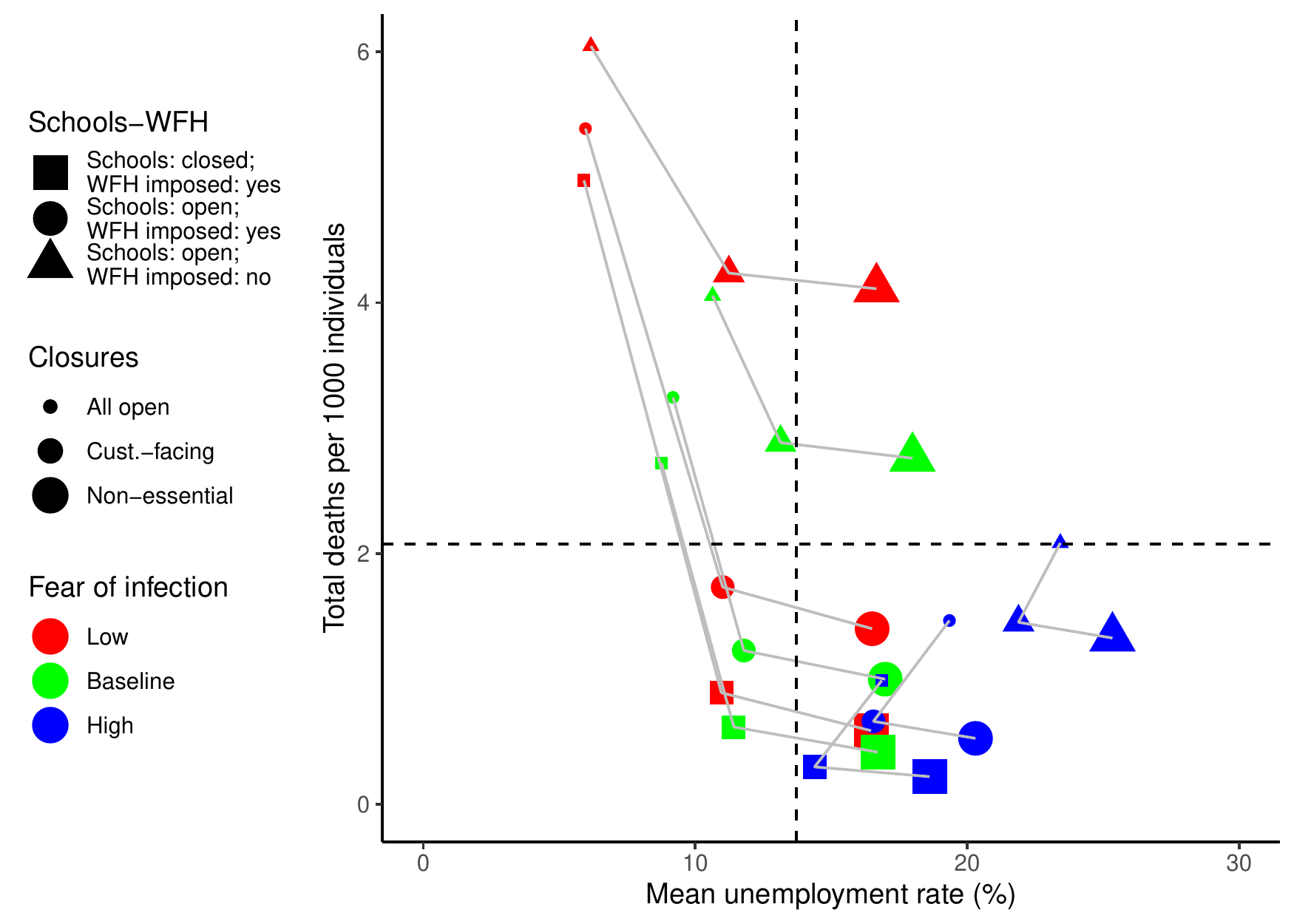}
    \caption{\textbf{Deaths and unemployment across counterfactual scenarios, varying school closures and imposition of work from home.} For each scenario, we show the aggregate unemployment rate as averaged throughout the simulation period, and the cumulative number of deaths. Dashed lines indicate the empirical unemployment and number of deaths. Scenarios are distinguished by the strength of behavior change, as exemplified by the fear of infection parameter (color); by the specific closure of economic activities (size); and by whether schools were closed and/or work from home imposed (symbol). In all cases, we impose protective measures on 2020-02-17 (Early), so that they have maximal effect and we can better see the effect of school closures and work from home.  Error bars indicate 2.5-97.5 percentiles. } 
        \label{fig:supp_counterfactuals_aggregate_all_schools_wfh}
\end{figure}

Figure \ref{fig:supp_counterfactuals_aggregate_all_schools_wfh} shows results when school and work from home policies are changed. Compared to the baseline in which schools are closed and work from home is imposed, we see a much increase in deaths due to relaxing the work from home mandate than to closing schools (the distance between squares and circles is much smaller than the distance between circles and triangles).

\begin{figure}[!h]
    \centering
\includegraphics[width = 0.5\textwidth]{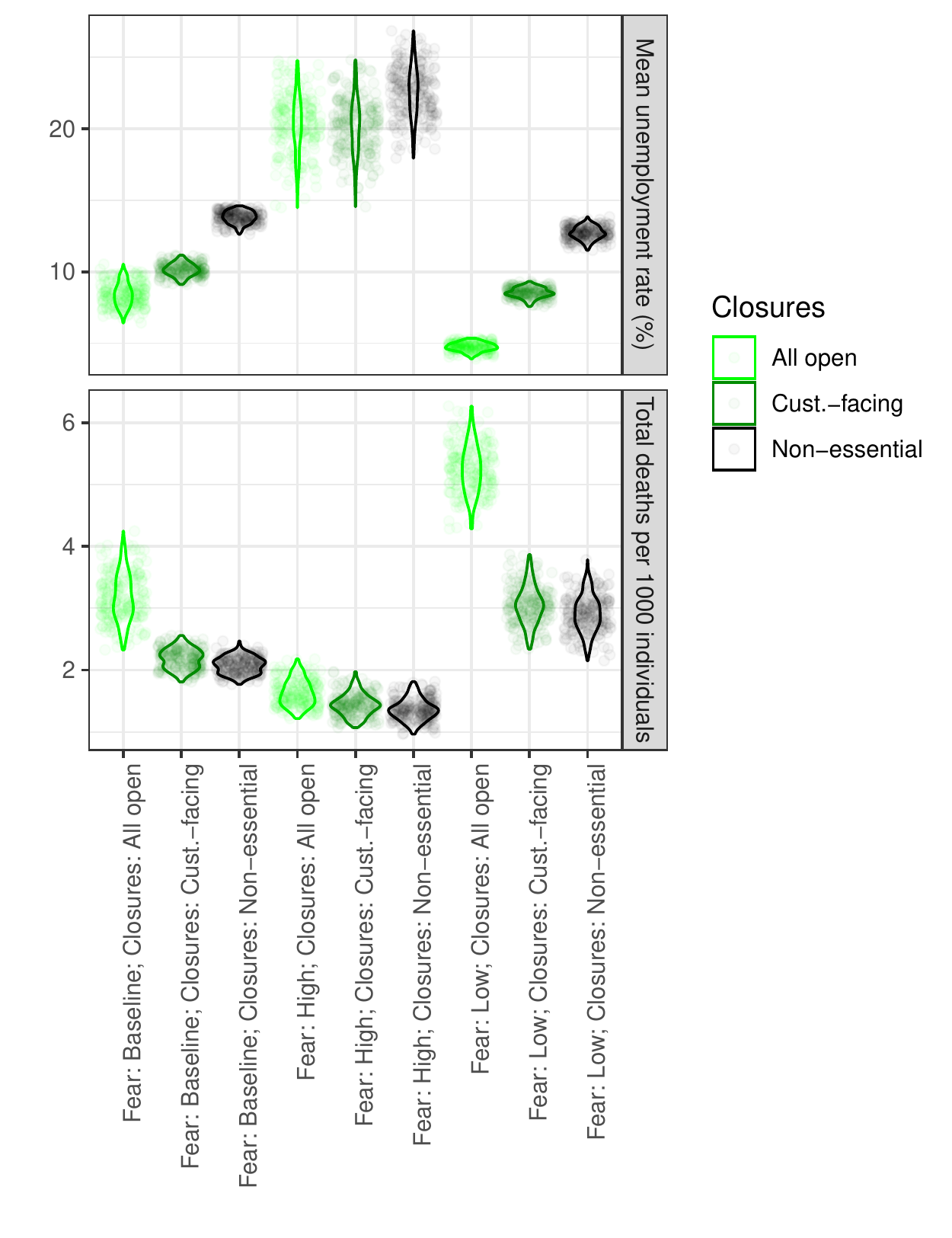}
    \caption{\textbf{Distribution of deaths and unemployment across simulation runs.} For each of the scenarios in Figure \ref{fig:main_counterfactuals_aggregate}A, we show a violin plot and a point distribution for both the mean unemployment rate and the cumulative number of deaths.} 
        \label{fig:supp_counterfactuals_aggregate_distributions}
\end{figure}

Finally, Figure \ref{fig:supp_counterfactuals_aggregate_distributions} shows how results are distributed across simulation runs for the scenarios shown in Figure \ref{fig:main_counterfactuals_aggregate}A. As discussed in the main text, when comparing different counterfactuals we only report mean results: reporting error bars would make the plot more complicated without adding much useful information, as what matters most is the difference across scenarios rather than absolute values. For instance, it would not be much informative to say that unemployment could vary between 7\% and 10\% under baseline fear of infection and no closures, and between 9\% and 11\% when fear of infection is again baseline but customer-facing industries are closed (as in the two leftmost examples in the top row of Figure \ref{fig:supp_counterfactuals_aggregate_distributions}). This would suggest that there is substantial overlap between the two scenarios. However, using a paired Welch t-test, one can reject the hypothesis that there is no difference in means with p-value $<10^{-6}$. More systematically, we find that all pairs of scenarios shown in Figure \ref{fig:supp_counterfactuals_aggregate_distributions} have highly statistically significant differences in means, despite overlaps between the distributions.

\subsection{Disaggregate economic results}
\label{apx:scenarios_economic_results}

\begin{figure}[htbp]
    \centering
\includegraphics[width = 1\textwidth]{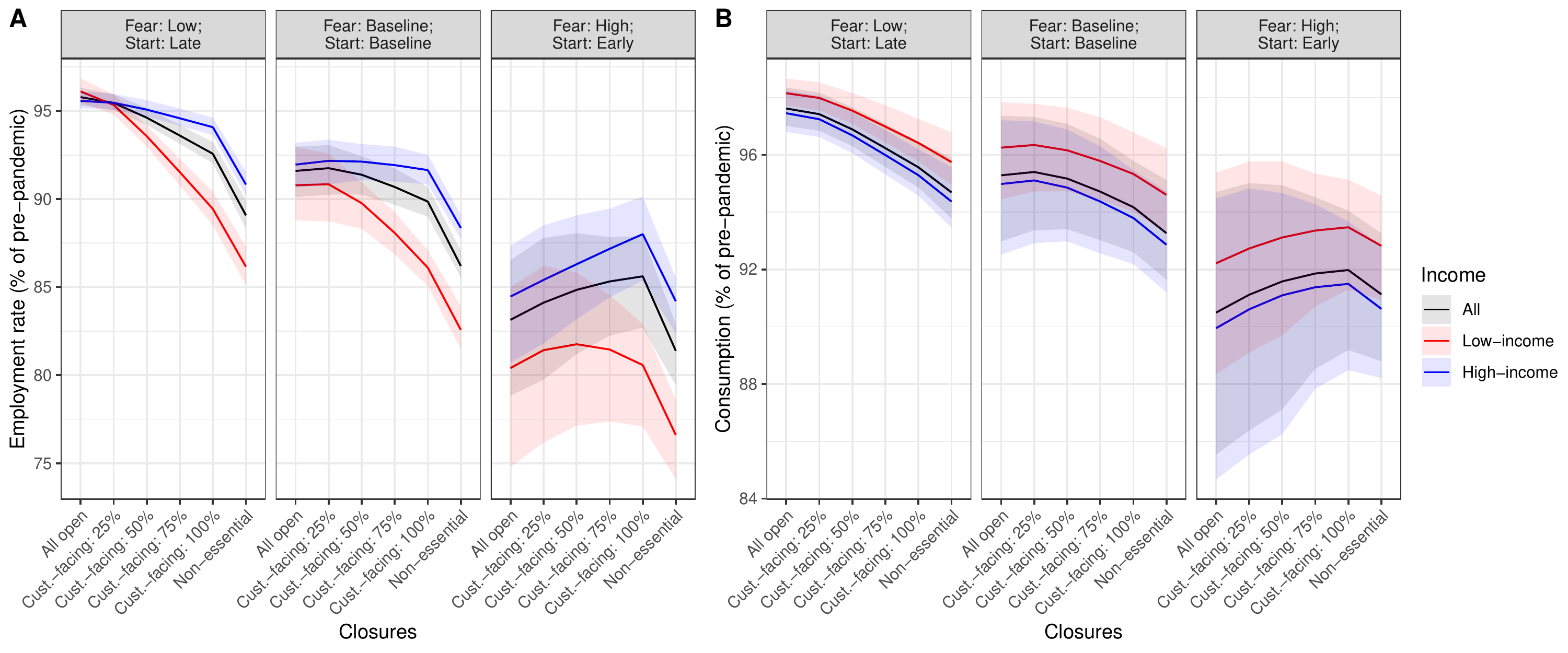}
    \caption{\textbf{Employment and consumption by income across counterfactual scenarios.} We select three combinations of fear of infection and start of protective measures, and for each of these combinations we study all six possibilities for closures of economic activities. Both employment and consumption are aggregated across industries and averaged throughout the simulation period, and then compared to their pre-pandemic level.  We distinguish between aggregate employment and consumption, and employment and consumption by low-income individuals and households (those earning less than 27k\$, to keep the same convention as in the validation part of the paper) and by high-income individuals and households (those earning more than 60k\$).  Error bands indicate 2.5-97.5 percentiles. } 
        \label{fig:supp_counterfactuals_disaggregate_employment_consumption_by_income}
\end{figure}

Figure \ref{fig:supp_counterfactuals_disaggregate_employment_consumption_by_income} focuses on three scenarios selected for illustration. In all cases, employment and consumption go down as fear of infection is increased; employment of low-income individuals is below employment of high-income individuals (except in the leftmost cases in the leftmost panel), and this difference widens with stricter closures of economic activities; consumption by low-income households decreases less than consumption by high-income households, with little difference when different closure policies are considered. When fear of infection is high, we see an ``inverted-V'' shape for both employment and consumption, in line with the aggregate result in Figure \ref{fig:supp_counterfactuals_aggregate_all_closures} that in this case opening all economic activities leads to more unemployment than keeping customer-facing activities closed.

\begin{figure}[htbp]
    \centering
\includegraphics[width = 1\textwidth]{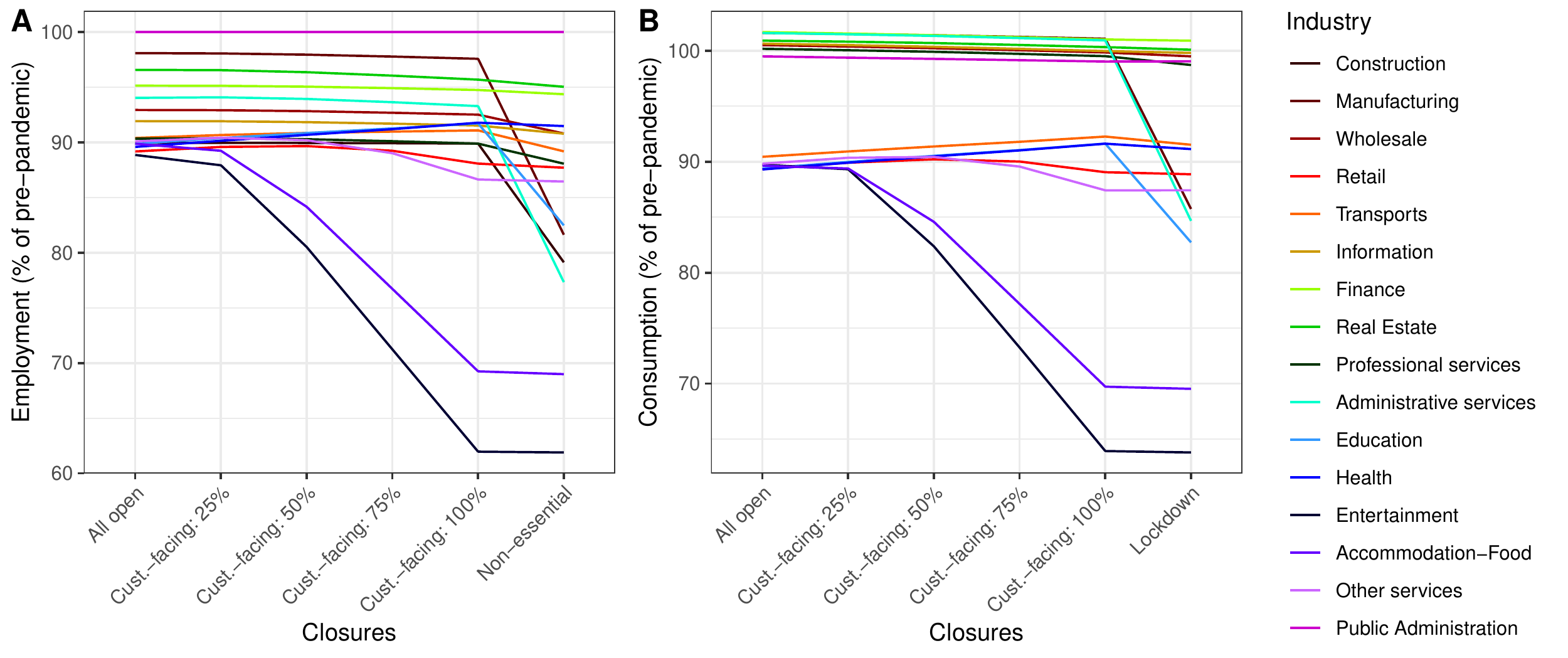}
    \caption{\textbf{Employment and consumption by industry across counterfactual scenarios.} We focus on the empirical scenario to select fear of infection (baseline) and the start of protective measures (baseline), and we study all six possibilities for closures of economic activities. Both employment and consumption are averaged throughout the simulation period, and then compared to their pre-pandemic level. Here and in following plots, we do not show error bands to enhance visibility.  } 
        \label{fig:supp_counterfactuals_disaggregate_employment_consumption_by_industry}
\end{figure}

Figure \ref{fig:supp_counterfactuals_disaggregate_employment_consumption_by_industry} focuses on the empirical scenario, only changing the possible closures of economic activities. It shows that the difference across industries in terms of employment and consumption becomes smaller when customer-facing industries are opened, but accommodation-food and entertainment remain among the ones most strongly hit. The reason why public administration employment remains equal to its pre-pandemic level is that most demand for public administration services comes from the government, which we assume to  increase its demand by 5\%. This means that the public administration industry is never demand-constrained.

\begin{figure}[htbp]
    \centering
\includegraphics[width = 0.9\textwidth]{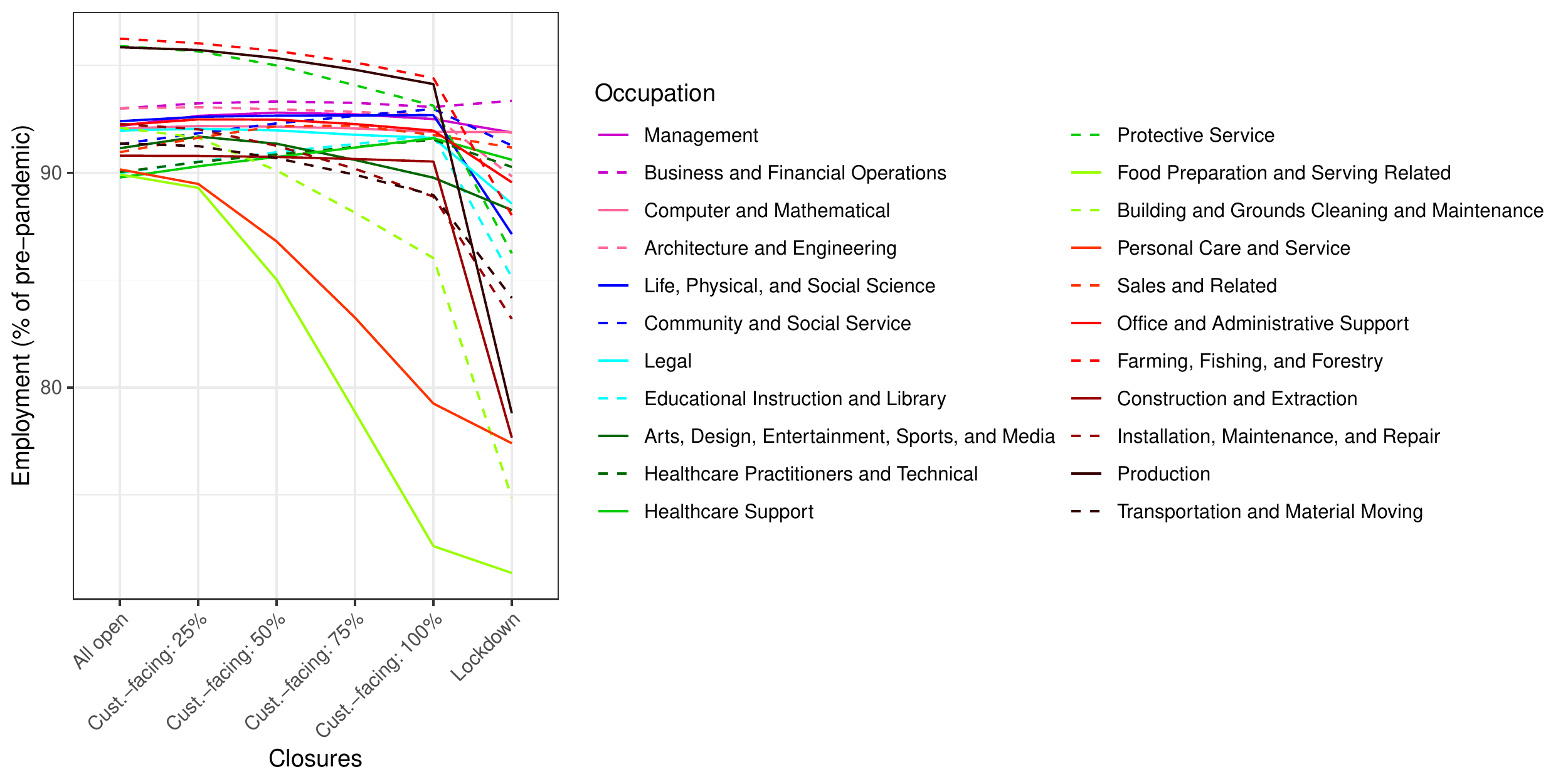}
    \caption{\textbf{Employment by occupation across counterfactual scenarios.} We focus on the empirical scenario to select fear of infection (baseline) and the start of protective measures (baseline), and we study all six possibilities for closures of economic activities. For each occupation, employment is averaged throughout the simulation period, and then compared to its pre-pandemic level. To enhance readability, we consider combinations of color and line type, so there is nothing specific to occupations represented by dashed vs. solid lines.} 
        \label{fig:supp_counterfactuals_disaggregate_employment_by_occupation}
\end{figure}

Figure \ref{fig:supp_counterfactuals_disaggregate_employment_by_occupation} is similar to Figure \ref{fig:supp_counterfactuals_disaggregate_employment_consumption_by_industry}, but it shows occupational employment. The results are in line with what can be expected given the occupational composition of industries (Figure \ref{fig:plots_joint_occ_ind}A). For instance, food preparation and serving related occupations are the worst hit across all six scenarios considered here, in line with the fact that the accommodation-food industry was the second most hit across all these scenarios.

\subsection{Disaggregate epidemic results}
\label{apx:scenarios_epidemic_results}

\begin{figure}[htbp]
    \centering
\includegraphics[width = 0.6\textwidth]{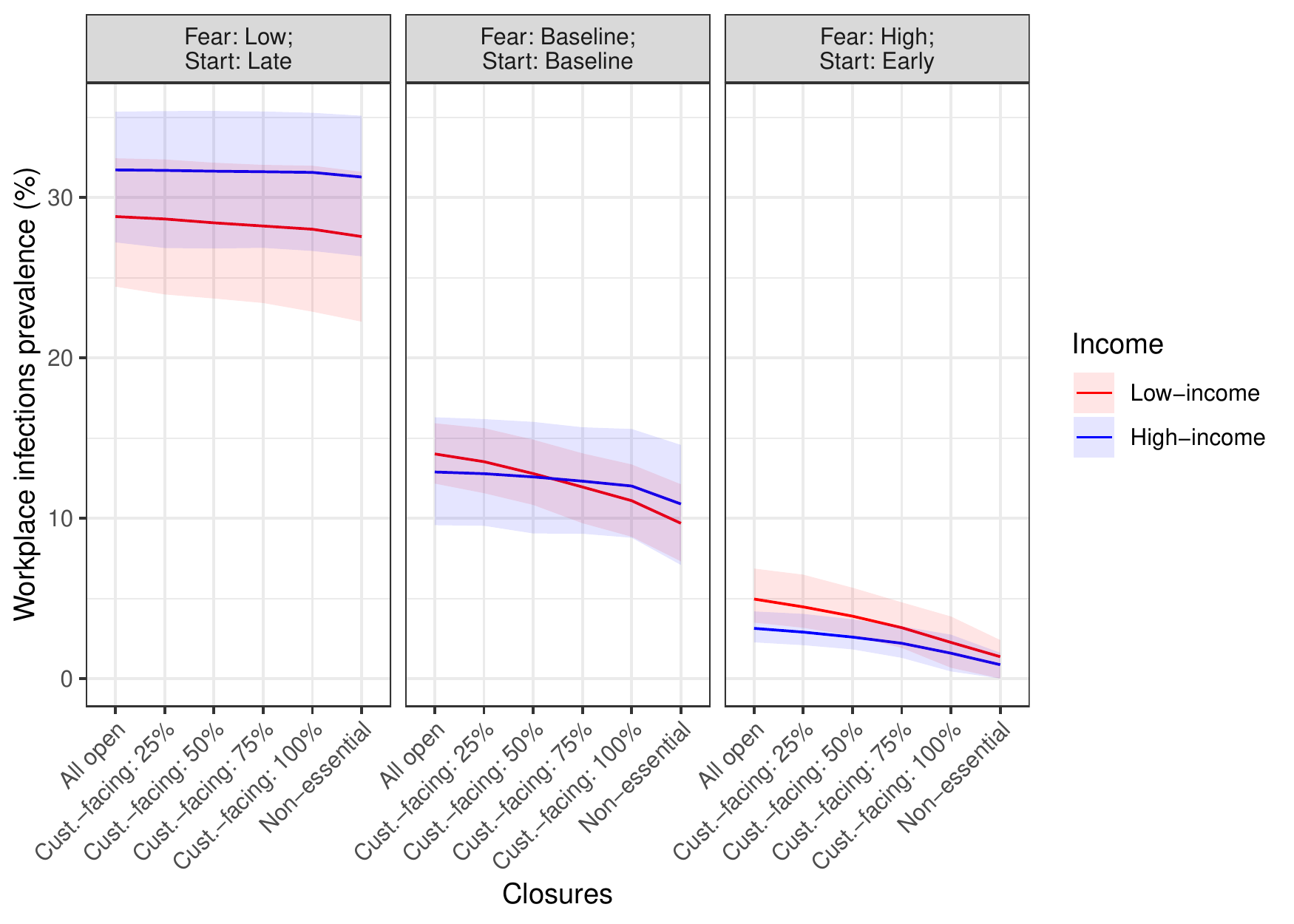}
    \caption{\textbf{Workplace infections prevalence by income across counterfactual scenarios.} We select three combinations of fear of infection and start of protective measures, and for each of these combinations we study all six possibilities for closures of economic activities. Workplace infections prevalence is defined as the percentage of all workers of a given type that have been infected in the workplace at the end of the simulation period. We distinguish between prevalence of low-income workers (those in the first quintile of the income distribution) and of high-income workers (those in the top quintile of the income distribution).  Error bands indicate 2.5-97.5 percentiles. } 
        \label{fig:supp_counterfactuals_disaggregate_infections_by_income}
\end{figure}

Figure \ref{fig:supp_counterfactuals_disaggregate_infections_by_income} shows that high-income workers get more infected in the workplace than low-income workers when fear of infection is low and work from home is imposed very late (left panel), while the reverse is true when work from home is imposed early (right panel). Closing customer-facing industries reduces infection rates among low-income workers the most.

\begin{figure}[htbp]
    \centering
\includegraphics[width = 0.6\textwidth]{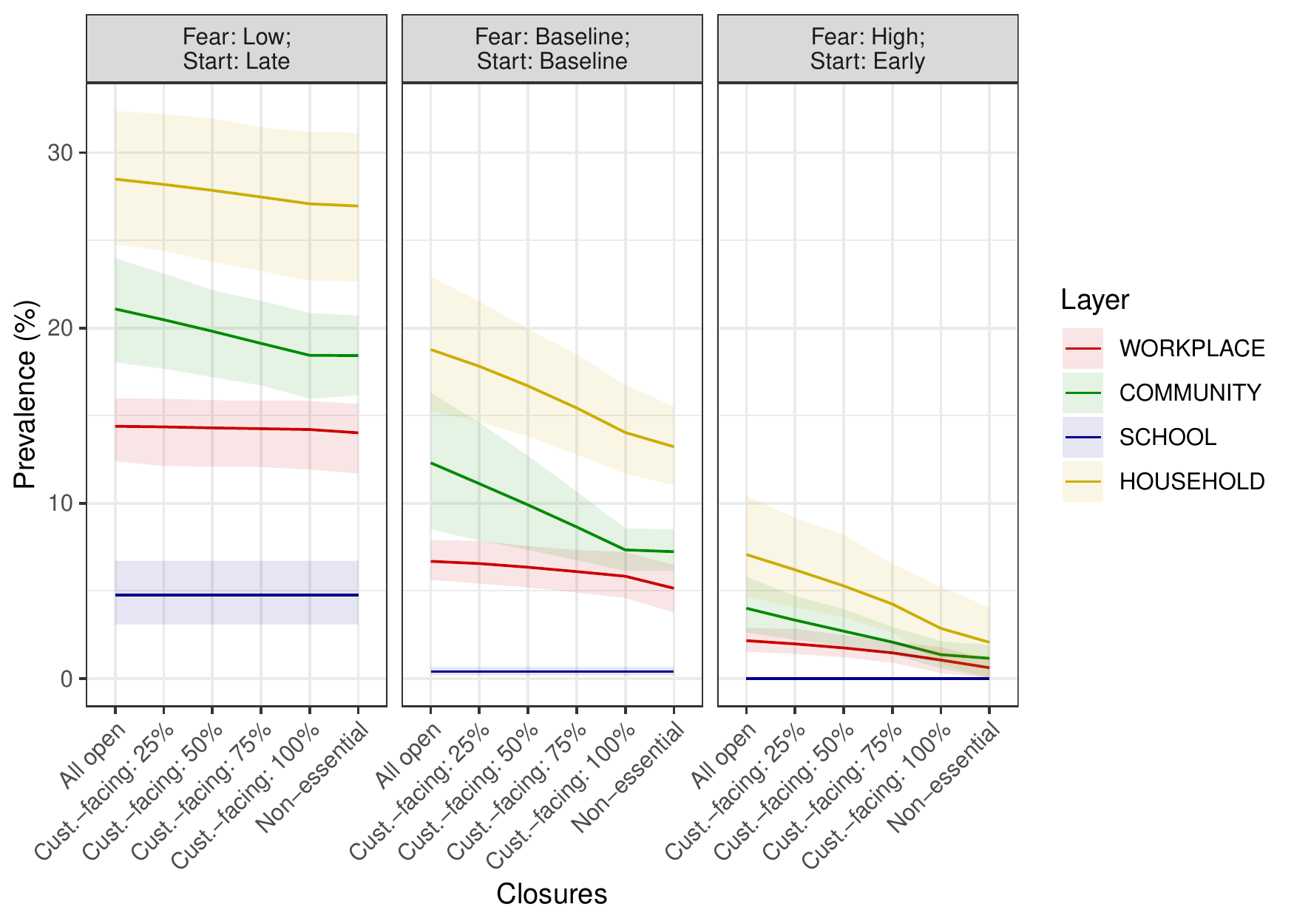}
    \caption{\textbf{Prevalence by origin across counterfactual scenarios.} We select three combinations of fear of infection and start of protective measures, and for each of these combinations we study all six possibilities for closures of economic activities. Prevalence of infections of a given origin is defined as the percentage of all individuals in the population that have been infected in the corresponding layer at the end of the simulation period. We distinguish between workplace, community, school and household infections.  Error bands indicate 2.5-97.5 percentiles.} 
        \label{fig:supp_counterfactuals_disaggregate_infections_by_origin}
\end{figure}

Figure \ref{fig:supp_counterfactuals_disaggregate_infections_by_origin} shows the origin of infections across the same scenarios as in Figure \ref{fig:supp_counterfactuals_disaggregate_infections_by_income}. In all cases the main contribution is from the household layer and the smallest contribution is from the school layer. Community infections are larger than workplace infections, although the difference is larger when customer-facing industries are open.

\begin{figure}[htbp]
    \centering
\includegraphics[width = 0.6\textwidth]{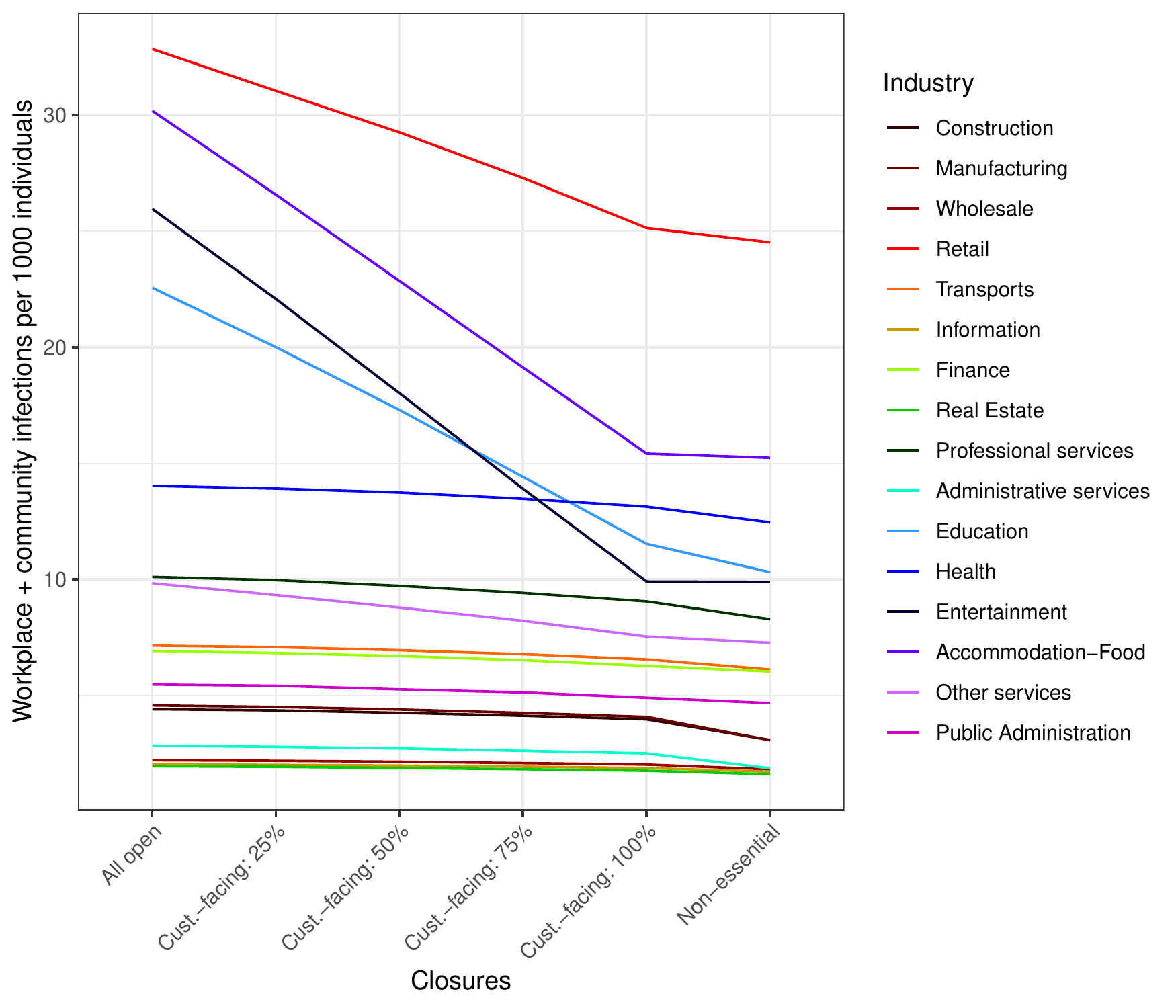}
    \caption{\textbf{Infections by industry of infection across counterfactual scenarios.}  We focus on the empirical scenario to select fear of infection (baseline) and the start of protective measures (baseline), and we study all six possibilities for closures of economic activities. We show the number of community and workplace infections per 1000 individuals that took place in any given industry for each scenario. } 
        \label{fig:supp_counterfactuals_disaggregate_infections_by_industry_of_infection}
\end{figure}

In line with this result, Figure \ref{fig:supp_counterfactuals_disaggregate_infections_by_industry_of_infection} shows that infection rates decline the most in retail, accommodation-food, education and entertainment when customer-facing industries are closed, although these industries remain the main overall source of infections.  

\begin{figure}[htbp]
    \centering
\includegraphics[width = 0.9\textwidth]{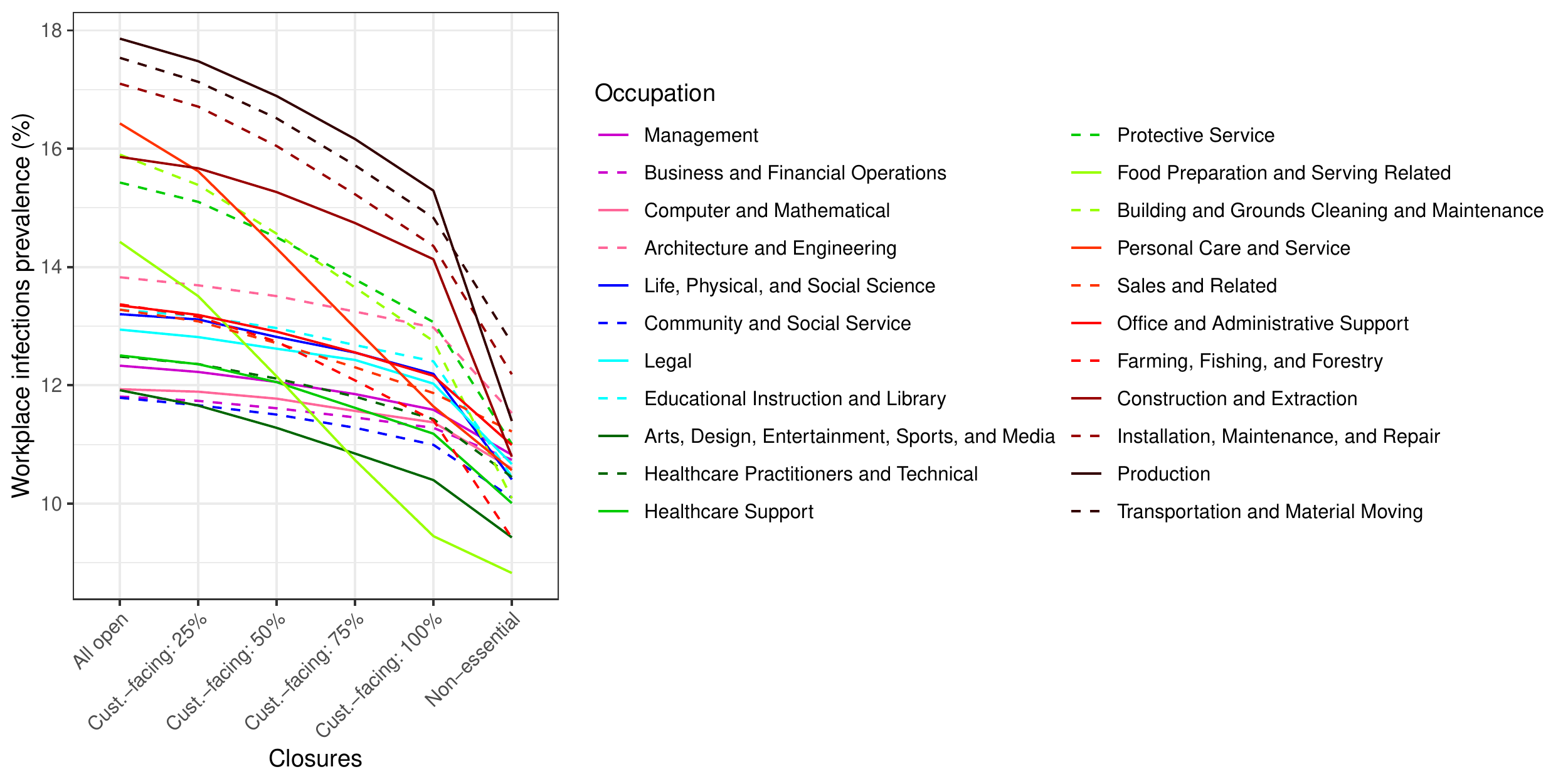}
    \caption{\textbf{Workplace infections prevalence by occupation across counterfactual scenarios.}  We focus on the empirical scenario to select fear of infection (baseline) and the start of protective measures (baseline), and we study all six possibilities for closures of economic activities. Workplace infections prevalence is defined as the percentage of all workers of a given occupation that have been infected in the workplace at the end of the simulation period.  To enhance readability, we consider combinations of color and line type, so there is nothing specific to occupations represented by dashed vs. solid lines.} 
        \label{fig:supp_counterfactuals_disaggregate_infections_by_occupation}
\end{figure}

Finally, Figure \ref{fig:supp_counterfactuals_disaggregate_infections_by_occupation} shows workplace infections by occupation. We see that production, installation, maintenance and repair, transportation and material moving occupations are among the most hit, as these occupations cannot be performed from home. Closing down construction and manufacturing, as in the ``non-essential'' scenarios, reduces infection rates among these occupations the most.

\FloatBarrier

\printbibliography[title=Methods and supplementary references,filter=appendixOnlyFilter]

\end{document}